\documentclass[amsmath,superscriptaddress, amssymb,aps,prd,reprint,longbibliography,nofootinbib,notitlepage,onecolumn,10pt]{revtex4-2}
\usepackage{natbib}

\RequirePackage[colorlinks,citecolor=blue,urlcolor=magenta,linkcolor=blue]{hyperref}
\usepackage{graphicx}% Include figure files
\usepackage{enumerate}
\usepackage[shortlabels]{enumitem}
\usepackage{url}
\usepackage{xcolor}
\usepackage{graphicx}
\usepackage{setspace}
\usepackage{dutchcal}
\usepackage{tabularx}
\usepackage{scalerel}
\usepackage{braket}
\usepackage{tikz}
\usepackage{tikz-cd}
\usetikzlibrary{svg.path}

\newcommand{\rd}{{\rm d}}

\newcommand{\e}{{\rm e}}

\usepackage{moreenum}
\usepackage[utf8]{inputenc}

\labelformat{equation}{Eq.~(#1)} 
\labelformat{figure}{Fig.~#1} 
\labelformat{subfigure}{Fig.~\thefigure#1} 
\labelformat{table}{Tab.~#1} 
\labelformat{appendix}{Appendix #1}
\definecolor{orcidlogocol}{HTML}{A6CE39}
\tikzset{
  orcidlogo/.pic={
    \fill[orcidlogocol] svg{M256,128c0,70.7-57.3,128-128,128C57.3,256,0,198.7,0,128C0,57.3,57.3,0,128,0C198.7,0,256,57.3,256,128z};
    \fill[white] svg{M86.3,186.2H70.9V79.1h15.4v48.4V186.2z}
                 svg{M108.9,79.1h41.6c39.6,0,57,28.3,57,53.6c0,27.5-21.5,53.6-56.8,53.6h-41.8V79.1z M124.3,172.4h24.5c34.9,0,42.9-26.5,42.9-39.7c0-21.5-13.7-39.7-43.7-39.7h-23.7V172.4z}
                 svg{M88.7,56.8c0,5.5-4.5,10.1-10.1,10.1c-5.6,0-10.1-4.6-10.1-10.1c0-5.6,4.5-10.1,10.1-10.1C84.2,46.7,88.7,51.3,88.7,56.8z};
  }
}

\newcommand\orcidicon[1]{\href{https://orcid.org/#1}{\mbox{\scalerel*{
\begin{tikzpicture}[yscale=-1,transform shape]
\pic{orcidlogo};
\end{tikzpicture}
}{|}}}}

\usepackage{hyperref}

\usepackage{bm}% bold math
\usepackage{xcolor}
\usepackage{bm}
\usepackage{slashed}
\usepackage{xparse}
\usepackage{dsfont}
\usepackage[mathscr]{euscript}
\usepackage{float}
\usepackage{etoolbox}
\patchcmd{\section}
  {\centering}
  {\raggedright}
  {}
  {}
\patchcmd{\subsection}
  {\centering}
  {\raggedright}
  {}
  {}
\DeclareUnicodeCharacter{2212}{-}
\allowdisplaybreaks
\begin{document}
\title{Exact path integrals on half-line in quantum cosmology with a fluid clock and aspects of operator ordering ambiguity}
\author{Vikramaditya Mondal}
\email{vikramaditya.academics@gmail.com}
\affiliation{School of Physical Sciences, Indian Association for the Cultivation of Science, Kolkata-700032, India}

\author{Harkirat Singh Sahota}
\email{harkirat221@gmail.com}
\affiliation{Department of Physical Sciences, Indian Institute of Science Education \& Research (IISER) Mohali, Sector 81 SAS Nagar,
Manauli PO 140306 Punjab India}
\affiliation{Department of Physics,  Indian Institute of Technology Delhi, Hauz Khas, New Delhi, 110016, India.}

\author{Kinjalk Lochan}
\email{kinjalk@iisermohali.ac.in}
\affiliation{Department of Physical Sciences, Indian Institute of Science Education \& Research (IISER) Mohali, Sector 81 SAS Nagar,
Manauli PO 140306 Punjab India}

% Force line breaks with  \nonumber\\
%\thanks{}%
%\date{\today}

%\doublespacing

\begin{abstract}
    We perform \textit{exact} half-line path integral quantization of flat, homogeneous cosmological models containing a perfect fluid acting as an internal clock, in a $D+1$ dimensional minisuperspace setup. We also discuss certain classes of operator ordering ambiguity inherent in such quantization procedures and argue that a particular ordering prescription in the quantum theory can preserve two symmetries, namely arbitrary lapse rescalings and general covariance, which are already present at the classical level. As a result of this imposition, a large class of quantum Hamiltonians differing by operator ordering produces the same inner products between quantum states. This imposition of the two symmetries of the classical minisuperspace models leads to a unique prescription for writing the quantum Hamiltonian for minisuperspace dimension $D>2$. Interestingly, in the case of $D=1$, the lapse rescaling symmetry is lost in the quantum theory, leading to an essentially ambiguous description of the canonical theory. We provide general proof of this in the context of both canonical quantization and path integrals. We supply concrete examples to validate our findings further.
\end{abstract}

\maketitle

%%%%%%%%%%%%%%%%%%%%%%%%%%%%%%%%%%%%%%%%%%%%%%%%%%%%%%%%%%%%%%%%%%%%%%%%%%%%%%%%%%%%%%%%%%%%%%%%%%%%%%%%%%%%%%%%%%%%%%%%%%%%%%%%%%%%%%%%%%%%%%%%%%%%%%%
{\singlespacing\tableofcontents}

\section{Introduction}

The laws of quantum mechanics are expected to play an important role in the origin of our Universe \cite{Lemaitre:1931zzb,Bojowald_2015}. The program of quantization of the Universe suffers---primarily due to the absence of a complete quantum theory of gravity \cite{Carlip:2001wq}---from many technical and conceptual difficulties, such as the description of the unitary quantum evolution of its states in time (the problem of time) \cite{Kuchar:1991qf,Isham:1992ms}, operator ordering ambiguity in passing from the classical to the quantum theory \cite{PhysRevD.37.888,Halliwell:1988wc,Moss:1988wk,PhysRevD.59.063513,Steigl:2005fk,PhysRevD.89.083510,He:2014kwa,He:2015wla,Haga_2017,PhysRevD.100.046008,Sahota:2021oid,Sahota:2022qbc,Sahota:2023vnm,Sahota:2023uoj,Sahota:2023kox,He:2022ekw}, resolution of the Big Bang singularity \cite{Thebault:2022dmv}, etc. There are, however, proposals for resolving the aforementioned issues in various ways. A particular approach suggests introducing a perfect fluid acting as the internal clock which can then describe the unitary evolution of the state of the Universe \cite{Lapchinsky:1977vb,Alvarenga:2001nm,PhysRevLett.108.141301}. The singularity resolution can be expressed in terms of the DeWitt criteria that the wave function of the universe vanishes at its singular configuration \cite{DeWitt_1967}. These avenues have been explored in the context of the canonical quantization approach. The present article aims to discuss the exact computation of path integrals in various early universe scenarios and connect the corresponding results with that obtained from the canonical formulation. Path integral quantization is a fascinating alternative to the canonical quantization approach, originally observed by Dirac \cite{Dirac:1933xn}, and later developed into a full-fledged formalism by Feynman \cite{RevModPhys.20.367}. The formalism's undeniable success rests on its vast applications from elementary quantum mechanics to the quantum field theories, such as non-Abelian gauge field theories. The application of the path integral quantization method to a gravitational field \cite{Teitelboim:1981ua,Teitelboim:1983fh,PhysRevD.28.310,PhysRevD.28.297,hartle1984path,PhysRevD.34.2323}, and more specifically to the Universe \cite{Barvinsky:1986qn,Barvinsky:1986bt,Halliwell:1988wc,Halliwell:1990qr} has been known for a long time. However, several subtleties associated with both canonical and path integral formalisms are scarcely discussed in the quantum cosmology literature, which we elucidate upon in this work.

Our central findings are as follows: it can be shown within both canonical and path integral formalisms that large classes of operator ordering ambiguities of the quantum Hamiltonian constraint do not affect the physical predictions of the theory---such as the inner product between states, quantum correlators, probability measure, etc.---if (a) there is a reference fluid working as the intrinsic clock degree of freedom describing the unitary evolution of the quantum state of the universe, (b) apart from the clock and the scale factor of the universe, there is at least two additional dynamical degree of freedom in the minisuperspace, such as a scalar field, that is, the dimension of the minisuperspace (excluding the clock) is greater than $2$, (c) the quantum Hamiltonian is written in a generally covariant form (the Laplace-Beltrami form), and (d) the potential receives a unique and well-defined quantum correction of the order $\mathcal{O}(\hbar^2)$ depending on the underlying geometry of the minisuperspace through the Ricci curvature scalar such that the quantum Hamiltonian constraint is preserved under arbitrary lapse rescalings. We provide proof of these results in both canonical and path integral formalisms.

Our motivation for the present paper is to be mindful of some of the subtleties, involved with the quantization program, which we describe in the following:

\begin{enumerate}[(a)]
    \item Path integrals in the context of quantum cosmology are generally computed in the semi-classical approximation using the steepest descent method due to analytical challenges associated with computing oscillatory integrals (see, for example, \cite{PhysRevD.39.2206}, and more recently \cite{PhysRevD.95.103508}). However, we show here that quite a few simpler cosmological scenarios exist in which exact path integration can be performed. We explicitly show the computations for the cases of a flat, homogeneous universe containing a perfect fluid with an equation of state parameter $\omega\neq 1$. We consider both the isotropic FLRW (Friedmann–Lema\^itre–Robertson–Walker) and the anisotropic Bianchi I cases. Moreover, we can also include a free massless scalar field. In these cases, the path integral is not of the Gaussian form, presenting technical challenges in computing the path integrals exactly. However, we get a handle on this issue using techniques available in the path integral literature of dealing with non-Gaussian integrals using coordinate transformations \cite{Steiner:1984uf,Grosche:1987ba,Grosche:1987gh,Grosche:1998yu}.
     \item Due to Hamiltonian constraint for a generally covariant system, quantum gravity has a problem of time, that is, the ``Scr\"odinger equation" for quantum gravity, which is called the Wheeler-DeWitt equation is timeless---in explicit notation, this reads as $\hat{\mathscr{H}}\Psi=0$ with no extrinsic time evolution. In the absence of time, it then becomes difficult to describe the unitary evolution of the universe with respect to time or even properly define the Hilbert space of states and interpret probabilities. Among many proposed solutions \cite{Hoehn:2019fsy}, we consider the relational approach to the problem of time, wherein a perfect fluid is introduced as a reference fluid, which acts as the clock degree of freedom \cite{Kuchar:1990vy,Kuchar:1991pq,Kuchar:1992au,PhysRevD.51.5600}. The fluid acts as an internal clock for the system against which the unitary evolution of quantum states of the universe can be described and this framework can be used to study early universe physics \cite{Peter:2005hm,Peter:2006hx,Peter:2008qz,Bergeron:2017ddo,Malkiewicz:2020fvy,Martin:2021dbz,Maniccia_2022,Martin_2024,Maniccia_2024}. %Then, the evolution of the universe's wave function can be described with respect to the time kept by this internal clock, and therefore, the aforementioned problems are avoided.  \dots
    \item The operator ordering ambiguity is at the center of the canonical approach to the quantization of the gravitational systems. Many previous works \cite{PhysRevD.37.888,Halliwell:1988wc,Moss:1988wk,PhysRevD.59.063513,Steigl:2005fk,PhysRevD.89.083510,He:2014kwa,He:2015wla,Haga_2017,PhysRevD.100.046008,Sahota:2021oid,Sahota:2022qbc,Sahota:2023vnm,Sahota:2023uoj,Sahota:2023kox,He:2022ekw} discuss this issue in some details, however, similar discussion in the context of path integral quantization of cosmological models is lacking. Many works evaluate quantum cosmological path integrals semiclassically, and in the limit $\hbar\to0$, the difference between different ordering choices becomes immaterial. Since in quite a few cases, we are able to perform the path integral exactly, we discuss how the ordering ambiguities manifest in the path integral quantization. Following \cite{Kiefer_2019}, a generalized ordering of the kinetic differential operator in the Wheeler-DeWitt equation is considered, where the ordering ambiguity is parameterized by $p$ whose different values correspond to different quantum theories. We explore the question of what definition for the Feynman path integral, appropriate for the context of minisuperspace of quantum cosmology, should be used such that the quantization is equivalent to that of the one-parameter family of Wheeler-DeWitt equations in the canonical approach.
    
    \item In the minisuperspace quantization of cosmological models, the scale factor $a(t)$ of the universe plays the role of dynamical variable $x(t)$ in ordinary classical/quantum mechanics. However, a significant difference lies in the fact that the physicality of the scale factor requires $a(t)>0$, as negative values for the scale factor are not allowed classically and have no physical interpretation. In the path integral formulation of quantum cosmology, generally, this restriction is not taken into account due to analytical difficulty (for example, see section IX in \cite{Halliwell:1988wc}). However, in the present work, we remain mindful of such restrictions. Half-line quantization has many subtleties associated with it; for example, the self-adjointness of differential operators, such as the momentum and kinetic energy operators, has to be dealt with carefully due to the restricted domain of the scale factor to the half-line.
    
    \item It is well known that the minisuperspace action or Hamiltonian for homogeneous cosmologies resembles those of a particle moving in a curved (pseudo)Riemannian manifold. Path integral quantization of such non-linear systems has to be dealt with carefully. Depending on the choice of lattice splitting in the discretized definition of the path integral, it is essential to add $\mathscr{O}(\hbar^2)$ correction to the quantum mechanical potential such that a legitimate ordering of the corresponding Hamiltonian is obtained (see, for example, \cite{Grosche:1987ba,Grosche:1998yu,Gervais:1976ws}).
    \item Another crucial question to address in the path integral formulation is that of the boundary condition to be used in defining the path integral, which in recent times has given rise to quite a bit of controversy in the context of Lorentzian quantum cosmology (LoQC) \cite{feldbrugge2017lorentzian,feldbrugge2017no,dorronsoro2017real,dorronsoro2018damped,vilenkin2018tunneling,feldbrugge2018no,feldbrugge2018inconsistencies,vilenkin2019tunneling}. In our case, for half-line quantization, it is convenient to use the Dirichlet boundary condition, which naturally leads to the DeWitt criteria---vanishing of the wave function at the singular point ($a = 0$) of the configuration space---a condition often considered for the quantum resolution of the big bang singularity as such configurations now have a vanishing probability to occur.
\end{enumerate}

In what follows, we write down the general Wheeler-DeWitt equation for a $D+1$ dimensional minisuperspace, where $D$ denotes the dimension of the minisuperspace without any clock, and $+1$ corresponds to the reference fluid acting as a clock. We then classify the possible ambiguities in writing the quantum Hamiltonian operator into three categories. We show, following \cite{Halliwell:1988wc,Moss:1988wk}, that if a particular class of the ambiguity that appears as an order $\mathcal{O}(\hbar^2)$ correction to the superpotential in the Wheeler-DeWitt equation is chosen to be proportional to the Ricci curvature scalar of the minisuperspace (with the proportionality constant being $+\frac{(D-2)}{8(D-1)}$), then the quantum theory becomes conformally invariant in the sense that the inner product and the quantum correlators do not get affected by arbitrary lapse function rescalings. When conformal and general covariance, which are symmetries of the classical theory, are both present in the quantum theory as well then (for $D> 2$) the other classes of the ambiguities in the quantum Hamiltonian do not affect the physical predictions. Following the proof in the canonical formalism, we also present proof of this result in the path integral approach. For the case of $D=1$, these symmetry-guided arguments fail and operator ambiguities become an essential feature of the quantum theory. Our work is different from \cite{Halliwell:1988wc,Moss:1988wk} in two key ways: that we have an additional clock in the system which allows us to define unitary evolution of the quantum states and positive definite probability measure, and we also extend the proofs in the context of path integral as well.

Thereafter, we proceed to give concrete examples of exactly solvable cases. We provide examples in $D=1$ and $D=4$ cases, where the exact path integral computations with careful considerations of all the $\mathcal{O}(\hbar^2)$ corrections from various sources lead to the stationary states that solve the corresponding Wheeler-DeWitt equations. We also discuss the fact that the inner products do indeed depend on the ambiguity parameter in the case $D=1$, whereas in $D=4$ case, due to conformal invariance the information of the ambiguity parameter washes away from the inner products and quantum correlators.

The paper is organized as follows. We start with the discussion on the general formalism for the quantization of a minisuperspace model of the Universe with a reference fluid, canonical quantization in Sec. \ref{sec:QGF}, and the path integral formalism in Sec. \ref{sec:QGF2}. Relating the lapse rescaling of the minisuperspace with a class of generalized ordering scheme for the WDW equation in \ref{sec:CQG}; we show that the minisuperspace of one-dimension with reference fluid suffers from ordering ambiguity, while the higher dimensional systems are impervious to it. Using the product form prescription for the path integral, we derive the propagator for the one-dimensional minisuperspace of flat-FLRW universe with a reference fluid in Sec. \ref{sec:ESC2} and Bianchi I universe with massless scalar field and a reference fluid in Sec. \ref{sec:ESC2}, taking care of various caveats discussed above.

\section{Canonical Quantization: general formalism}\label{sec:QGF}

Before going into specific examples, we develop a general formalism that will work for a wide variety of problems. After the formalism is developed, we shall consider the specific cases of quantization of a flat FLRW and a Bianchi I spacetime containing a perfect fluid as a reference clock.

\subsection{Canonical quantum gravity and Wheeler-DeWitt equation}\label{sec:QGF1}

The configuration space of general relativity, called the superspace---space of all 3-metrics (modulo the gauge invariant configurations) and of matter field configurations---is infinite-dimensional, and as a result, the quantization of gravity requires a field-theoretic treatment. However, such a problem is not tractable and one makes considerable symmetry assumptions. In the case of quantum cosmology, the homogeneity assumption freezes most of the degrees of freedom, and only finitely many degrees of freedom remain. Such a symmetry-reduced finite-dimensional configuration space is called the minisuperspace. In the classical picture, the evolution of the universe in time is represented by a parametrized curve in the minisuperspace, with each point in space representing the configuration of spacetime and matter fields of the universe at an instant. Thus, the description of the evolution of a homogeneous universe is effectively that of a mechanical particle moving in the minisuperspace, which is a finite-dimensional (pseudo)Riemannian manifold of, say, dimension $D$. Therefore, the corresponding quantum problem is that of doing quantum mechanics for a particle moving on a curved manifold.

In the minisuperspace model of quantum cosmology, the first-order action assumes the form \cite{Halliwell:1988wc,Teitelboim:1981ua}
\begin{align}
    S = \int \rd t \left[p_{A} \dot{q}^A - N \mathscr{H}(\boldsymbol{q},\boldsymbol{p})\right],
\end{align}
after the integration over the spatial coordinates has been carried out. Here, $q^A$ are the minisuperspace coordinates consisting of dynamical variables of the 3-geometry and matter fields. The index $A$ runs from $1$ up to $D$, the dimension of minisuperspace. For example, in the homogeneous and isotropic FLRW spacetime with a scalar field, the minisuperspace coordinates are the scale factor of the universe $a(t)$, and homogeneous scalar field configuration $\phi(t)$, with $D$ being $2$. The conjugate momenta corresponding to the coordinates $q^A$ is denoted by $p_{A}$ and $\mathscr{H}$ is the Hamiltonian of the system. The lapse function $N$ appears as a Lagrange multiplier of the theory and its variation leads to the well-known Hamiltonian constraint $\mathscr{H} = 0$. As Dirac prescribes that the quantum operator corresponding to the first class constraints should annihilate the physical wave function \cite{dirac2001lectures}, one arrives at the Wheeler-DeWitt equation $\hat{\mathscr{H}}\Psi = 0$ \cite{DeWitt_1967}. Two problems become apparent immediately---(a) the Hamiltonian annihilates the wave function owing to the time reparameterization invariance of the system and thus there is no extrinsic time parameter to describe the evolution in the quantum paradigm; (b) the form of the operator $\hat{\mathscr{H}}$ may have ordering ambiguity if in the classical Hamiltonian $\mathscr{H}(\boldsymbol{q},\boldsymbol{p})$ the conjugate variables $q^A$ and $p_{A}$ appear in product form.

To address the first issue, we deparametrize the system by adding a perfect fluid which will play the role of intrinsic time in the quantum theory \cite{Kuchar:1991qf,Isham:1992ms} (and thus enhancing the minisuperspace dimension to $D+1$). However, in the following discussions, we shall keep referring to $D$ as the dimension of the minisuperspace, even though we mean by it the dimension of the \textit{physical minisuperspace}: the space of non-clock degrees of freedom. The Hamiltonian of the fluid is linear in its momentum, and the action for the perfect fluid dominated universe takes the form (see, for example, \cite{Lapchinsky:1977vb,Alvarenga:2001nm})
\begin{align} \label{eq:mini_action_generic}
    S = \int \rd t \left[p_{A} \dot{q}^A + p_{T} \dot{T} - N \left(\mathscr{H}(\boldsymbol{q},\boldsymbol{p}) + p_{T} \right)\right].
\end{align}
In appendix \ref{sec:schutz_fluid}, we shall provide more details on how the fluid Hamiltonian comes about. Then the new Hamiltonian constraint becomes $\mathscr{H}+p_{T} = 0$ and leads to the new Wheeler-DeWitt equation
\begin{align}\label{eq:Schrodinger_generic}
     i \hbar \frac{\partial\Psi}{\partial T} = \hat{\mathscr{H}} \Psi,
\end{align}
where we have used the identification $p_{T}\to-i\hbar\frac{\partial}{\partial T}$, and $\mathscr{H}\to \hat{\mathscr{H}}$. Due to the linear dependence of the Hamiltonian of the fluid on its momentum, the fluid variable $T$ now acts as the time parameter for the evolution of the wave function of the universe. The precise form of the quantum Hamiltonian operator $\hat{\mathscr{H}}$ depends on the system, and it generates the translations in the fluid variable $T$. Therefore, ensuring the self-adjointness of this Hamiltonian implies a unitary evolution with respect to the fluid clock. Moreover, as mentioned before, the form of the operator $\hat{\mathscr{H}}$ can have ordering ambiguities, and we discuss this issue in the following.

We note that `fluid' is an effective description of matter, and at a more fundamental level, all matter should be described as renormalizable quantum fields. Therefore, the above treatment works insofar as the Schutz fluid description (see appendix \ref{sec:schutz_fluid}) is valid, which may break at high energies. However, we would like to emphasize that the precise nature of the fluid description is not essential for our treatment. What is required is that we have a degree of freedom whose Hamiltonian can be brought to have a linear dependence on its momentum using canonical transformation, such that upon quantization, this degree of freedom can play the role of a clock. A perfect fluid is an ideal demonstration of this principle; however, by no means is it a unique choice for the clock, and other degrees of freedom may play a similar role. The formalism discussed here applies to any degree of freedom with Hamiltonian depending linearly on the momentum. Furthermore, in minisuperspace approximation, it is always assumed that relevant energy scales are much below the Planck scale so that arbitrary metric fluctuations (e.g. inhomogeneities) can be ignored or treated perturbatively.

\subsection{Breaking of the lapse rescaling symmetry as operator ordering ambiguity}

In quantum cosmology, we primarily encounter the Hamiltonian of the quadratic form
\begin{align} \label{eq:mini_hamiltonian_generic}
    \mathscr{H}(\boldsymbol{q},\boldsymbol{p}) = \frac{1}{2}\mathscr{G}^{AB}(\boldsymbol{q}) p_{A} p_{B} + U(\boldsymbol{q}),
\end{align}
where $\mathscr{G}_{AB}$ is the minisuperspace metric with hyperbolic signature $(-++\cdots)$ and $\mathscr{G}^{AB}$ is its inverse. A function of the minisuperspace coordinates $U(\boldsymbol{q})$ denotes the superpotential. The corresponding quantum Hamiltonian has an operator ordering ambiguity \cite{PhysRevD.59.063513,Steigl:2005fk} due to the fact that all the quantum Hamiltonians of the following form describe inequivalent quantum theories but correspond to the same classical theory in the $\hbar\to0$ limit:
%%%%%%%%%%%%%%%%%%%%%%%%%%%%%%%%%%%%%%%%%%%%%%%%%%%%%%%%%%%%%%%%%%%%%%%
\begin{align} \label{eq:ambiguous_hamiltonian}
    \hat{\mathscr{H}} = - \frac{\hbar^2}{2} \frac{1}{F_{1}(\boldsymbol{q})F_{2}(\boldsymbol{q})} \nabla_{A} \mathscr{G}^{AB} F_{1}(\boldsymbol{q})\nabla_{B}F_{2}(\boldsymbol{q}) + U(\boldsymbol{q}) + \hbar^2 F_{3}(\boldsymbol{q}),
\end{align}
%%%%%%%%%%%%%%%%%%%%%%%%%%%%%%%%%%%%%%%%%%%%%%%%%%%%%%%%%%%%%%%%%%%%%%%%%%%%%%%%%%%%%%%%%%
where, we have used the identification $p_{A} \to -i\hbar\nabla_{A}$, with $\nabla_{A}$ being the covariant derivative compatible with the metric $\mathscr{G}_{AB}$, and $F_{1,2,3}(\boldsymbol{q})$ are some non-vanishing arbitrary real functions of the minisuperspace coordinates $q^A$. At this stage, even though the potential term proportional to $\hbar^2$ can arise from the first term as well, we are introducing the arbitrary quantum potential term $F_3(\boldsymbol{q})$, to accommodate for the different sources that could lead to it, e.g., many authors have found that this extra term could be proportional to the Ricci scalar corresponding to the metric $\mathscr{G}_{AB}$ \cite{DeWitt:1952js,DeWitt_1957,Mclaughlin:1971td,Cheng:1972yb,Hartle:1976tp,Parker_1979,Halliwell:1988wc,Grosche:1987ba}. It would be convenient for us to call the ambiguities characterized by the functions $F_{1}(\boldsymbol{q})$, $F_{2}(\boldsymbol{q})$, and $F_{3}(\boldsymbol{q})$ as \textit{conformal-class}, \textit{factor-class}, and \textit{potential-class} ambiguities, respectively. The reasons for adopting this nomenclature are as follows: we shall show below that (a) the ambiguity created by $F_{1}$, can be identified with the freedom of lapse rescaling in the classical theory, (b) the wave function can be suitably rescaled so that the function $F_{2}$ is factored out and this overall factor in the wave function will not affect the inner products, and (c) $F_{3}$ acts as a quantum correction to the potential.

Suppose that the Wheeler-DeWitt equation \ref{eq:Schrodinger_generic} with the Hamiltonian \ref{eq:ambiguous_hamiltonian} is solved by the wave function $\Psi$ when $F_{2} = 1$, then it can be shown that a new wave function $\Psi/F_{2}$ solves the Wheeler-DeWitt equation when $F_{2}\neq 1$. We shall later show that this difference between the two wave functions by an overall factor does not affect the inner products as the factor gets canceled by the integration measure. This measure has to be chosen by demanding the self-adjointness of the quantum Hamiltonian \ref{eq:ambiguous_hamiltonian}. As a result of this simplification, we shall ignore the ambiguity introduced by $F_{2}$ (by setting $F_{2}=1$), and rather focus on the other classes of ambiguity---the conformal-class and potential-class. In the following, we argue that the ambiguity introduced by $F_{1}$ can be understood in a different way as well. We shall start by writing a generally covariant version of the quantum theory corresponding to the classical Hamiltonian \ref{eq:mini_hamiltonian_generic}, and then show that the arbitrary lapse rescaling freedom in the classical theory is responsible for the ambiguity of the form generated by $F_{1}$.

From the action \ref{eq:mini_action_generic} it is clear that the classical minisuperspace model has two very crucial symmetries---general covariance with respect to the minisuperspace metric\footnote{The general covariance is under the arbitrary (minisuperspace) coordinate transformations $q^A \to \tilde{q}^A(q^B)$. These transformations are restricted so as not to violate the assumption of homogeneity of the universe.} and conformal invariance under arbitrary lapse rescalings $N\to \tilde{N} \Omega^{-2}$. Notice that under the lapse rescaling, the classical Hamiltonian changes by an overall factor, and the classical Hamiltonian constraint remains unchanged. This lapse rescaling is equivalent to time reparameterization $\rd t \to \Omega^{-2} \rd t$, and thus the symmetry under lapse rescaling is equivalent to the time reparametrization invariance. Note that the factor $\Omega(\boldsymbol{q},\boldsymbol{p})$ can, in principle, be any arbitrary non-vanishing function of the phase space since recall from the form of the action in \ref{eq:mini_action_generic}, that we are working with the first order form of the action which depends on both $\boldsymbol{q}$ and $\boldsymbol{p}$. However, in the following, we restrict ourselves to the case where $\Omega(\boldsymbol{q})$ is only a function of the configuration space since this form will be useful for our discussions on the operator ordering ambiguities.

% In the following, we argue that if the quantum theory has to retain both of these symmetries, then in the minisuperspaces of dimension $D\geq 2$, the form of the quantum Hamiltonian operator is unique, and there is no operator ordering ambiguity. In $D=1$, however, the conformal invariance cannot be retained in the quantum theory and the operator ordering ambiguity will persist.

The canonical quantization of the Hamiltonian of the form \ref{eq:mini_hamiltonian_generic} is analogous, as discussed above, to the quantization of a mechanical particle on a $D$-dimensional (pseudo-)Riemannian manifold \cite{DeWitt_1957,DeWitt:1952js,Omote:1972ba,audretsch1978quantum,Parker_1979,Parker_1980_PRD,Parker_1980_PRL}. Using a generally covariant form for the quantum Hamiltonian corresponding to \ref{eq:mini_hamiltonian_generic}, we can write the Wheeler-DeWitt equation as follows
\begin{align}\label{eq:quantum_Hamiltonian_generic}
    \left( - \frac{\hbar^2}{2} \Delta_{\sf LB} + U + \hbar^2 \xi \mathcal{R}\right)\Psi = i\hbar \frac{\partial}{\partial T}\Psi,
\end{align}
where $\mathcal{R}$ is the Ricci scalar defined with the minisuperspace metric $\mathscr{G}_{AB}$ and the Laplacian $\Delta_{\sf LB}$ is the Laplace-Beltrami operator on the curved manifold defined as
\begin{align}
    \Delta_{\sf LB} & = \frac{1}{\sqrt{|\mathscr{G}|}} \partial_{A} \sqrt{|\mathscr{G}|} \mathscr{G}^{AB} \partial_{B}.
\end{align}
Due to the generally covariant form of the quantum Hamiltonian \ref{eq:quantum_Hamiltonian_generic}, the quantum theory remains equivalent under point canonical transformations (coordinate transformations in the minisuperspace) of the form $q^A \to \tilde{q}^A(q^B)$ \cite{DeWitt:1952js,DeWitt_1957,Omote:1972ba}. Notice that the form of the potential-class term has been restricted to curvature invariants because of the requirement of invariance under the point canonical transformation. We have made the simplest choice for this invariant to be the Ricci scalar, consistent with prior works \cite{DeWitt:1952js,DeWitt_1957,Mclaughlin:1971td,Cheng:1972yb,Hartle:1976tp,Parker_1979,Halliwell:1988wc,Grosche:1987ba}.  However, this choice has not yet removed the ambiguity from potential-class terms fully as, at this moment, as the constant $\xi$ is completely arbitrary. General covariance alone cannot resolve this ambiguity. The additional conformal invariance under lapse rescalings will fix the constant $\xi$, which we will return to at the end of this section. Note that choosing to respect the classical symmetry under the point canonical transformations $q^A \to \tilde{q}^A$ naturally leads to Laplace-Beltrami ordering for the kinetic differential operator; however, this also does not completely fix the operator ordering, or the conformal-class ambiguity, due to the freedom existing associated with lapse rescalings.

Consider the following transformation of the lapse function $N\to \tilde{N} \Omega^{-2}(q)$. This changes the classical action of the minisuperspace \ref{eq:mini_action_generic} to the following
%%%%%%%%%%%%%%%%%%%%%%%%%%%%%%%%%%%%%%%%%%%%%%%%%%%%%%%%%%%%%%%%%%%%%%%%%%%%%%%%%%%%%%%%%%%%%%%%%%%%%%%%%%%%%%%%%%%%%%%%%%%%%%%%%%%%%%%%%%%%%%
\begin{align}\label{eq:generic_conformal_rescaled_action}
    \tilde{S} = \int \rd t \left[p_{A}\dot{q}^{A} + p_{T}\dot{T} -\tilde{N}\left(\frac{1}{2} \tilde{\mathscr{G}}^{AB}p_{A}p_{B}+\tilde{U}(q)+\Omega^{-2} p_{T}\right)\right],
\end{align}
%%%%%%%%%%%%%%%%%%%%%%%%%%%%%%%%%%%%%%%%%%%%%%%%%%%%%%%%%%%%%%%%%%%%%%%%%%%%%%%%%%%%%%%%%%%%%%%%%%%%%%%%%%%%%%%%%%%%%%%%%%%%%%%%%%%%%%%%%%%%%%%%%%%%%%%
where we have defined $\tilde{\mathscr{G}}^{AB}=\Omega^{-2}\mathscr{G}^{AB}$, and $\tilde{U}=\Omega^{-2}U$. Here, notice that restricting $\Omega(\boldsymbol{q})$ to be only a function of the configuration space (and not the full phase space) allows us to reinterpret this lapse rescaling as the Weyl transformation of the minisuperspace metric. These transformations are less general than the time reparametrization invariance, which is characterized by $\Omega(\boldsymbol{q},\boldsymbol{p})$, a function of the full phase space. As the classical theory remains unchanged by such rescalings, this new action is as good as the previous one in \ref{eq:mini_action_generic}. We could in principle start to quantize this action in \ref{eq:generic_conformal_rescaled_action} instead of the one in \ref{eq:mini_action_generic}. However, the resulting Wheeler-DeWitt equation for the new theory is
%%%%%%%%%%%%%%%%%%%%%%%%%%%%%%%%%%%%%%%%%%%%%%%%%%%%%%%%%%%%%%%%%%%%%%%%%%%%%%%%%%%%%%%%%%%%%%%%%%%%%%%%%%%%%%%%%%%%%%%%%%%%%%%%%%%%%%%%%%%%%%%%%%%%%%%
\begin{align}\label{eq:Schrodinger_conformal}
    \left(\hat{\tilde{\mathscr{H}}} + \hat{\tilde{p}}_{T}\right)\tilde{\Psi} = \left(-\frac{\hbar^2}{2\sqrt{|\tilde{\mathscr{G}}|}} \frac{\partial}{\partial q^{A}} \sqrt{|\tilde{\mathscr{G}}|} \tilde{\mathscr{G}}^{AB} \frac{\partial}{\partial q^{B}} + \tilde{U} + \hbar^2 \xi \tilde{\mathcal{R}} - i \hbar \Omega^{-2} \frac{\partial}{\partial T}\right)\tilde{\Psi}=0,
\end{align}
%%%%%%%%%%%%%%%%%%%%%%%%%%%%%%%%%%%%%%%%%%%%%%%%%%%%%%%%%%%%%%%%%%%%%%%%%%%%%%%%%%%%%%%%%%%%%%%%%%%%%%%%%%%%%%%%%%%%%%%%%%%%%%%%%%%%%%%%%%%%%%%%%%%%%%%
where we have the new minisuperspace metric as $\tilde{\mathscr{G}}_{AB} = \Omega^{2} \mathscr{G}_{AB}$, and $\tilde{\mathcal{R}}$ is the curvature scalar defined with the new metric. The conformal covariance of the classical theory seems to be lost in the quantum theory as such conformal transformations lead to inequivalent quantum Hamiltonians. Moreover, notice that the quantum theories given by both \ref{eq:quantum_Hamiltonian_generic} and \ref{eq:Schrodinger_conformal} are invariant under point canonical transformations. However, to realize that the lapse rescaling has made these theories differ by operator ordering, we rewrite the above Schr\"{o}dinger equation in terms of the old metric $\mathscr{G}_{AB}$ of the minisuperspace, as follows
%%%%%%%%%%%%%%%%%%%%%%%%%%%%%%%%%%%%%%%%%%%%%%%%%%%%%%%%%%%%%%%%%%%%%%%%%%%%%%%%%%%%%%%%%%%%%%%%%%%%%%%%%%%%%%%%%%%%%%%%%
\begin{align}\label{eq:wheelerdewitt_conformal}
    \left(-\frac{\hbar^2}{2\Omega^{D-2}\sqrt{|{\mathscr{G}}|}} \frac{\partial}{\partial q^{A}} \Omega^{D-2} \sqrt{|{\mathscr{G}}|} {\mathscr{G}}^{AB} \frac{\partial}{\partial q^{B}} + U + \hbar^2 \xi \Omega^{2} \tilde{\mathcal{R}} \right)\tilde{\Psi}=i \hbar \frac{\partial}{\partial T}\tilde{\Psi}.
\end{align}
%%%%%%%%%%%%%%%%%%%%%%%%%%%%%%%%%%%%%%%%%%%%%%%%%%%%%%%%%%%%%%%%%%%%%%%%%%%%%%%%%%%%%%%%%%%%%%%%%%%%%%%%%%%%%%%%%%%%%%%%%%%%%%%%%%%%%%%%%%%%%%%%%%%%
We see that the arbitrary factor $\Omega^{D-2}$ can be identified with $F_{1}$ and the term $\xi\Omega^2\tilde{\mathcal{R}}$ with $F_{3}$, and hence justifying the names conformal-class and potential-class, respectively. Therefore, if one constructs a quantum theory with invariance under point canonical transformations as well as the conformal invariance, the operator ordering ambiguity of the type \ref{eq:ambiguous_hamiltonian} should leave no imprint on the physical observables, i.e., this preservation of conformal invariance should manifest in the quantum theory as invariance of inner products, correlators, etc. The potential-class ambiguity term can be tuned such that the quantum theory is conformally invariant and all the quantum Hamiltonians differing by further conformal-class orderings will all predict the same inner products between the quantum states.

The fact that $\Omega^{D-2}$ can be identified with $F_{1}$ works when $D\neq2$. Therefore, in $D=2$, building a quantum theory that is conformally invariant (this requires the choice $\xi=0$) does not give any handle on the ambiguity functions $F_{1}$ and, therefore, the operator ordering ambiguity cannot be tackled based on symmetries.

\subsection{Conformal invariance and resolution of the operator ordering ambiguity}

As the breaking of the lapse rescalings symmetry in quantum theories is a source for operator ordering ambiguity in otherwise point canonical transformation invariant theories, we have to construct a quantum Hamiltonian that preserves both the symmetries. Following \cite{Halliwell:1988wc}, if we assume that under the lapse rescaling the wave function changes by a factor $\Psi\to\tilde{\Psi}=\Omega^{\gamma} \Psi$, then the Wheeler-DeWitt equation after lapse rescaling can be expressed as 
%%%%%%%%%%%%%%%%%%%%%%%%%%%%%%%%%%%%%%%%%%%%%%%%%%%%%%%%%%%%%%%%%%%%%%%%%%%%%%%%%%%%%%%%%%%%%%%%%%%%%%%%%%%%%%%%%%%%%%%%%%%%%%%%%%%%%%%%%%%%%%%%%%%%
\begin{align}
    \left(\hat{\tilde{\mathscr{H}}} + \hat{\tilde{p}}_{T}\right) \tilde \Psi = &  \Omega^{\gamma-2} \left(\hat{\mathscr{H}} + \hat{p}_{T} \right) \Psi - \frac{\hbar^2}{2} \bigg[(\gamma(\gamma + D - 3) + 2 (D-1)(D-4) \xi)\Omega^{\gamma-4}(\nabla\Omega)^2  \nonumber\\
    & + (\gamma + 4 \xi (D-1)) \Omega^{\gamma-3} (\nabla^2\Omega)\bigg]\Psi - \frac{\hbar^2}{2} (2\gamma+D-2) \Omega^{\gamma-3} (\nabla\Omega)\cdot(\nabla\Psi)
\end{align}
%%%%%%%%%%%%%%%%%%%%%%%%%%%%%%%%%%%%%%%%%%%%%%%%%%%%%%%%%%%%%%%%%%%%%%%%%%%%%%%%%%%%%%%%%%%%%%%%%%%%%%%%%%%%%%%%%%%%%%%%%%%%%%%%%%%%%%%%%%%%%%%%%%%%%%%
If we choose $\gamma = \frac{2-D}{2}$, then the choice of
%%%%%%%%%%%%%%%%%%%%%%%%%%%%%%%%%%%%%%%%%%%%%%%%%%%%%%%%%%%%%%%%%%%%%%%%%%%%%%%%%%%%%%%%%%%%%%%%%%%%%%%%%%%%%%%%%%%%%%%%%%%%%%%%%%%%%%%%%%%%%%%%%%
\begin{align}
    \xi = + \frac{D-2}{8(D-1)}, ~~~~ (D\neq 1),
\end{align}
%%%%%%%%%%%%%%%%%%%%%%%%%%%%%%%%%%%%%%%%%%%%%%%%%%%%%%%%%%%%%%%%%%%%%%%%%%%%%%%%%%%%%%%%%%%%%%%%%%%%%%%%%%%%%%%%%%%%%%%%%%%%%%%%%%%%%%%%%%%%%%%%
renders the Wheeler-DeWitt equation conformally invariant
%%%%%%%%%%%%%%%%%%%%%%%%%%%%%%%%%%%%%%%%%%%%%%%%%%%%%%%%%%%%%%%%%%%%%%%%%%%%%%%%%%%%%%%%%%%%%%%%%%%%%%%%%%%%%%%%%%%%%%%%%%%%%%%%%%%%%%%%%%%%%%%%%%%%%%
\begin{align}
    \left(\hat{\tilde{\mathscr{H}}} + \hat{\tilde{p}}_{T}\right) \tilde \Psi = &  \Omega^{\gamma-2} \left(\hat{\mathscr{H}} + \hat{p}_{T} \right) \Psi = 0.
\end{align}
%%%%%%%%%%%%%%%%%%%%%%%%%%%%%%%%%%%%%%%%%%%%%%%%%%%%%%%%%%%%%%%%%%%%%%%%%%%%%%%%%%%%%%%%%%%%%%%%%%%%%%%%%%%%%%%%%%%%%%%%%%%%%%%%%%%%%%%%%%%%%%%%%%%%%%%%%
Notice that we have a different sign for $\xi$ than what \cite{Halliwell:1988wc} suggests. On the other hand, our sign matches with \cite{Moss:1988wk}. Furthermore, the key difference here from the analysis in \cite{Halliwell:1988wc} or in \cite{Moss:1988wk} is that in our formalism we have introduced `clock' (our minisuperspace is $D+1$ dimensional), and as a result, we have well-defined notions of the unitary quantum evolution of the states of the universe and inner products between them. We see that the Hamiltonian in \ref{eq:wheelerdewitt_conformal} is hermitian if the kinetic differential operator  of the following form
%%%%%%%%%%%%%%%%%%%%%%%%%%%%%%%%%%%%%%%%%%%%%%%%%%%%%%%%%%%%%%%%%%%%%%%%%%%%%%%%%%%%%%%%%%%%%%%%%%%%%%%%%%%%%%%%
\begin{align}
    \mathbb{D}\equiv -\frac{\hbar^2}{2\Omega^{D-2}\sqrt{|{\mathscr{G}}|}} \frac{\partial}{\partial q^{A}} \Omega^{D-2} \sqrt{|{\mathscr{G}}|} {\mathscr{G}}^{AB} \frac{\partial}{\partial q^{B}},
\end{align}
%%%%%%%%%%%%%%%%%%%%%%%%%%%%%%%%%%%%%%%%%%%%%%%%%%%%%%%%%%%%%%%%%%%%%%%%%%%%%%%%%%%%%%%%%%%%%%%%%%%%%%%%%%%%%%%%%%%%%%%%%
is Hermitian as well. The operator $\mathbb{D}$ is hermitian with respect to the inner product defined as follows
\begin{align}
    \langle \tilde{\psi}|\mathbb{D}\tilde{\chi}\rangle = \int \rd^D q \, \tilde{\mu}(\boldsymbol{q}) \, \tilde{\psi}^* \mathbb{D} \tilde{\chi},
\end{align}
where the integration measure has to be $\tilde{\mu}(\boldsymbol{q}) = \Omega^{-2}(\boldsymbol{q}) \sqrt{|\tilde{\mathscr{G}}(\boldsymbol{q})|},$ and we have assumed that the wave functions $\tilde{\psi},\tilde{\chi}$ vanish at the boundary of the minisuperspace. Thus, under the conformal transformation, the wave function and the integration measure for the inner product should change as
\begin{subequations}
\begin{align}
    \Psi & \to \tilde{\Psi} = \Omega^{\frac{2-D}{2}}\Psi,  \\
    \mu(\boldsymbol{q}) = \sqrt{|\mathscr{G}(\boldsymbol{q})|} & \to \tilde{\mu}(\boldsymbol{q}) = \frac{\sqrt{|\mathscr{G}(\boldsymbol{q})|}}{\Omega(\boldsymbol{q})^{2-D}}, \label{eq:inner_product_measure}
\end{align}    
\end{subequations}
%%%%%%%%%%%%%%%%%%%%%%%%%%%%%%%%%%%%%%%%%%%%%%%%%%%%%%%%%%%%%%%%%%%%%%%%%%%%%%%%%%%%%%%%%%%%%%%%%%%%%%%%%%%%%%%%%%%%%%%%%%%%%%%%%%%%%%%%%%%%%%%%%%%%%%
which, again relates the wave functions and inner product measures in quantum theories differing by conformal-class ambiguity. As a result of these transformation properties, the inner products remain invariant under such conformal transformations
%%%%%%%%%%%%%%%%%%%%%%%%%%%%%%%%%%%%%%%%%%%%%%%%%%%%%%%%%%%%%%%%%%%%%%%%%%%%%%%%%%%%%%%%%%%%%%%%%%%%%%%%%%%%%%%%%%%%%%%%%%%%%%%%%%%%%%%%%%%%%%%%%%%%%%%%%
\begin{align}
    \langle \tilde\psi|\tilde\chi\rangle = \int \rd^D q \frac{\sqrt{|\tilde{\mathscr{G}}(\boldsymbol{q})|}}{\Omega^2(\boldsymbol{q})} \, \tilde{\psi}^* \tilde{\chi} = \int \rd^D q \sqrt{|\mathscr{G}(\boldsymbol{q})|} \psi^* \chi = \langle\psi|\chi\rangle.
\end{align}
%%%%%%%%%%%%%%%%%%%%%%%%%%%%%%%%%%%%%%%%%%%%%%%%%%%%%%%%%%%%%%%%%%%%%%%%%%%%%%%%%%%%%%%%%%%%%%%%%%%%%%%%%%%%%%%%%%%%%%%%%%%%%%%%%%%%%%%%%%%%%%%
Furthermore, the Wheeler-DeWitt equation \ref{eq:Schrodinger_generic} allows us to define a current
%%%%%%%%%%%%%%%%%%%%%%%%%%%%%%%%%%%%%%%%%%%%%%%%%%%%%%%%%%%%%%%%%%%%%%%%%%%%%%%%%%%%%%%%%%%%%%%%%%%%%%%%%%%%%%%%%%%%%%%%%%%%%%%%%%%%%%%%%%%%%%%%%%%
\begin{align}
    j^A = \frac{i\hbar}{2} \mathscr{G}^{AB}\left(\Psi^* \nabla_{B}\Psi - \Psi\nabla_{B}\Psi^*\right),
\end{align}
which satisfies a conservation equation
\begin{align}\label{eq:continuity_of_probability}
    \partial_{T} j^T + \nabla_{A} j^A = 0,
\end{align}
where $j^T=|\Psi|^2$. Therefore, the probability measure at a time $T$ can be defined as follows
\begin{align}
    \int j^T \sqrt{|\mathscr{G}|} \rd^Dq = \int |\Psi|^2 \sqrt{|\mathscr{G}|} \, \rd^D q .
\end{align}
Therefore, the probability measure remains invariant under lapse rescalings as well
%%%%%%%%%%%%%%%%%%%%%%%%%%%%%%%%%%%%%%%%%%%%%%%%%%%%%%%%%%%%%%%%%%%%%%%%%%%%%%%%%%%%%%%%%%%%%%%%%%%%%%%%%%%%%%%%%%%%
\begin{align}
    \int j^T \sqrt{|\mathscr{G}|} \rd^Dq \to \int |\tilde{\Psi}|^2 \frac{\sqrt{|\tilde{\mathscr{G}}|}}{\Omega^2} \, \rd^D q = \int j^T \sqrt{|\mathscr{G}|} \rd^Dq.
\end{align}
%%%%%%%%%%%%%%%%%%%%%%%%%%%%%%%%%%%%%%%%%%%%%%%%%%%%%%%%%%%%%%%%%%%%%%%%%%%%%%%%%%%%%%%%%%%%%%%%%%%%%%%%%%%%%%%%%%%%%%%%%%%%
Notice that the invariance of the inner products and the probability measure depended on the fact that under the lapse rescaling, the integration measure changes \textit{not} by $\sqrt{|\mathscr{G}|}\to \sqrt{|\tilde{\mathscr{G}}|}$, but by an additional factor of $\Omega^{-2}$, that is, correctly the measure transforms as $\sqrt{|\mathscr{G}|}\to \sqrt{|\tilde{\mathscr{G}}|} \Omega^{-2}$. This additional factor appears due to the transformation $\partial/\partial T \to \Omega^{-2}\partial/\partial T$, and is lost in the case of the timeless formalisms to quantum cosmology. Thus, having a clock in the theory is crucial to ensure the fact that the inner products or the physical predictions of the theory are ambiguity-free and do not depend on the operator ordering choices (of conformal-class) of the quantum Hamiltonian. We note that it is possible to define a probability measure that respects the conformal invariance by construction, even in the absence of a clock. For example, in \cite{Kiefer:2019bxk} the probability measure $|\Psi|^{\frac{2D}{D-2}} \,\, {\rm dvol} = |\Psi|^{\frac{2D}{D-2}} \sqrt{|g|} \rd^D q$ is conformally invariant by construction. However, in our case, we did not construct the probability measure by demanding conformal invariance. We simply used the Sturm-Liouville procedure to determine the weight factor in the usual inner product in non-relativistic quantum mechanics, demanding only the self-adjointness of the Hamiltonian operator. Surprisingly, this natural inner product also turns out to be a conformally invariant inner product as well.

Further, recall that we have so far ignored the ambiguity caused by the factor $F_{2}$, in which case the wave function gets additionally scaled by $\Psi/F_{2}$ and the corresponding measure of integration for the inner products such that the Hamiltonian remains hermitian becomes
%%%%%%%%%%%%%%%%%%%%%%%%%%%%%%%%%%%%%%%%%%%%%%%%%%%%%%%%%%%%%%%%%%%%%%%%%%%%%%%%%%%%%%
\begin{subequations}
\begin{align}
    \Psi & \to \tilde{\Psi} = \frac{\Omega^{\frac{2-D}{2}}}{F_{2}}\Psi \\
    \mu(\boldsymbol{q}) = \sqrt{|\mathscr{G}(\boldsymbol{q})|} & \to \tilde{\mu}(\boldsymbol{q}) = F_{2}^2 \frac{\sqrt{|\mathscr{G}(\boldsymbol{q})|}}{\Omega(\boldsymbol{q})^{2-D}},
\end{align}    
\end{subequations}
%%%%%%%%%%%%%%%%%%%%%%%%%%%%%%%%%%%%%%%%%%%%%%%%%%%%%%%%%%%%%%%%%%%%%%%%%%%%%%%%%%%%%%%%%%%%%%%%%%%%%%%%%%%%%%%%%%%%%%%%%%%%%%%%%%%%%%%%%%%%%%%%%%%%%%
and it is easy to see that the function $F_{2}$ does not affect the inner products. A similar measure has also been advocated in recent works \cite{Partouche:2021epk,Kaimakkamis:2024fmy,Kaimakkamis:2024lkb}.

Therefore, guided by the symmetry principle that the quantum theory should preserve the symmetries of the classical theory---under arbitrary field redefinitions $q^A\to\tilde{q}^A(q^B)$ and lapse rescalings $N\to \tilde{N}\Omega^{-2}$---we have arrived at quantum Hamiltonians that are equivalent in the sense that the quantum theories that differ by these transformations have the same inner products between the states. This requires choosing a specific form for the function $F_{3} = + \frac{D-2}{8(D-1)}\mathcal{R}$ where $D\neq 1$. The choice for the potential-class function is sufficient to render the conformal-class ambiguity immaterial for the inner products, whereas the factor-class ambiguity is already inconsequential for the inner products due to the hermiticity of the quantum Hamiltonian. In what follows, we shall discuss the path integral realization of the above conclusion. We however only discuss the potential and conformal class ambiguities as the factor-class is already taken care of by the hermiticity requirement alone.

\section{Path integral quantization: general formalism}\label{sec:QGF2}
Now, let us construct the path integral corresponding to the minisuperspace model \ref{eq:mini_action_generic}. The discrete definition of the path integral can be written by following many different conventions depending on at which discrete point should the metric, momenta, and superpotential be evaluated (for example, mid-point, pre-point, post-point, product form conventions, etc. See \cite{Grosche:1998yu,Grosche:1987ba} for more details). All of these prescriptions are equivalent and lead to the same result, that is, to the propagator corresponding to the Schr\"odinger-like equation \ref{eq:Schrodinger_generic} with the quantum Hamiltonian \ref{eq:quantum_Hamiltonian_generic}. Given a Sch\"{o}dinger equation, the corresponding definition for the path integral is not unique (see, specifically \cite{Parker_1979} and also \cite{Grosche:1998yu,Grosche:1987ba}); however, for convenience, in the present work, it will be useful for us to use the product form prescription for the path integral \cite{Grosche:1987gh}, and we shall stick to this convention throughout this article. In the original work \cite{Grosche:1987gh}, the author considered the case of a Riemannian manifold. Here, however, we are working with a pseudo-Riemannian manifold with a hyperbolic signature. The generalization is, nonetheless, quite straightforward as we shall show.

In the product form prescription, the metric is expressed in terms of the product of tetrads as $\mathscr{G}_{AB} = \eta_{ab} e^a_A e^b_B$, where $\eta_{ab}$ is the Minkowski metric. Then, for a diagonal metric $\mathscr{G}_{AB}$, we have $e^a_{A} = \sqrt{|\mathscr{G}_{AA}|} \delta^a_{A}$. Moreover, the determinant of the metric is related to the determinant of the tetrads as $\sqrt{|\mathscr{G}|}=\det{e}\equiv e$. The quantum Hamiltonian \ref{eq:quantum_Hamiltonian_generic} may be expressed in terms of the tetrads as
%%%%%%%%%%%%%%%%%%%%%%%%%%%%%%%%%%%%%%%%%%%%%%%%%%%%%%%%%%%%%%%%%%%%%%%%%%%%%%%%%%%%%%%%%%%%%%%%%%%%%%%%%%%%%%%%%%%%%%%%%%%%%%%%%%%%%%%%%%%%%%%%%%%%%%%
\begin{align}
    \hat{\mathscr{H}} = - \frac{\hbar^2}{2} \Delta_{\sf LB} + U + \hbar^2 \xi \mathcal{R} = \frac{1}{2} \eta^{ab} e^{A}_{a} \hat{p}_{A} \hat{p}_{B} e^{B}_{b}+ U + \hbar^2 \xi \mathcal{R} + \Delta V_{Q}^{\sf PF},
\end{align}
%%%%%%%%%%%%%%%%%%%%%%%%%%%%%%%%%%%%%%%%%%%%%%%%%%%%%%%%%%%%%%%%%%%%%%%%%%%%%%%%%%%%%%%%%%%%%%%%%%%%%%%%%%%%%%%%%%%%%%%%%%%%%%%%%%%%%%%%%%%%%%%%%%%%%%%
where $e^{A}_{a}$ are the inverse tetrads satisfying $e^{A}_{a} e^{b}_{A} = \delta^b_{a}$ and $e^{A}_{a} e^{a}_{B} = \delta^A_{B}$ and the momentum operators are defined \cite{DeWitt_1957} as
\begin{align}
    \hat{p}_{A} \equiv \frac{\hbar}{i}|\mathscr{G}|^{-\frac{1}{4}} \partial_{A} |\mathscr{G}|^{\frac{1}{4}} = \frac{\hbar}{i}\left(\partial_{A} + \frac{1}{2} \partial_{A} \log \sqrt{|\mathscr{G}|} \right) = \frac{\hbar}{i}\left(\partial_{A} + \frac{1}{2} \partial_{A} \log e \right).
\end{align}
Because of the way we have chosen to express the kinetic differential operator, an additional well-defined quantum potential term $\Delta V_{Q}^{\sf PF}$ has appeared (see \cite{Grosche:1987gh} for a Riemannian version), where 
\begin{align}\label{eq:quantum_potential}
    \Delta V_{Q}^{\sf PF} = \frac{\hbar^2}{8} \eta^{ab}\left[4 e^{A}_{a} (\partial_{A} \partial_{B} e^{B}_{b}) + 2 e^{A}_{a} e^{B}_{b} \frac{(\partial_{A} \partial_{B}e)}{e} + 2 e^{A}_{a} \left((\partial_{A}e^{B}_{b})\frac{(\partial_{B} e)}{e} + (\partial_{B}e^{B}_{b}) \frac{(\partial_{A} e)}{e}\right) - e^{A}_{a}e^{B}_{b}\frac{(\partial_{A}e)( \partial_{B}e)}{e^2} \right].
\end{align}
The minisuperspace model we are dealing with is a constrained system, and as a result, defining the path integral kernel has to be done carefully. The constraints correspond to gauge symmetries in the system, which leads to overcounting of the paths (or histories) that are equivalent due to the symmetry. Therefore, constructing the path integral is quite non-trivial. The task of constructing the path integral in the context of gravitational systems was taken upon by Teitelboim (presently Bunster) \cite{Teitelboim:1981ua,Teitelboim:1983fh,PhysRevD.28.310,PhysRevD.28.297}, Barvinsky and Ponomariov \cite{Barvinsky:1986qn,Barvinsky:1986bt}, Kucha\v r and Hartle \cite{Kuchar_Skeletonizations,hartle1984path,PhysRevD.34.2323}. However, all of these works dealt with the problem only at a formal level. Later, Halliwell \cite{Halliwell:1988wc} explicitly constructed the path integral for the minisuperspace cosmological models (also see Halliwell and Hartle \cite{Halliwell:1990qr}), and that framework will be the basis for the discussions in the current paper. Halliwell showed that the imposition of the gauge fixing condition $\dot{N} = \chi(a,p_{a},N)$ can be done using the BFV (Batalin-Fradkin-Vilkovisky) formalism in which one adds to the gauge fixed action additional anti-commuting ghost degrees of freedom such that the BRST (Becchi-Rouet-Stora-Tyutin) symmetry is respected. Due to the BRST symmetry, the path integral will not depend on the choice of the function $\chi$. Even though \cite{Halliwell:1988wc} does not consider the additional fluid $(T,p_{T})$, the arguments presented in that work can be straightforwardly generalized to the present case. In the proper time gauge $\dot{N}=0$, the Feynman propagator has the form
\begin{align}
    \mathscr{K}(\boldsymbol{q}_{\sf f},T_{\sf f}; \boldsymbol{q}_{\sf i}, T_{\sf i}) = \int \rd N(t_{\sf f} - t_{\sf i}) \langle \boldsymbol{q}_{\sf f}, T_{\sf f} | \exp\left[-\frac{i}{\hbar} N(t_{\sf f}-t_{\sf i}) (\hat{\mathscr{H}}+\hat{p}_{T})\right] | \boldsymbol{q}_{\sf i}, T_{\sf i} \rangle,
\end{align}
where $| \boldsymbol{q}_{\sf i}, T_{\sf i} \rangle$ and $| \boldsymbol{q}_{\sf f}, T_{\sf f} \rangle$ are the initial and final states of the system. The propagator does not depend on the coordinate time as it is integrated out and depends only on the field configurations. However, we shall see that due to the term $\hat{p}_{T}$, the fluid canonical variable $T$ will play the role of time. The integral $\int \rd N(t_{\sf f}-t_{\sf i})$ over the full range $(-\infty,\infty)$ ensures that the kernel satisfies the constraint $(\hat{\mathscr{H}}_{\boldsymbol{q}_{\sf f}}+\hat{p}_{T})\mathscr{K}(\boldsymbol{q}_{\sf f},T_{\sf f}; \boldsymbol{q}_{\sf i}, T_{\sf i}) = 0$. Whereas, the lapse integral over the half-range $(0,\infty)$ would produce a Green's function for the total Hamiltonian operator. Now, let us find the discrete path integral definition in the present case.

We discretize the coordinate time interval into $M$ infinitesimal steps with step size $\epsilon = (t_{\sf f}-t_{\sf i})/M$. Let $\boldsymbol{q}^{(j)}$ denote the coordinate at time $t_{j} = t_{\sf i}+j \epsilon$ with $j$ being an integer index that runs from 0 to $M$ and $t_{M}=t_{\sf f}$. Then by using the transitive property of the propagator, we can write the full propagator in terms of infinitesimal ones
\begin{align}
    \mathscr{K}(\boldsymbol{q}_{\sf f},T_{\sf f} ; \boldsymbol{q}_{\sf i},T_{\sf i}) & = \int \rd N (t_{\sf f}-t_{\sf i}) \left(\prod_{j=1}^{M-1} \int \sqrt{|\mathscr{G}^{(j)}|} \rd^D q^{(j)} \, \rd T^{(j)} \right) \prod_{j=1}^{M} \langle \boldsymbol{q}^{(j)},T^{(j)}| \exp\left[-\frac{i}{\hbar} \epsilon N (\hat{\mathscr{H}} + \hat{p}_{T} ) \right] | \boldsymbol{q}^{(j-1)},T^{(j-1)} \rangle,
\end{align}
where the infinitesimal propagators themselves can be approximated in the limit $\epsilon\to 0$ as 
\begin{align}
     & \langle \boldsymbol{q}^{(j)},T^{(j)}| \exp\left[-\frac{i}{\hbar} \epsilon N (\hat{\mathscr{H}} + \hat{p}_{T} ) \right] | \boldsymbol{q}^{(j-1)},T^{(j-1)} \rangle \simeq \langle \boldsymbol{q}^{(j)},T^{(j)}| \left(1-\frac{i}{\hbar} \epsilon N (\hat{\mathscr{H}} + \hat{p}_{T} ) \right) | \boldsymbol{q}^{(j-1)},T^{(j-1)} \rangle + \mathcal{O}(\epsilon^2) \nonumber\\
    & = \frac{[\mathscr{G}^{(j)}\mathscr{G}^{(j-1)}]^{-\frac{1}{4}}}{(2\pi\hbar)^{D+1}} \int \rd^D p^{(j)} \, \rd p_{T}^{(j)} \exp \left[\frac{i}{\hbar} p^{(j)}_{A} (q^{A(j)} - q^{A(j-1)}) + \frac{i}{\hbar} p^{(j)}_{T} (T^{(j)}-T^{(j-1)})\right] \Big(1 - \frac{i\epsilon N}{2\hbar} \eta^{ab} e^{A(j)}_{a} p^{(j)}_{A} p^{(j)}_{B} e^{B(j-1)}_{b} \nonumber\\
    & - \frac{i\epsilon N}{\hbar}\left( U(\boldsymbol{q}^{(j)})+\Delta V_{Q}^{\sf PF}(e^{A(j)}_a)+\hbar^2 \xi \mathcal{R}(\boldsymbol{q}^{(j)})+p_{T}^{(j)}\right) \Big). 
\end{align}
In deriving the above expression, we have used the following properties related to the orthogonality of the coordinate eigenstates and overlap between the coordinate-momentum eigenstates
\begin{subequations}
    \begin{align}
    \langle \boldsymbol{q}'',T''| \boldsymbol{q}',T' \rangle & = \left(\mathscr{G}(\boldsymbol{q}'') \mathscr{G}(\boldsymbol{q}')\right)^{-\frac{1}{4}} \delta^{(D)}(\boldsymbol{q}''-\boldsymbol{q}') \delta(T''-T'), \\
    \langle \boldsymbol{q},T| \boldsymbol{p},p_{T} \rangle & = \frac{\exp\left[\frac{i}{\hbar} (p_{A}q^A+p_{T}T)\right]}{(2\pi\hbar)^{\frac{D+1}{2}}}.
    \end{align}
\end{subequations}
Notice that we are using a convention in which the normalization for the Dirac delta function is symmetrized with respect to both boundaries, $\boldsymbol{q}'$ and $\boldsymbol{q}''$. The expression for the infinitesimal propagator can be re-expressed up to order $\mathcal{O}(\epsilon)$ as 
\begin{align}
     & \langle \boldsymbol{q}^{(j)},T^{(j)}| \exp\left[-\frac{i}{\hbar} \epsilon N (\hat{\mathscr{H}} + \hat{p}_{T} ) \right] | \boldsymbol{q}^{(j-1)},T^{(j-1)} \rangle \simeq \frac{[\mathscr{G}^{(j)}\mathscr{G}^{(j-1)}]^{-\frac{1}{4}}}{(2\pi\hbar)^{D+1}} \int \rd^D p^{(j)} \, \rd p_{T}^{(j)} \exp \Bigg[\frac{i}{\hbar} p^{(j)}_{A} (q^{A(j)} - q^{A(j-1)}) \nonumber\\
     & + \frac{i}{\hbar} p^{(j)}_{T} (T^{(j)}-T^{(j-1)}) - \frac{i\epsilon N}{\hbar} \left(\frac{1}{2} \eta^{ab} e^{A(j)}_{a} p^{(j)}_{A} p^{(j)}_{B} e^{B(j-1)}_{b} - U(\boldsymbol{q}^{(j)})+\Delta V_{Q}^{\sf PF}(e^{A(j)}_a)+\hbar^2 \xi \mathcal{R}(\boldsymbol{q}^{(j)}) + p_{T}^{(j)} \right) \Bigg]. 
\end{align}
Finally, inserting the expression for the infinitesimal propagators and utilizing Trotter's formula $\lim_{M\to\infty} \left(\e^{A/M} \e^{B/M}\right)^M = \e^{A+B}$, we can write the full propagator as
\begin{align}\label{eq:path_integral_discrete}
    \mathscr{K}(\boldsymbol{q}_{\sf f},T_{\sf f} ; \boldsymbol{q}_{\sf i},T_{\sf i})= & \int \rd N (t_{\sf f}-t_{\sf i}) [\mathscr{G}(\boldsymbol{q}_{\sf f})\mathscr{G}(\boldsymbol{q}_{\sf i})]^{-\frac{1}{4}} \left(\prod_{j=1}^{M-1} \int \rd^D q^{(j)} \, \rd T^{(j)} \times \prod_{j=1}^{M} \frac{\rd^D p^{(j)}}{(2\pi\hbar)^D} \frac{\rd p_{T}^{(j)}}{2\pi\hbar} \right) \nonumber\\
    & \times \exp \Bigg[\frac{i}{\hbar} \sum_{j=1}^{M} \Bigg( p^{(j)}_{A} (q^{A(j)} - q^{A(j-1)}) + p^{(j)}_{T} (T^{(j)}-T^{(j-1)}) - \epsilon N \Big(\frac{1}{2} \eta^{ab} e^{A(j)}_{a} p^{(j)}_{A} p^{(j)}_{B} e^{B(j-1)}_{b} \nonumber\\
    &  - U(\boldsymbol{q}^{(j)})+\Delta V_{Q}^{\sf PF}(e^{A(j)}_a)+\hbar^2 \xi \mathcal{R}(\boldsymbol{q}^{(j)}) + p_{T}^{(j)} \Big)\Bigg) \Bigg].
\end{align}
This is the discrete definition of path integral on a curved pseudo-Riemannian manifold in the ``product form'' definition \cite{Grosche:1987gh}. In most parts of the article, however, for formal manipulations, we shall use the following continuous notation for the above definition of the path integral
\begin{align}\label{eq:generic_kernel}
    \mathscr{K}(\boldsymbol{q}_{\sf f},T_{\sf f} ; \boldsymbol{q}_{\sf i},T_{\sf i})= & \int \rd N (t_{\sf f}-t_{\sf i}) [\mathscr{G}(\boldsymbol{q}_{\sf f})\mathscr{G}(\boldsymbol{q}_{\sf i})]^{-\frac{1}{4}} \int \mathscr{D}q^A \mathscr{D} p_{A} \int \mathscr{D}T \mathscr{D} p_{T} \exp \Bigg[\frac{i}{\hbar} \int_{t_{\sf i}}^{t_{\sf f}} \rd t \Bigg( p_{A} \dot{q}^A + p_{T}\dot{T} \nonumber\\
    &  - N \Big(\frac{1}{2} \eta^{ab} e^{A}_{a} p_{A} p_{B} e^{B}_{b} - U(\boldsymbol{q})+\Delta V_{Q}^{\sf PF} + \hbar^2 \xi \mathcal{R} + p_{T}\Big)\Bigg) \Bigg].
\end{align}
Although for computational convenience we have used the product form definition of the path integral, it is not necessary to use this particular construction. One can use other discrete definitions as well. As the definitions depend on at which discrete point in time various functions have to be evaluated, going from one construction to the other requires Taylor expansion of the functions keeping all terms up to $\mathcal{O}(\epsilon)$. Moreover, the expansion of the kinetic term is quite tricky as in Feynman path integrals the typical path difference varies as the square root of the infinitesimal time interval $\Delta\boldsymbol{q}^{(j)}\equiv(\boldsymbol{q}^{(j)}-\boldsymbol{q}^{(j-1)})\sim\sqrt{\epsilon}$ \cite{RevModPhys.20.367}. Then in the Taylor expansion of the kinetic term up to the quartic order term contributes as $\Delta^4\boldsymbol{q}^{(j)}/\epsilon\sim\mathcal{O}(\epsilon)$. Also, recall that for different discrete definitions of the path integral, the expression for the quantum potential $\Delta V_{Q}$ is different \cite{Grosche:1987ba}. Once these subtleties are taken care of, all discrete definitions of the path integral should be equivalent, in the sense that all of these different conventions produce the propagator for the Schr\"{o}dinger equation \ref{eq:Schrodinger_generic} with the Hamiltonian \ref{eq:quantum_Hamiltonian_generic}.

We note, however, that a completely different perspective is possible which exploits the skeletonization ambiguity of path integrals. If we do not purposefully add the compensating quantum correction $\Delta V_{Q}$, which keeps the quantum Hamiltonian in the Laplace-Beltrami form, then different choices for the infinitesimal splitting of the path integral will result in different ordering choices for the quantum Hamiltonian \footnote{See \cite{Cohen:1970kv} for an example of how different splitting choices in the path integral generates different quantum Hamiltonians. Also see \cite{kerner1970unique}, which gives an example in quantum mechanics on how defining path integral with the coordinate average of the function $\mathscr{H}$ generates the so called Born-Jordan ordering.}. Therefore, it could be, in principle, possible to choose a path integral skeletonization such that the quantum Hamiltonian corresponding to the classical theory in \ref{eq:generic_conformal_rescaled_action} matches exactly that of the theory in \ref{eq:mini_action_generic}, where the latter theory has to be quantized following an independent skeletonization scheme than the former case. The major drawback of this approach remains in finding these appropriate skeletonizations that render a particular theory to become equivalent to a lapse-rescaled version of the same. Therefore, we follow a different route, in which we fix the skeletonization scheme once and for all. Then, all the quantum theories differing by conformal-class ambiguity (which, in principle, would have been a reflection of different skeletonizations) can be made equivalent by demanding conformal invariance under that fixed skeletonization. This is achieved simply by choosing a specific form for the potential-class term, depending only on the underlying geometry of the minisuperspace. 

In the following subsections, we shall first, integrate out the clock degree of freedom and then proceed to discuss how the requirement of conformal invariance of the theory requires $\xi$ to have a specific value depending on the dimension $D$ of the minisuperspace.

\subsection{Integrating out the fluid degree of freedom}

In the canonical approach (in \ref{eq:Schrodinger_generic}), we have seen that due to the linear dependence of the fluid's Hamiltonian on its momentum, upon quantization the quantum Hamiltonian constraint takes the form of a Schr\"odinger equation with the fluid canonical variable $T$ playing the role of time. In the path integral approach, this fact will become apparent when we integrate over the clock degrees of freedom. First, notice that the integrals on the fluid canonical variables $(T,p_{T})$ are separable from the rest as follows 
\begin{align}
    \mathscr{K}(\boldsymbol{q}_{\sf f},T_{\sf f} ; \boldsymbol{q}_{\sf i},T_{\sf i})= & \int \rd N (t_{\sf f}-t_{\sf i}) [\mathscr{G}(\boldsymbol{q}_{\sf f})\mathscr{G}(\boldsymbol{q}_{\sf i})]^{-\frac{1}{4}} \left(\prod_{j=1}^{M-1} \int \rd^D q^{(j)}\times \prod_{j=1}^{M} \frac{\rd^D p^{(j)}}{(2\pi\hbar)^D} \right) \exp \Bigg[\frac{i}{\hbar} \sum_{j=1}^{M} \Bigg( p^{(j)}_{A} (q^{A(j)} - q^{A(j-1)}) \nonumber\\
    & - \epsilon N \Big(\frac{1}{2} \eta^{ab} e^{A(j)}_{a} p^{(j)}_{A} p^{(j)}_{B} e^{B(j-1)}_{b} - U(\boldsymbol{q}^{(j)})+\Delta V_{Q}^{\sf PF}(e^{A(j)}_a)+\hbar^2 \xi \mathcal{R}(\boldsymbol{q}^{(j)}) \Big)\Bigg) \Bigg] \nonumber\\
    & \times \left(\prod_{j=1}^{M-1} \int \rd T^{(j)} \times \prod_{j=1}^{M} \frac{\rd p_{T}^{(j)}}{2\pi\hbar} \right) \exp \Bigg[\frac{i}{\hbar} \sum_{j=1}^{M} \Bigg(p^{(j)}_{T} (T^{(j)}-T^{(j-1)}) - \epsilon N p_{T}^{(j)} \Big)\Bigg) \Bigg]
\end{align}
Then, we can first integrate over the momentum variables $p_{T}^{(j)}$, which will lead to $M$ number of nested Dirac delta functions. Further, we can perform $M-1$ number of integrals over the coordinates $T^{(j)}$. As there is one more delta function than there are coordinate integrals, the final result will be a Dirac delta function, that is
\begin{align}
    & \left(\prod_{j=1}^{M-1} \int \rd T^{(j)} \times \prod_{j=1}^{M} \frac{\rd p_{T}^{(j)}}{2\pi\hbar} \right) \exp \Bigg[\frac{i}{\hbar} \sum_{j=1}^{M} \Bigg(p^{(j)}_{T} (T^{(j)}-T^{(j-1)}) - \epsilon N p_{T}^{(j)} \Big)\Bigg) \Bigg] \nonumber\\
    & = \left(\prod_{j=1}^{M-1} \int \rd T^{(j)} \right) \prod_{j=1}^{M}\delta(T^{(j)} - T^{(j-1)} - N \epsilon ) \nonumber\\
    & = \left(\prod_{j=2}^{M-1} \int \rd T^{(j)} \right) \left(\prod_{j=3}^{M-1}\delta(T^{(j)} - T^{(j-1)} - N \epsilon )\right) \delta(T^{(2)} - T_{\sf i} - 2\epsilon N) \nonumber\\
    & \vdots \nonumber\\
    & = \int \rd T^{(M-1)} \delta(T^{(M)}-T^{(M-1)}-N\epsilon) \delta(T^{(M-1)}-T_{\sf i}-N\epsilon(M-1)) \nonumber\\
    & = \delta(T_{\sf f} -T_{\sf i} - N(t_{\sf f}-t_{\sf i})),
\end{align}
where we have used the fact that $M \epsilon = t_{\sf f} - t_{\sf i}$. Thus the path integral over the clock degrees of freedom produces a Dirac delta function which enforces the fact that the proper time difference between the quantum transition $N(t_{\sf f} - t_{\sf i})$ is to be measured by the difference in the configuration of the clock fluid $T_{\sf f}-T_{\sf i}\equiv \mathscr{T}$. After the integration over the lapse function, over the full range $(-\infty,\infty)$, we finally get the following relation
\begin{align}
    \mathscr{K}(\boldsymbol{q}_{\sf f}, \boldsymbol{q}_{\sf i};\mathscr{T})= & [\mathscr{G}(\boldsymbol{q}_{\sf f})\mathscr{G}(\boldsymbol{q}_{\sf i})]^{-\frac{1}{4}} \int \mathscr{D}q^A \mathscr{D} p_{A} \exp \Bigg[\frac{i}{\hbar} \int_{0}^{\mathscr{T}} \rd t \Bigg( p_{A} \dot{q}^A - \Big(\frac{1}{2} \eta^{ab} e^{A}_{a} p_{A} p_{B} e^{B}_{b} - U(\boldsymbol{q})+\Delta V_{Q}^{\sf PF} + \hbar^2 \xi \mathcal{R}\Big)\Bigg) \Bigg].
\end{align}
To arrive at the above expression, we have, without loss of any generality, chosen $t_{\sf i}$ as the origin of the time coordinate and rescaled the time coordinate as $t\to t/N$. The resulting propagator has the form such that it describes the quantum evolution of the degrees of freedom $\boldsymbol{q}$ from the coordinates $\boldsymbol{q}_{\sf i}$ to $\boldsymbol{q}_{\sf f}$ in time interval $ \mathscr{T}=T_{\sf f}-T_{\sf i}$. On the other hand, if the lapse integration had been carried out over the half range $(0,\infty)$, then, assuming $t_{\sf f}-t_{\sf i}>0$, the result is only non-zero if $\mathscr{T}>0$. In that case, we will have a factor of Heaviside theta function $\Theta(\mathscr{T})$. Therefore, the path integral kernel has a notion of the direction of time.

\subsection{Conformal invariance under lapse rescaling}

Now, let us turn our attention to the path integral propagator after the conformal transformation
%%%%%%%%%%%%%%%%%%%%%%%%%%%%%%%%%%%%%%%%%%%%%%%%%%%%%%%%%%%%%%%%%%%%%%%%%%%%%%%%%%%%%%%%%%%%%%%%%%%%%%%%%%%%%%%%%%%%%%%
\begin{align}\label{eq:conformal_path_integral_kernel}
    \tilde{\mathscr{K}}(\boldsymbol{q}_{\sf f},T_{\sf f} ; \boldsymbol{q}_{\sf i},T_{\sf i})= & \int \rd \tilde{N} (t_{\sf f}-t_{\sf i}) [\tilde{\mathscr{G}}(\boldsymbol{q}_{\sf f})\tilde{\mathscr{G}}(\boldsymbol{q}_{\sf i})]^{-\frac{1}{4}} \int \mathscr{D}q^A \mathscr{D} p_{A} \int \mathscr{D}T \mathscr{D} p_{T} \exp \Bigg[\frac{i}{\hbar} \int_{t_{\sf i}}^{t_{\sf f}} \rd t \Bigg( p_{A} \dot{q}^A + p_{T}\dot{T} \nonumber\\
    &  - \tilde{N} \Big(\frac{1}{2} \eta^{ab} \tilde{e}^{A}_{a} p_{A} p_{B} \tilde{e}^{B}_{b} - \tilde{U}(\boldsymbol{q})+\widetilde{\Delta V}_{Q}^{\sf PF} + \hbar^2 \xi \tilde{\mathcal{R}} + \tilde{p}_{T}\Big)\Bigg) \Bigg],
\end{align}
%%%%%%%%%%%%%%%%%%%%%%%%%%%%%%%%%%%%%%%%%%%%%%%%%%%%%%%%%%%%%%%%%%%%%%%%%%%%%%%%%%%%%%%%%%%%%%%%%%%%%%%%
here the relation between the quantities before and after the conformal transformation is as follows
%%%%%%%%%%%%%%%%%%%%%%%%%%%%%%%%%%%%%%%%%%%%%%%%%%%%%%%%%%%%%%%%%%%%%%%%%%%%%%%%%%%
\begin{align}
    \tilde{e}^a_A = \Omega e^a_A, \qquad
    \tilde{e}^A_a = \Omega^{-1} e^A_a, \qquad \tilde{\mathscr{G}} = \Omega^{2D} \mathscr{G}, \qquad
    \tilde{U} = \Omega^{-2} U, \qquad \tilde{p}_T = \Omega^{-2} p_{T},
\end{align}
%%%%%%%%%%%%%%%%%%%%%%%%%%%%%%%%%%%%%%%%%%%%%%%%%%%%%%%%%%%%%%%%%%%%%%%%%%%%%%%%%%%%%%%%%%%%%%%%%%%%
The above quantum theory \ref{eq:conformal_path_integral_kernel} differs from the quantum theory \ref{eq:generic_kernel} by a conformal-class ambiguity. Their equivalence can be maintained by requiring the conformal invariance of the inner products and quantum correlators. In such a case, we show below that the path integral kernel changes by overall conformal factors (as do the wave functions). It can be easily shown that (see Appendix \ref{app:conformal_quantum_potential}) the quantum potential and the Ricci scalar term changes due to the conformal transformation as
\begin{align}\label{eq:quantum_correction_transformed}
     \widetilde{\Delta V}_{Q}^{\sf PF} + \hbar^2 \xi \tilde{\mathcal{R}} = &  \Omega^{-2} \left(\Delta V_{Q}^{\sf PF} + \hbar^2 \xi \mathcal{R} \right) + \frac{\hbar^2}{2 \Omega^3} \left(\frac{D}{2} - 1 - 4 \xi (D-1) \right)  \eta^{ab} \Big( e^A_a (\partial_{B} e^{B}_{b})(\partial_{A}\Omega)   \nonumber \\
    & + e^A_a (\partial_{A} e^{B}_{b})(\partial_{B}\Omega) + e^A_a e^{B}_{b} (\partial_{A}\partial_{B}\Omega) + \frac{e^A_a e^B_b}{e}(\partial_A e)(\partial_B \Omega) \Big) \nonumber\\
    & + \frac{\hbar^2}{\Omega^4} \left(\frac{(D-4)(D-2)}{8} - \xi (D-1)(D-4) \right) \eta^{ab} e^A_a e^B_b (\partial_{A} \Omega)(\partial_{B}\Omega).
\end{align}
As $\xi$ is arbitrary, choosing $\xi = +\frac{D-2}{8(D-1)}$ implies the quantum potential will transform conformally, that is,
\begin{align}
    \widetilde{\Delta V}_{Q}^{\sf PF} + \hbar^2 \frac{D-2}{8(D-1)} \tilde{\mathcal{R}} =  \Omega^{-2} \left(\Delta V_{Q}^{\sf PF}+\hbar^2 \frac{D-2}{8(D-1)}\mathcal{R}\right).
\end{align}
Then it is possible that in the path integral, $\Omega^{-2}$ can be factored out from the total Hamiltonian (see in the first-order action below)
\begin{align}\label{eq:conformal_kernel_intermediate}
    \tilde{\mathscr{K}}(\boldsymbol{q}_{\sf f},T_{\sf f} ; \boldsymbol{q}_{\sf i},T_{\sf i})= & \int \rd \tilde{N} (\Omega(\boldsymbol{q}_{\sf f}) \Omega(\boldsymbol{q}_{\sf i}))^{-\frac{D}{2}} [\mathscr{G}(\boldsymbol{q}_{\sf f})\mathscr{G}(\boldsymbol{q}_{\sf i})]^{-\frac{1}{4}} \int \mathscr{D}q^A \mathscr{D} p_{A} \int \mathscr{D}T \mathscr{D} p_{T} \exp \Bigg[\frac{i}{\hbar} \int_{0}^{\tilde{N}} \rd t \Bigg( p_{A} \dot{q}^A + p_{T}\dot{T} \nonumber\\
    &  - \Omega^{-2} \Big(\frac{1}{2} \eta^{ab} e^{A}_{a} p_{A} p_{B} e^{B}_{b} - U(\boldsymbol{q})+\Delta V_{Q}^{\sf PF} + \hbar^2 \frac{D-2}{8(D-1)} \mathcal{R} + p_{T}\Big)\Bigg) \Bigg],
\end{align}
where without the loss of any generality, we have set the initial and final coordinate time to be $t_{\sf i} = 0$ and $t_{\sf f} = 1$ and performed a constant rescaling of the coordinate time $t\to t/\tilde{N}$. The extra factor of $\Omega^{-2}$ can be tackled using a time redefinition. Now, we define a new time coordinate $u\in[0,u_{\sf f}]$ as follows
\begin{align}\label{eq:time_coordinate_transformation}
    \Omega^{-2} \rd t = \rd u,~~\text{or}~~ t(u)=\int_{0}^u \rd u' \, \Omega^{2}(q(u')),
\end{align}
such that the boundary values are respected
\begin{align}
    t(0) = 0,~ t(u_{\sf f}) = \tilde{N},~~ \Omega(q(t(0))) = \Omega(\boldsymbol{q}_{\sf i}), ~~ \Omega(q(t(u_{\sf f}))) = \Omega(\boldsymbol{q}_{\sf f}).
\end{align}
Given that we have a functional integration over the minisuperspace coordinate $\boldsymbol{q}(t(u))$, for every such function, we assume that there exists a unique solution for $u_{\sf f}$ such that the above boundary conditions are met. At first glance, ensuring this seems to be a Herculean task. However, to achieve this one can enforce a constraint in the functional integral of the scale factor by means of the conjurer's trick of writing identity in terms of the Dirac delta function (see, for example, \cite{Steiner:1984uf})
\begin{align}\label{eq:identity_delta_function}
    1 & = (\Omega(\boldsymbol{q}_{\sf f})\Omega(\boldsymbol{q}_{\sf i})) \int_{0}^{\infty} \rd u_{\sf f} \, \delta \left(\tilde{N} - \int_{0}^{u_{\sf f}} \Omega(\boldsymbol{q}(t(u)))^{2} \rd u\right) \nonumber\\
    & = (\Omega(\boldsymbol{q}_{\sf f})\Omega(\boldsymbol{q}_{\sf i})) \int_{-\infty}^{\infty} \frac{\rd \mathcal{E}}{2\pi\hbar} \e^{\frac{i}{\hbar} \mathcal{E} \tilde{N}} \int_{0}^{\infty} \rd u_{\sf f} \exp \left(- \frac{i}{\hbar} \int_{0}^{u_{\sf f}}  \rd u \, \Omega(\boldsymbol{q}(t(u)))^{2} \mathcal{E} \right).
\end{align}
In the last line, we have used the integral representation of the Dirac delta function, and the convention for the normalization of the delta function has been done with a symmetric combination of both initial and final scale factors, which is consistent with the product form definition of the path integral. 

Recalling the discrete definition of the path integral \ref{eq:path_integral_discrete}, we see that the time transformation \ref{eq:time_coordinate_transformation} has to be defined in a symmetric manner over the pre and post points, that is, $(\Omega^{(j)})^{-1}(\Omega^{(j-1)})^{-1} \Delta t = \Delta u^{(j)}$. Where $\Delta t=\epsilon$ is the discrete time interval that remains uniform throughout the lattice. On the other hand, the new time interval $\Delta u^{(j)}$ is not uniform and itself becomes time-dependent. The overall factor $\Omega^{-2}$ with the Hamiltonian in \ref{eq:conformal_kernel_intermediate} can also be written as $(\Omega^{(j)}\Omega^{(j-1)})^{-1}$, since the difference of this quantity from $(\Omega^{(j)})^{-2}$ appears in the path integral at a greater order than $\mathcal{O}(\epsilon)$ and can be neglected in the limit $\epsilon\to 0$. In the following, we continue to use the continuous version of the expressions.

Inserting the identity \ref{eq:identity_delta_function}, the Feynman propagator takes the following form
\begin{align}\label{eq:time_transformed_kernel}
    \tilde{\mathscr{K}}(\boldsymbol{q}_{\sf f},T_{\sf f} ; \boldsymbol{q}_{\sf i},T_{\sf i}) & = (\Omega(\boldsymbol{q}_{\sf f}) \Omega(\boldsymbol{q}_{\sf i}))^{\frac{2-D}{2}} \int \rd \tilde{N}  [\mathscr{G}(\boldsymbol{q}_{\sf f})\mathscr{G}(\boldsymbol{q}_{\sf i})]^{-\frac{1}{4}} \int \mathscr{D}q^A \mathscr{D} p_{A} \int \mathscr{D}T \mathscr{D} p_{T} \int_{-\infty}^{\infty} \frac{\rd \mathcal{E}}{2\pi\hbar}  \e^{\frac{i}{\hbar} \mathcal{E} \tilde{N}} \int_{0}^{\infty} \rd u_{\sf f}  \nonumber\\
    &  \times \exp \Bigg[\frac{i}{\hbar} \int_{0}^{u_{\sf f}} \rd u \Bigg( p_{A} \dot{q}^A + p_{T}\dot{T} - \Big(\frac{1}{2} \eta^{ab} e^{A}_{a} p_{A} p_{B} e^{B}_{b} - U(\boldsymbol{q})+\Delta V_{Q}^{\sf PF} + \hbar^2 \frac{D-2}{8(D-1)} \mathcal{R} + p_{T}\Big) - \Omega^2 \mathcal{E} \Bigg) \Bigg].
\end{align}
The overhead dot now means derivative with respect to the new time coordinate $u$. We have introduced this time coordinate transformation in the continuous notation of the path integral. It is interesting to note that the above propagator can be rewritten as
\begin{align}
    \tilde{\mathscr{K}}(\boldsymbol{q}_{\sf f},T_{\sf f} ; \boldsymbol{q}_{\sf i},T_{\sf i}) = \int \rd \tilde{N} \int_{-\infty}^{\infty} \frac{\rd \mathcal{E}}{2\pi\hbar}  \e^{\frac{i}{\hbar} \mathcal{E} \tilde{N}} G_{\mathcal{E}} (\boldsymbol{q}_{\sf f},T_{\sf f} ; \boldsymbol{q}_{\sf i},T_{\sf i}),
\end{align}
where $G_{\mathcal{E}} (\boldsymbol{q}_{\sf f},T_{\sf f} ; \boldsymbol{q}_{\sf i},T_{\sf i})$ is the fixed energy propagator
\begin{align}
    G_{\mathcal{E}} (\boldsymbol{q}_{\sf f},T_{\sf f} ; \boldsymbol{q}_{\sf i},T_{\sf i}) & = (\Omega(\boldsymbol{q}_{\sf f}) \Omega(\boldsymbol{q}_{\sf i}))^{\frac{2-D}{2}} [\mathscr{G}(\boldsymbol{q}_{\sf f})\mathscr{G}(\boldsymbol{q}_{\sf i})]^{-\frac{1}{4}} \int \mathscr{D}q^A \mathscr{D} p_{A} \int \mathscr{D}T \mathscr{D} p_{T} \int_{0}^{\infty} \rd u_{\sf f} \exp \Bigg[\frac{i}{\hbar} \int_{0}^{u_{\sf f}} \rd u \Bigg( p_{A} \dot{q}^A + p_{T}\dot{T} \nonumber\\
    & - \Big(\frac{1}{2} \eta^{ab} e^{A}_{a} p_{A} p_{B} e^{B}_{b} - U(\boldsymbol{q})+\Delta V_{Q}^{\sf PF} + \hbar^2 \frac{D-2}{8(D-1)} \mathcal{R} + p_{T}\Big) - \Omega^2 \mathcal{E} \Bigg) \Bigg].
\end{align}
Thus the parameter $\mathcal{E}$ has the interpretation of the total energy of the system. Performing the $\tilde{N}$ integration produces a Dirac delta function which enforces the vanishing of the total energy $\mathcal{E}=0$, which is essentially the Hamiltonian constraint. After performing the path integrals over $(T,p_{T})$, we get a Dirac delta function $\delta(T_{\sf f}-T_{\sf i} - u_{\sf f})$. Then performing the $u_{\sf f}$ integration over the half-line we finally get
\begin{align}\label{eq:propagator_conformal_transformation}
    \tilde{\mathscr{K}}(\boldsymbol{q}_{\sf f},T_{\sf f} ; \boldsymbol{q}_{\sf i},T_{\sf i}) & = \Theta(\mathscr{T})(\Omega(\boldsymbol{q}_{\sf f}) \Omega(\boldsymbol{q}_{\sf i}))^{\frac{2-D}{2}} [\mathscr{G}(\boldsymbol{q}_{\sf f})\mathscr{G}(\boldsymbol{q}_{\sf i})]^{-\frac{1}{4}} \int \mathscr{D}q^A \mathscr{D} p_{A} \exp \Bigg[\frac{i}{\hbar} \int_{0}^{\mathscr{T}} \rd t \Bigg( p_{A} \dot{q}^A \nonumber\\
    & - \Big(\frac{1}{2} \eta^{ab} e^{A}_{a} p_{A} p_{B} e^{B}_{b} - U(\boldsymbol{q})+\Delta V_{Q}^{\sf PF} + \hbar^2 \frac{D-2}{8(D-1)} \mathcal{R}\Big)\Bigg) \Bigg] \nonumber\\
    & = (\Omega(\boldsymbol{q}_{\sf f}) \Omega(\boldsymbol{q}_{\sf i}))^{\frac{2-D}{2}} \mathscr{K}(\boldsymbol{q}_{\sf f},T_{\sf f} ; \boldsymbol{q}_{\sf i},T_{\sf i}), ~~~~~~ (\mathscr{T}>0).
\end{align}
which matches the expectation from the discussion of the canonical case above. In the last equality, we have assumed $\mathscr{T}>0$. We see that since the integration over $u_{\sf f}$ was restricted to the half-real line, the integral of the form
\begin{align}
    \int_{0}^{\infty} \rd u_{\sf f} \, \delta(\mathscr{T}-u_{\sf f}) f(u_{\sf f})
\end{align}
is nonzero only if $\mathscr{T}>0$, hence the result is a Heaviside theta function $\Theta(\mathscr{T})$. Then, the Feynman propagator is only nonzero if $\mathscr{T}=T_{\sf f}-T_{\sf i}>0$, \textit{i.e.}, it is now a causal propagator. As a result, the fluid variable $T$ can be interpreted as a meaningful clock and we have an arrow of time. Notice that this is a direct result of the restriction of the range of $u_{\sf f}$ to the half-line in Eq. \ref{eq:identity_delta_function}. Had we considered a full-line integration of $u_{\sf f}$, the result would not have been a causal propagator, justifying our choice for the range of integration.

Then, using the path integral kernel, we can write the final wave function from the initial wave function of the system as
\begin{align}
    \tilde{\Psi}(\boldsymbol{q}_{\sf f},T_{\sf f}) & = \int \tilde{\mathscr{K}}(\boldsymbol{q}_{\sf f},T_{\sf f} ; \boldsymbol{q}_{\sf i},T_{\sf i}) \tilde{\Psi} (\boldsymbol{q}_{\sf i},T_{\sf i}) \frac{\sqrt{\tilde{\mathscr{G}}(\boldsymbol{q}_{\sf i})}}{\Omega^2(\boldsymbol{q}_{\sf i})} \rd^D \boldsymbol{q}_{\sf i} \nonumber\\
    & = \Omega(\boldsymbol{q}_{\sf f})^{\frac{2-D}{2}} \left(\int \mathscr{K}(\boldsymbol{q}_{\sf f},T_{\sf f} ; \boldsymbol{q}_{\sf i},T_{\sf i}) \left(\Omega(\boldsymbol{q}_{\sf i})^{-\frac{2-D}{2}} \tilde{\Psi} (\boldsymbol{q}_{\sf i},T_{\sf i})\right) \sqrt{\mathscr{G}(\boldsymbol{q}_{\sf i})} \rd^D \boldsymbol{q}_{\sf i}\right)
\end{align}
The consistency of the transitive property of the path integral suggests that the wave function must transform as follows under the conformal transformation or lapse rescaling
\begin{align}\label{eq:wave_function_conformal_transformation}
    \tilde{\Psi}(\boldsymbol{q},T) = \Omega(\boldsymbol{q})^{\frac{2-D}{2}}\Psi(\boldsymbol{q},T).
\end{align}
In \cite{Halliwell:1988wc}, the author had derived the above result in the context of canonical quantization and expected that a path integral proof could also be provided. Here, we have presented such a proof for path integrals albeit for the case of $D$-dimensional minisuperspace with a reference fluid.

The transformation of the wave function under lapse rescaling can also be argued from a different point of view. From the spectral decomposition representation of the Feynman kernel, we can write
\begin{align} \label{eq:spectral_decomposition}
    \mathscr{K}(\boldsymbol{q}_{\sf f},T_{\sf f}; \boldsymbol{q}_{\sf i},T_{\sf i}) & = \int \rd E_{\sigma} \, \e^{-\frac{i}{\hbar} E_{\sigma} (T_{\sf f} - T_{\sf i})} \Psi_{\sigma}^* (\boldsymbol{q}_{\sf f}) \Psi_{\sigma}(\boldsymbol{q}_{\sf i}),
\end{align}
where the integration $\int \rd E_{\sigma}$ represents a sum over both continuous and discrete states. For the conformally transformed propagator \ref{eq:propagator_conformal_transformation} then we may write
\begin{align}
    \tilde{\mathscr{K}}(\boldsymbol{q}_{\sf f},T_{\sf f}; \boldsymbol{q}_{\sf i},T_{\sf i}) & = \int \rd E_{\sigma} \left(\Omega(\boldsymbol{q}_{\sf f})^{\frac{2-D}{2}} \Psi_{\sigma} (\boldsymbol{q}_{\sf f})\e^{\frac{i}{\hbar} E_{\sigma} T_{\sf f}}\right)^* \left( \Omega(\boldsymbol{q}_{\sf i})^{\frac{2-D}{2}} \Psi_{\sigma}(\boldsymbol{q}_{\sf i})\e^{\frac{i}{\hbar} E_{\sigma} T_{\sf i}}\right),
\end{align}
indicating the scaling property \ref{eq:wave_function_conformal_transformation} of the wave function under lapse rescaling.

Furthermore, it is important to note that the observables, such as the $n$-point correlation function
\begin{align}
    \langle q^{A_{1}}(t_{1}) q^{A_{2}}(t_{2}) \dots q^{A_{n}}(t_{n}) \rangle \equiv \frac{\int \rd N (t_{\sf f} - t_{\sf i})\int \mathscr{D}\boldsymbol{q} \mathscr{D}\boldsymbol{p} \, q^{A_{1}}(t_{1}) q^{A_{2}}(t_{2}) \dots q^{A_{n}}(t_{n}) \e^{i S_{\rm eff}}}{\int \rd N (t_{\sf f} - t_{\sf i}) \int \mathscr{D}\boldsymbol{q} \mathscr{D}\boldsymbol{p} \, \e^{i S_{\rm eff}}},
\end{align}
remains invariant under lapse rescaling as the preceding steps are unaffected by the insertion $q^{A_{1}}(t_{1}) q^{A_{2}}(t_{2}) \dots q^{A_{n}}(t_{n})$ into the path integral, and the boundary factors $(\Omega(\boldsymbol{q}_{\sf f}) \Omega(\boldsymbol{q}_{\sf i}))^{\frac{2-D}{2}}$ cancel from both numerator and denominator. The effective action $S_{\rm eff}$, here, is the one defined with the $\mathcal{O}(\hbar^2)$ correction, $\Delta V_{Q}^{\sf PF}$, added to the potential in the conventional action. A similar conclusion holds for the insertion of any string of coordinates or momenta
\begin{align}
    \langle p_{A_{1}}(t_{1}) p_{A_{2}}(t_{2}) \dots p_{A_{n}}(t_{n}) q^{A_{n+1}}(t_{n+1}) q^{A_{n+2}}(t_{n+2}) \dots q^{A_{n+m}}(t_{n+m}) \rangle.
\end{align}
The functions $q^{A}(t)$ and $p_{A}(t)$ can be further rearranged in the string by the time at which these are evaluated. Recall that the ordering of the operators in the canonical formalism corresponds to an ordering in time at which these functions are evaluated in the functional approach \cite{RevModPhys.20.367}. Correlators of coordinates and momenta of all possible order, still, remain the same since boundary factors depending on the conformal parameter $\Omega(\boldsymbol{q}_{\sf i/f})$ cancel from the numerator and denominator.

Then the upshot is that $\xi$ cannot be completely arbitrary and when one chooses $\xi = \frac{D-2}{8(D-1)}$, for a $D$-dimensional minisuperspace, thereby fixes the potential-class ambiguity term, then the new wave function conformally transforms as $\Psi\to\tilde{\Psi}=\Omega^{(2-D)/2}\Psi$, such that the Wheeler-DeWitt equation satisfies $(\hat{\tilde{\mathscr{H}}}+\hat{\tilde{p}}_T)\tilde{\Psi} = \Omega^{-(D+2)/2}(\hat{\mathscr{H}}+\hat{p}_{T})\Psi=0$, and the path integral kernel changes as ${\mathscr{K}}(\boldsymbol{q}_{\sf f},T_{\sf f} ; \boldsymbol{q}_{\sf i},T_{\sf i}) \to (\Omega(\boldsymbol{q}_{\sf f}) \Omega(\boldsymbol{q}_{\sf i}))^{\frac{2-D}{2}} {\mathscr{K}}(\boldsymbol{q}_{\sf f},T_{\sf f} ; \boldsymbol{q}_{\sf i},T_{\sf i})$. These changes do not affect the inner products or the correlation functions. Since theories differing by lapse rescalings correspond to Hamiltonians differing by operator ordering ambiguity (of conformal-class), such ambiguities play no role in physical predictions. The above argument, however, works only for $D>1$, and breaks down in the case for a one-dimensional minisuperspace since $\mathcal{R}$ is identically zero in one-dimension, and thus, there is no extra term to counterbalance the change in kinetic differential operator term due to the lapse rescaling. For $D=1$, the theories differing by conformal-class ambiguity do have different physical predictions (see for example \cite{Sahota:2021oid,Sahota:2022qbc,Sahota:2023kox,Sahota:2023uoj,Sahota:2023vnm}).

To summarize, \textit{building quantum theories demanding general covariance under point canonical transformations and conformal invariance under the lapse rescaling (in $D>2$) leads to unique physical predictions for a large class of operator-ordering ambiguous quantum Hamiltonians.}
In $D=1$ minisuperspace, the symmetry-guided arguments constructed above do not work, and there is an ambiguity associated with lapse rescaling, which we will show in the next section, with an explicit example, is related to the operator ordering ambiguity.

\section{Examples of exactly solvable cases: canonical formalism}\label{sec:CQG}
In this section, we show with examples that the quantization of lapse-rescaled Hamiltonians with Laplace-Beltrami ordering is equivalent to a one-parameter family of different orderings of the original classical Hamiltonian. We consider two cases: flat-FLRW spacetime with perfect fluid where the dimension of the minisuperspace is one and Bianchi I spacetime with a massless scalar field and perfect fluid clock where the dimension of minisuperspace is 4. 

\subsection{Case I: Flat FLRW spacetime with a perfect fluid as clock}\label{sec:CQG1}

The metric of the flat FLRW spacetime in the spherical polar coordinate system can be written as
\begin{align} \label{eq:metric_flrw}
    \rd s^2 = - N(t)^2 \rd t^2 + a(t)^2 \left[ \rd r^2 + r^2 \left( \rd \theta^2 + \sin^2 \theta \rd \phi^2 \right)\right],
\end{align}
where $a(t)$ is the scale factor of the Universe, and $N(t)$ is the lapse function. The classical dynamics of this spacetime will be determined by the Einstein-Hilbert action. From the action, we shall see that the lapse function only appears as a Lagrange's multiplier corresponding to the Hamiltonian constraint. Thus, in this minisuperspace model, the only coordinate is the scale factor of the universe $a(t)$, \textit{i.e.}, $D=1$. As a result, its quantization is analogous to a one-dimensional non-relativistic quantum mechanical problem.

\subsubsection{Classical action}\label{sec:CQG11}

The Einstein-Hilbert action without a cosmological constant and GHY boundary term \cite{York:1972sj,Gibbons:1976ue} is
\begin{align}\label{eq:Einstein-Hilbert}
    S_{\text{E-H}} = \frac{1}{16\pi G}\int_{\mathscr{M}} \rd^4 x \sqrt{-g} R - \frac{1}{8\pi G} \int_{\partial \mathscr{M}} \rd^3 y \sqrt{h} \mathcal{K},
\end{align}
where $G$ is Newton's universal gravitational constant, $h$ is the determinant of the induced metric $h_{ij}$ on the boundary 3-hypersurfaces $\partial \mathscr{M}$, and $\mathcal{K}$ is the trace of the extrinsic curvature tensor $\mathcal{K}_{ij} = \frac{1}{2N}\frac{\partial h_{ij}}{\partial t}$ taken with the induced metric. The above action then evaluates to the following given the metric in \ref{eq:metric_flrw}
\begin{align}
    S_{\text{E-H}} & = \frac{3V_{3}}{8\pi G} \int \rd t \left[ - \frac{\dot{a}^2 a}{N} \right],
\end{align}
where $V_{3}$ is the comoving 3-volume of the spacetime patch. The action of general relativity depends on the second-order time derivative of the dynamical metric degrees of freedom (in the present case, $\ddot{a}$). However, this dependence is not of an essential nature, and we have integrated by parts to do away with the second-order derivative with the cost of boundary terms which are precisely canceled by the GHY terms, leading to a consistent variational problem with Dirichlet boundary conditions.

Then it is easy to see that, the first-order action for the gravitational sector is
\begin{align}
    S_{\text{E-H}} & = \int \rd t \left[p_{a} \dot{a} - N \mathscr{H} \right] = \int \rd t \left[p_{a} \dot{a} - N \left(-\frac{2\pi G}{3V_{3}}\frac{p_{a}^2}{a} \right) \right],
\end{align}
where the momentum conjugate to the scale factor is $p_{a} = -\frac{3V_{3}a}{4\pi G N}\dot{a}$. Notice that the gravitational kinetic term appears with an unusual negative sign. Then the total action of the minisuperspace including a clock fluid reads as
\begin{align}\label{eq:one_dimensional_action}
    S_{\sf Total} = S_{\text{E-H}} + S_{\sf clock} = \int \rd t \left[p_{a} \dot{a}+ p_{T}\dot{T} - N (\mathscr{H} + h^{-\omega/2} p_{T}) \right] = \int \rd t \left[p_{a} \dot{a} + p_{T}\dot{T} - N \left(-\frac{2\pi G}{3V_{3}}\frac{p_{a}^2}{a} + a^{-3\omega} p_{T} \right) \right].
\end{align}
See, appendix \ref{sec:schutz_fluid} for a derivation of the Hamiltonian. There is a factor of $a^{-3\omega}$ with the momentum of the fluid, which differs from the form of the Hamiltonian assumed in sections \ref{sec:QGF}, and \ref{sec:QGF}. However, this assumption was made for the simplicity of the discussions and the presence of this factor does not alter much the arguments of the preceding sections. One can always make a time coordinate transformation $\rd t \to \rd s ~ a^{3\omega}$ to remove the factor with $p_{T}$ before integrating out the $(T,p_{T})$ variables. However, one difference from the above discussions we need to be mindful of is the change in that of the measure for inner products. Notice that the Wheeler-DeWitt equation corresponding to the action in \ref{eq:one_dimensional_action} has the form
%%%%%%%%%%%%%%%%%%%%%%%%%%%%%%%%%%%%%%%%%%%%%%%%%%%%%%%%%%%%%%%%%%%%%%%%%%%%%%%%%%%%%%%%%%%%%%%%%%%
\begin{align}
    \hat{\mathscr{H}} {\Psi} = i h^{-\frac{\omega}{2}} \frac{\partial {\Psi}}{\partial T}.
\end{align}
%%%%%%%%%%%%%%%%%%%%%%%%%%%%%%%%%%%%%%%%%%%%%%%%%%%%%%%%%%%%%%%%%%%%%%%%%%%%%%%%%%%%%%%%%%%%%%%%%%%%
If both sides are multiplied with the factor $h^{\frac{\omega}{2}}$, then the right hand side of the above equation assumes the conventional form of a Schr\"{o}dinger equation with the Hamiltonian $h^{\frac{\omega}{2}}\hat{\mathscr{H}}$. As a result, the integration measure with respect to which this operator is Hermitian picks up an additional factor $\mu \to \mu h^{-\frac{\omega}{2}}$, where $\mu$ is the measure for which $\hat{\mathscr{H}}$ is Hermitian.

In the following, we shall discuss the issue of the quantization of this one (or, rather $1+1$) dimensional minisuperspace model.
\subsubsection{Quantization}\label{sec:CQG12}
For the case of one-dimensional minisuperspace, we expect the conformal invariance to break down. Since by lapse rescaling, we can bring about arbitrary factors of $a(t)$ in the kinetic term, lapse rescaling should be a source for the operator ordering ambiguity in the quantum Hamiltonian---the conformal-class as discussed above. To quantify this ambiguity, instead of starting with the classical action \ref{eq:one_dimensional_action}, we first perform a lapse rescaling $N\to N a^{-2p+1}$. The resulting rescaled action is

\begin{align}\label{eq:one_dimensional_rescaled_action}
    S_{\sf Total} = \int \rd t \left[p_{a} \dot{a} - N \left(-\frac{2\pi G}{3V_{3}}\frac{p_{a}^2}{a^{2p}} + a^{-3\omega-2p+1} p_{T} \right) \right].
\end{align}
The above expression represents a large class of parametrized action, all of which lead to inequivalent quantum theories. Here, $p=\frac{1}{2}$ corresponds to the minisuperspace action \ref{eq:one_dimensional_action} due to the FLRW metric \ref{eq:metric_flrw}. On the other hand, $p=-(3\omega-1)/2$ corresponds to a minisuperspace model in which the fluid momentum $p_{T}$ is decoupled from the scale factor degree of freedom. The above action \ref{eq:one_dimensional_rescaled_action} can also be thought of as a conformally transformed action starting from the action in the case $p=-(3\omega-1)/2$, that is,
%%%%%%%%%%%%%%%%%%%%%%%%%%%%%%%%%%%%%%%%%%%%%%%%%%%%%%%%%%%%%%%%%%%%%%%%%%%%%%%%%%%%%%%%%%%%%
\begin{align}\label{eq:intermediate_action}
    S_{\sf Total} = \int \rd t \left[p_{a} \dot{a} - N \left(-\frac{2\pi G}{3V_{3}}\frac{p_{a}^2}{a^{-3\omega+1}} + p_{T} \right) \right].
\end{align}
%%%%%%%%%%%%%%%%%%%%%%%%%%%%%%%%%%%%%%%%%%%%%%%%%%%%%%%%%%%%%%%%%%%%%%%%%%%%%%%%%%%%%%%%%%%%%%%%%%%%
Then the conformal factor $\Omega^{-2} = a^{-3\omega-2p+1}$ brings the action in \ref{eq:intermediate_action} into the form in \ref{eq:one_dimensional_rescaled_action}.

The Wheeler-DeWitt equation corresponding to the action \ref{eq:one_dimensional_rescaled_action}, obtained with Laplace-Beltrami prescription, reads as
\begin{align}\label{eq:WDW_1D}
    \left(\frac{\ell^2_{\rm Pl}}{12V_{3}} a^{3\omega-1+p} \frac{\partial}{\partial a} a^{-p} \frac{\partial}{\partial a} - i \frac{\partial}{\partial T}\right)\Psi = 0,
\end{align}
where $\ell_{\rm Pl}=\sqrt{8\pi G \hbar}$ is the reduced Planck length. Thus, we have a one-parameter family of inequivalent quantum theories, the parameter $p$ characterizing the operator ordering ambiguity inherent in the quantization procedure. Below, we summarize the above discussion in a flow diagram.
{\linespread{0}
\begin{center}
    \begin{tikzcd}[row sep=2cm, column sep=2cm]
{H = N \left(-\frac{2\pi G}{3V_{3}}\frac{p_{a}^2}{a} + \frac{p_{T}}{a^{3\omega}} \right)} \arrow[d, "\text{\parbox{2.5cm}{\centering quantization: Laplace-Beltrami ordering}}"'] \arrow[swap]{rr}{\text{\parbox{4.4cm}{\centering Classically equivalent, that is, leads to the same Friedmann equation}}}[swap]{\text{\parbox{4.2cm}{\centering lapse rescaling $N\to N a^{-2p+1}$}}}                                                               &  & {{\tilde{H} = \tilde{N} \left(-\frac{2\pi G}{3V_{3}}\frac{p_{a}^2}{a^{2p}} + p_{T} a^{-3\omega-2p+1} \right)}} \arrow[d, "\text{\parbox{2.5cm}{\centering quantization: Laplace-Beltrami ordering}}"]                                   \\
{\left( \frac{\ell_{\rm Pl}^2}{12 V_{3}} a^{3\omega-\frac{1}{2}} \frac{\partial }{\partial a} a^{-\frac{1}{2}} \frac{\partial}{\partial a} + i \frac{\partial}{\partial T}\right)\psi = 0 } \arrow[leftrightarrow,swap,"\not\equiv" description]{rr}{\text{\parbox{3.6cm}{\centering differs by operator ordering (conformal-class)}}}[swap]{\text{\parbox{3.6cm}{\centering quantum mechanically inequivalent, when $(p\neq 1/2)$}}} &  & { \left( \frac{\ell_{\rm Pl}^2}{12 V_{3}} a^{3\omega-1+p} \frac{\partial }{\partial a} a^{-p} \frac{\partial}{\partial a} + i\frac{\partial}{\partial T}\right)\tilde{\psi} = 0 }
\end{tikzcd}
\end{center}
}
We remark that the form of the Wheeler-DeWitt equation above is equivalent to the one suggested in \cite{Kiefer_2019,Sahota:2023kox} with $q=0$. A non-zero $q$ corresponds to having factor-class ambiguity. However, we will ignore this as the factor-class ambiguity is resolved by simply demanding the hermiticity of the quantum Hamiltonian. In \cite{Kiefer_2019,Sahota:2023kox} it was noticed that the inner products do not depend on the parameter $q$.
The Hamiltonian operator in \ref{eq:WDW_1D} is Hermitian with inner product
\begin{align}\label{eq:one_dimensional_inner_product}
    \braket{\psi|\chi}=\int_0^\infty da\;\sqrt{\frac{3V_{3}}{4\pi G}}a^{1-3\omega-p} \psi^*(a,T)\chi(a,T).
\end{align}
Notice that the inner product measure for the quantum theory resulting from \ref{eq:intermediate_action} is given by $\sqrt{|\mathscr{G}|} = \sqrt{\frac{3V_{3}}{4\pi G}} a^{\frac{-3\omega+1}{2}}$. However, we know, the quantum theory given by the Wheeler-DeWitt equation \ref{eq:WDW_1D} results from the action \ref{eq:one_dimensional_rescaled_action}, which has been obtained from \ref{eq:intermediate_action} by a conformal transformation. Therefore, according to \ref{eq:inner_product_measure}, the inner product measure for the quantum theory corresponding to the classical action \ref{eq:one_dimensional_rescaled_action}, in $D=1$ case, can be written as
%%%%%%%%%%%%%%%%%%%%%%%%%%%%%%%%%%%%%%%%%%%%%%%%%%%%%%%%%%%%%%%%%%%%
\begin{align}
    \sqrt{|\mathscr{G}|} \to \sqrt{|{\mathscr{G}|}} \Omega^{-1} = \sqrt{\frac{3V_{3}}{4\pi G}} a^{-3\omega+1-p},
\end{align}
%%%%%%%%%%%%%%%%%%%%%%%%%%%%%%%%%%%%%%%%%%%%%%%%%%%%%%%%%%%%%%%%%
which is in agreement with \ref{eq:one_dimensional_inner_product}.

The stationary states for the system with ansatz $\psi(a,T)=\psi_E(a)e^{iET}$, and $E>0$ takes the form
\begin{align} \label{eq:Eigenstate}
    \psi_{E}(a) & = \frac{\left(2E\right)^{\frac{1}{4}}}{\sqrt{\hbar} } a^{\frac{1+p}{2}}  J_{\frac{|1+p|}{3(1-\omega)}} \left(\frac{2\sqrt{2 E \left(\frac{3V_{3}}{4\pi G}\right)}}{3\hbar (1-\omega) } a^{\frac{3(1-\omega)}{2}}\right).
\end{align}
At this stage, we can see the ambiguity parameter $p$ leaves its imprint on the probability distribution $a^{1-3\omega-p}|\psi_E|^2$
\begin{align}
    a^{1-3\omega-p}|\psi_E|^2=\frac{\sqrt{2E}}{\hbar}a^{2-3\omega} \left[J_{\frac{|1+p|}{3(1-\omega)}} \left(\frac{2\sqrt{2 E \left(\frac{3V_{3}}{4\pi G}\right)}}{3\hbar (1-\omega) } a^{\frac{3(1-\omega)}{2}}\right)\right]^2,
\end{align}
through the index of the Bessel function, leading to ambiguity-ridden quantum dynamics \cite{Sahota:2021oid,Sahota:2022qbc,Sahota:2023kox}. We will return to this discussion again after evaluating the path integral in this case.

\subsection{Case II: Bianchi I spacetime with a free massless scalar field and perfect fluid clock}\label{sec:CQG2}

Next, we consider a minisuperspace of higher dimension, the Bianchi I spacetime, which describes a spatially flat, homogeneous, but anisotropic universe. The anisotropic nature of the spacetime may play an important role in understanding the quantum nature of the early universe. Even in the classical treatment of cosmology, there are indications that in the case of a bounce, the anisotropy may grow during the initial contracting phase of the universe and make the bounce unstable (see however \cite{Rajeev:2021yyl}). Therefore, considering an anisotropic universe in quantum cosmology remains an interesting problem.

A Bianchi I spacetime is described with the following metric \cite{Pereira:2007yy,Grain:2020wro}
\begin{align} \label{eq:metric_bianchi}
    \rd s^2 = - N(t)^2 \rd t^2 + a^2(t) \left[\e^{\beta_{1}(t) + \sqrt{3}\beta_{2}(t)} \rd x^2 + \e^{\beta_{1}(t) - \sqrt{3}\beta_{2}(t)} \rd y^2 + \e^{-2 \beta_{1}(t)} \rd z^2\right],
\end{align}
where $a(t)$ is called the isotropized scale factor of the universe and $\beta_{1,2}(t)$ are the parameters capturing the anisotropic nature of the spacetime.

\subsubsection{Classical action}\label{sec:CQG21}

The Einstein-Hilbert gravity action \ref{eq:Einstein-Hilbert} calculated for the metric \ref{eq:metric_bianchi} takes the form
\begin{align}
    S_{\text{E-H}} = \frac{3V_{3}}{8\pi G} \int \rd t \left[ - \frac{\dot{a}^2 a}{N} + \frac{a^3}{4 N}\left(\dot{\beta}_{1}^2 + \dot{\beta}_{2}^2\right) \right],
\end{align}
where again the boundary term coming from the integration by parts of the second-order time derivative of the isotropized scale factor is exactly canceled by the GHY boundary term. Again, it will be convenient to express the action in the first-order form as
\begin{align}
    S_{\text{E-H}} & =  \int {\rm d}t \left[p_{a} \dot{a} + \boldsymbol{p_{\beta}} \cdot \dot{\boldsymbol{\beta}} - N \left(- \frac{2\pi G}{3V_{3}}  \frac{p_{a}^2}{a} + \frac{8\pi G}{3V_{3}} \frac{\boldsymbol{p_{\beta}}^2}{a^3} \right) \right]
\end{align}
where we have introduced, for notational convenience, the boldface notation to denote a vector in the anisotropic parameter space $\boldsymbol{\beta}=(\beta_{1}(t),\beta_{2}(t))$. The conjugate momenta corresponding to the isotropized scale factor is $p_{a} = -\frac{3V_{3}a}{4\pi G N}\dot{a}$ and similarly, for the anisotropic parameters the momenta are defined as $\boldsymbol{p_{\beta}} = (\frac{3V_{3} a^3}{16 \pi G N}\dot{\beta}_{1},\frac{3V_{3} a^3}{16 \pi G N}\dot{\beta}_{2})$. The inner product $\boldsymbol{p_{\beta}}^2$ is defined with respect to an Euclidean metric. Note that $\boldsymbol{\beta}$ is not a vector in the physical spacetime. Moreover, we also note that the flat, homogeneous, and isotropic FLRW spacetime is a special case of the Bianchi I spacetime occurring when the anisotropic parameter vanishes identically, \textit{i.e.} $\boldsymbol{\beta}=\boldsymbol{0}$.

We can also consider a massless, free, homogeneous scalar field as an additional matter component of the universe. For the metric \ref{eq:metric_bianchi}, the action for the scalar field reads as follows
\begin{align}
    S_{\phi} & = -  \frac{1}{2} \int \rd^4 x \sqrt{-g}g^{\mu\nu} \partial_{\mu}\phi \partial_{\nu} \phi \nonumber\\
    & =  V_{3} \int \rd t \, a^{3} \frac{\dot{\phi}^2}{2N}.
\end{align}
Expressing in terms of the Hamiltonian we find the action in the first-order form to be
\begin{align}
    S_{\phi} & =  V_{3} \int \rd t \left[ p_{\phi} \dot{\phi} -  \frac{N}{2 V_{3}} \frac{p_{\phi}^2}{ a^{3}}\right],
\end{align}
where the conjugate momentum to $\phi$ is defined as $p_{\phi} = \frac{V_{3}a^{3}}{N}\dot{\phi}$. Putting everything together, we find that the total action of the system, with Bianchi I universe with scalar field, and fluid clock, is given by
\begin{align} \label{eq:mini_action_bianchi}
    S_{\sf Total} & = S_{\text{E-H}} + S_{\phi} + S_{\sf clock} = \int {\rm d}t \left[p_{a} \dot{a} + \boldsymbol{p_{\beta}} \cdot \dot{\boldsymbol{\beta}} + p_{T}\dot{T} + p_{\phi} \dot{\phi} - N \left(- \frac{2\pi G}{3V_{3}}  \frac{p_{a}^2}{a} + \frac{8\pi G}{3V_{3}} \frac{\boldsymbol{p_{\beta}}^2}{a^3} + \frac{p_{\phi}^2}{2V_{3}a^3} + \frac{p_{T}}{a^{3\omega}} \right) \right].
\end{align}
Now we have a four-dimensional ($D=4$) minisuperspace $(a,\beta_{1},\beta_{2},\phi)$. Like before let us introduce the $p$-parametrization by rescaling the lapse function as $N\to N a^{-2p+1}$ and the rescaled action takes the form
\begin{align} \label{eq:mini_action_bianchi_rescaled}
    S_{\sf Total} & = \int {\rm d}t \left[p_{a} \dot{a} + \boldsymbol{p_{\beta}} \cdot \dot{\boldsymbol{\beta}} + p_{T}\dot{T} + p_{\phi} \dot{\phi} - N \left(- \frac{2\pi G}{3V_{3}}  \frac{p_{a}^2}{a^{2p}} + \frac{8\pi G}{3V_{3}} \frac{\boldsymbol{p_{\beta}}^2}{a^{2p+2}} + \frac{p_{\phi}^2}{2V_{3}a^{2p+2}} + \frac{p_{T}}{a^{3\omega+2p-1}} \right) \right].
\end{align}
Again, one may think that the above action is obtained from starting with the $p = -(3\omega-1)/2$ case and lapse rescaling with the factor $\Omega^{-2} = a^{-3\omega-2p+1}$.

In the following, we start by discussing the quantization procedure and the ambiguities associated with it. 
%%%%%%%%%%%%%%%%%%%%%%%%%%%%%%%%%%%%%%%%%%%%%%%%%%%%%%%%%%%%%%%%%%%%%%%%%%%%%%%%%%%%%%%%%%%%%%%
\subsubsection{Quantization}\label{sec:CQG22}
%%%%%%%%%%%%%%%%%%%%%%%%%%%%%%%%%%%%%%%%%%%%%%%%%%%%%%%%%%%%%%%%%%%%%%%%%%%%%%%%%%%%%%%%%%%%%%%%%%%%%%%%
For the Hamiltonian constraint
\begin{align}
    - \frac{2\pi G}{3V_{3}}  \frac{p_{a}^2}{a^{2p}} + \frac{8\pi G}{3V_{3}} \frac{\boldsymbol{p_{\beta}}^2}{a^{2p+2}} + \frac{p_{\phi}^2}{2V_{3}a^{2p+2}} + \frac{p_{T}}{a^{3\omega+2p-1}}\approx 0,
\end{align}
the Wheeler-DeWitt equation
%%%%%%%%%%%%%%%%%%%%%%%%%%%%%%%%%%%%%%%%%%%%%%%%%%%%%%%%%%%%%%%%%%%%%%%%%%%%%%%%%%%%%%%
\begin{align}
    \left(-\frac{\hbar^2}{2}\frac{1}{\sqrt{|\mathscr{G}|}}\partial_{A}\sqrt{|\mathscr{G}|}\mathscr{G}^{AB}\partial_{B} + \hbar^2 \frac{D-2}{8(D-1)} \mathcal{R} + E a^{-3\omega-2p+1}\right)\psi = 0,
\end{align}
%%%%%%%%%%%%%%%%%%%%%%%%%%%%%%%%%%%%%%%%%%%%%%%%%%%%%%%%%%%%%%%%%%%%%%%%%%%%
with ansatz $\psi(a,\phi,\boldsymbol{\beta},T)=\Psi_E(a,\phi,\boldsymbol{\beta})e^{iET/\hbar}$, takes the form
\begin{align}\label{eq:bianchi_wheeler_dewitt}
    \left(\frac{4\pi G \hbar^2}{6V_{3}}a^{3\omega-4-2p}\partial_{a} a^{3+2p} \partial_{a} -a^{3\omega-3}\left( \frac{\hbar^2}{2V_{3}}\partial^2_{\phi} + \frac{8\pi G \hbar^2}{3V_{3}a^{2+2p}} \left(\partial^2_{\beta_{1}} + \partial^2_{\beta_{2}}\right)\right) - \frac{2\pi G (1+p)^2 \hbar^2}{3V_3 a^{2+2p}} + E\right)\Psi_{E} = 0.
\end{align}
Where the Hamiltonian operator is Hermitian with the inner product
\begin{align}\label{BianchiInnerProduct}
    \braket{\psi|\chi}=\int_{-\infty}^\infty d\phi\int_{-\infty}^\infty d\beta_1\int_{-\infty}^\infty d\beta_2\int_{}^\infty da\;\frac{3\sqrt{3}V_{3}^2}{32(\pi G)^{\frac{3}{2}}}a^{4+2p-3\omega}\psi^*(a,\phi,\boldsymbol{\beta},T)\chi(a,\phi,\boldsymbol{\beta},T).
\end{align}
For the eigenstates of $\hat{\boldsymbol{p}}_{\boldsymbol{\beta}}$, i.e. $\frac{\e^{\pm i \boldsymbol{\Pi\cdot\beta}}}{{2\pi\hbar}}$, the stationary states for this system take the form
\begin{align}\label{eq:wave_function_bianchi}
    \Psi_{E}(a,\phi,\boldsymbol{\beta}) & = \frac{4\left(2E\right)^{\frac{1}{4}}\sqrt{G\pi}}{\sqrt{3\hbar} V_{3}^{\frac{3}{4}} (2\pi\hbar)^{\frac{3}{4}}} \e^{\pm\frac{i}{\hbar}\left(\boldsymbol{\Pi}\cdot\boldsymbol{\beta} + \mathsf{p}_{\phi} \phi \right)}  a^{-(1+p)}  J_{\nu} \left(\frac{2\sqrt{2 E \left(\frac{3V_{3}}{4\pi G}\right)}}{3\hbar (1-\omega) } a^{\frac{3(1-\omega)}{2}}\right).
\end{align}
where the index of the Bessel function is  
\begin{align}
    \nu = \pm i\frac{2\sqrt{2}}{3(1-\omega)}\sqrt{ 3\left(\frac{\mathsf{p}_{\phi}}{\sqrt{8\pi G}\hbar}\right)^2 + 2 \left(\frac{\boldsymbol{\Pi}}{\hbar}\right)^2}.
\end{align}
In this case, it is easy to see that the probability distribution
\begin{align}
    a^{4+2p-3\omega}|\Psi_{E}(a,\phi,\boldsymbol{\beta})|^2\propto a^{2-3\omega}\left[J_{\nu} \left(\frac{2\sqrt{2 E \left(\frac{3V_{3}}{4\pi G}\right)}}{3\hbar (1-\omega) } a^{\frac{3(1-\omega)}{2}}\right)\right]^2
\end{align}
is independent of the ambiguity parameter $p$. Unlike the case of $D=1$, the index of the Bessel function, here, does not depend on the parameter $p$. The parameter $p$ appears in the wave function only as an overall factor $a^{-(1+p)}$, and gets canceled by the integration measure in the inner product. In Appendix \ref{app:Bianchi}, we show that the expectation value of the geometric observables is also independent of the ambiguity parameter, implying that the quantum cosmology in $D>2$ or in this particular example $D=4$, is ambiguity-free: infinitely many different quantum Hamiltonians parametrized by $p$ (conformal-class ambiguous Hamiltonians) have been rendered equivalent by a simple choice of the potential-class ambiguity term.

% In what follows, we shall first provide a short discussion on how to obtain the fluid Hamiltonian from Schutz formalism, and then we shall proceed to deal with two explicit examples of path integral computation, which will illustrate our conclusion.

\section{Examples of exactly solvable cases: path integral formalism}\label{sec:ESC}

In the following, we shall discuss the exact path integral quantization of two flat, homogeneous cosmologies: (a) the isotropic FLRW, and (b) the anisotropic Bianchi I spacetime, along with the action for a homogeneous, massless, free scalar field. The case (a) will constitute an example of the quantization of minisuperspace of dimension $D=1$. In this case, after the quantization, the lapse rescaling symmetry is lost and we shall show that we can recover the solutions corresponding to a one-parameter family of Wheeler-Dewitt equations. For conformally invariant quantum theory we need to have a minisuperspace of dimension $D>2$. The case (b) will constitute an example of $D=4$ minisuperspace. We shall show that, in this case, the wave function conformally transforms under lapse rescalings.

\subsection{Case I: Flat FLRW spacetime with a perfect fluid as clock}\label{sec:ESC1}

Now, let us proceed to compute the path integral kernel corresponding to the action \ref{eq:one_dimensional_rescaled_action}. As the action \ref{eq:one_dimensional_rescaled_action} is obtained by lapse rescaling with $\Omega^{-3\omega-2p+1}$ from \ref{eq:intermediate_action}, we can directly use the definition for the rescaled kernel in \ref{eq:conformal_path_integral_kernel}, and write in the present case
\begin{align}
    \tilde{\mathscr{K}}(a_{\sf f},T_{\sf f} ; a_{\sf i},T_{\sf i})= & \int \rd \tilde{N} (t_{\sf f}-t_{\sf i}) \sqrt{\frac{4\pi G}{3V_{3}}}(a_{\sf f} a_{\sf i})^{-\frac{p}{2}} \int \mathscr{D}a \mathscr{D} p_{a} \int \mathscr{D}T \mathscr{D} p_{T} \exp \Bigg[\frac{i}{\hbar} \int_{t_{\sf i}}^{t_{\sf f}} \rd t \Bigg( p_{a} \dot{a} + p_{T}\dot{T} \nonumber\\
    &  - \tilde{N} \Big(- \frac{2\pi G}{3V_{3}} a^{-p} p_{a}^2 a^{-p} + \widetilde{\Delta V}_{Q}^{\sf PF} + a^{-3\omega-2p+1} p_{T}\Big)\Bigg) \Bigg],
\end{align}
where the quantum correction $\widetilde{\Delta V}_{Q}^{\sf PF}$ has to be calculated from the expression \ref{eq:quantum_potential}, with the conformally transformed tetrad $\tilde{e}^a_{A} = \sqrt{3V_{3}/4\pi G} a^{p} \delta^a_{A}$, and the result is
\begin{align}
    \widetilde{\Delta V}^{\sf PF}_{Q} = - \frac{\hbar^2 \pi G}{6 V_{3}}  \frac{p(p+2)}{a^{2(1+p)}} .
\end{align}
Before we can perform the integrals over clock degrees of freedom $(T,p_{T})$ we have to decouple the fluid's Hamiltonian from the scale factor by means of time redefinition. Like before, we choose $t_{\sf f} = 1$, $t_{\sf i}=0$, rescale $t\to t/\tilde{N}$ and again define a new time coordinate $u\in[0,u_{\sf f}]$ satisfying proper boundary conditions as follows
\begin{align}
    a^{-3\omega-2p+1} \rd t = \rd u,~~\text{or}~~ t(u)=\int_{0}^u \rd u' \, a(t(u'))^{3\omega+2p-1},\\
    t(0) = 0,~ t(u_{\sf f}) = \tilde{N},~~ a(t(0)) = a_{\sf i}, ~~ a(t(u_{\sf f})) = a_{\sf f}.
\end{align}
The above constraint can be imposed by utilizing the following identity
\begin{align}
    1 & = (a_{\sf f} a_{\sf i})^{\frac{3\omega+2p-1}{2}} \int_{0}^{\infty} \rd u_{\sf f} \, \delta \left(\tilde{N} - \int_{0}^{u_{\sf f}} a(u)^{3\omega+2p-1} \rd u\right) \nonumber\\
    & = (a_{\sf f} a_{\sf i})^{\frac{3\omega+2p-1}{2}} \int_{-\infty}^{\infty} \frac{\rd \mathcal{E}}{2\pi\hbar} \e^{\frac{i}{\hbar} \mathcal{E} \tilde{N}} \int_{0}^{\infty} \rd u_{\sf f} \exp \left(- \frac{i}{\hbar} \int_{0}^{u_{\sf f}}  \rd u \, a(u)^{3\omega+2p-1} \mathcal{E} \right).
\end{align}
Then the expression for the time redefined path integral kernel is as follows
\begin{align}
    \tilde{\mathscr{K}}(a_{\sf f},T_{\sf f} ; a_{\sf i},T_{\sf i}) & = \sqrt{\frac{4\pi G}{3V_{3}}}(a_{\sf f} a_{\sf i})^{\frac{3\omega+p-1}{2}} \int \rd \tilde{N} \int \mathscr{D}a \mathscr{D} p_{a}  \int_{-\infty}^{\infty} \frac{\rd \mathcal{E}}{2\pi\hbar}  \e^{\frac{i}{\hbar} \mathcal{E} \tilde{N}} \int_{0}^{\infty} \rd u_{\sf f} \exp \Bigg[\frac{i}{\hbar} \int_{0}^{u_{\sf f}} \rd u \Bigg( p_{a} \dot{a} + \frac{2\pi G}{3V_{3}} a^{\frac{3\omega-1}{2}} p_{a}^2 a^{\frac{3\omega-1}{2}}  \nonumber\\
    &  - a^{3\omega+2p-1} \widetilde{\Delta V}_{Q}^{\sf PF} - a^{-3\omega-2p+1} \mathcal{E} \Bigg) \Bigg] \times \int \mathscr{D}T \mathscr{D} p_{T}  \exp \Bigg[\frac{i}{\hbar} \int_{0}^{u_{\sf f}} \rd u \Big(p_{T}\dot{T} - p_{T}\Big)\Bigg].
\end{align}
Now it is straightforward to compute the path integrals over $(T,p_{T})$, and ordinary integrals over $\tilde{N}$, $\mathcal{E}$, and $u_{\sf f}$. The result is a path integral propagator which gives the quantum amplitude for the transition of the gravitational degree of freedom from $a_{\sf i}$ to $a_{\sf f}$ in a finite time $\mathscr{T}=T_{\sf f} - T_{\sf i}$ measured by the configuration of the clock-fluid
\begin{align}
    \tilde{\mathscr{K}}(a_{\sf f},T_{\sf f} ; a_{\sf i},T_{\sf i}) & = \Theta(\mathscr{T})\sqrt{\frac{4\pi G}{3V_{3}}}(a_{\sf f} a_{\sf i})^{\frac{3\omega+p-1}{2}} \int \mathscr{D}a \mathscr{D} p_{a} \exp \Bigg[\frac{i}{\hbar} \int_{0}^{\mathscr{T}} \rd u \Bigg( p_{a} \dot{a} + \frac{2\pi G}{3V_{3}} a^{\frac{3\omega-1}{2}} p_{a}^2 a^{\frac{3\omega-1}{2}}  - a^{3\omega+2p-1} \widetilde{\Delta V}_{Q}^{\sf PF} \Bigg) \Bigg].
\end{align}
Now, it is our turn to compute the path integral over the scale factor and its conjugate momentum $(a,p_{a})$. Our strategy for this computation would be to introduce coordinate transformations such that we reduce the above path integral to the form of a known radial path integral problem. The first difficulty we notice is the fact that the kinetic term is non-canonical; that is, a factor of $a^{3\omega-1}$ appears with $p_{a}^2$. We can get rid of this factor with yet another time redefinition. The new time coordinate $s\in[0,s_{\sf f}]$ is defined as follows along with the consistency boundary conditions
\begin{subequations}
    \begin{align}
    a^{3\omega-1} \rd u = \rd s,~~\text{or}~~ u(s)=\int_{0}^s \rd s' \, a(s')^{-3\omega+1}, \\ u(0) = 0,~ u(s_{\sf f}) = \mathscr{T},~~ a(u(0)) = a(0) = a_{\sf i}, ~~ a(u(s_{\sf f})) = a(\mathscr{T}) = a_{\sf f}.
\end{align}
\end{subequations}
Like before we have to introduce a Dirac delta function to enforce the constraint $u(s_{\sf f}) = \mathscr{T}$. After the transformation, the Feynman propagator is modified to
\begin{align}
    \tilde{\mathscr{K}}(a_{\sf f},T_{\sf f} ; a_{\sf i},T_{\sf i}) = &  \Theta(\mathscr{T})\sqrt{\frac{4\pi G}{3V_{3}}}(a_{\sf f} a_{\sf i})^{\frac{p}{2}}\int_{-\infty}^{\infty} \frac{\rd E}{2\pi \hbar}\e^{\frac{i}{\hbar} E \mathscr{T}} \int_{0}^{s_{\sf f}} \rd s  \int \mathscr{D}a \mathscr{D} p_{a} \exp \Bigg[\frac{i}{\hbar} \int_{0}^{s_{\sf f}} \rd s \Bigg( p_{a} \dot{a}- \Big(-\frac{2\pi G}{3V_{3}} p_{a}^2  \nonumber\\
    & + a^{2p} \widetilde{\Delta V}_{Q}^{\sf PF}\Big) - E a^{-3\omega+1} \Bigg) \Bigg].
\end{align}
The overhead dot now means time derivative with respect to $s$. After carrying out the Gaussian integrals over the $p_{a}$ variables, we are left with
\begin{align}\label{eq:inverse_square_path_integral}
    \tilde{\mathscr{K}}(a_{\sf f},T_{\sf f} ; a_{\sf i},T_{\sf i}) = &  \Theta(\mathscr{T})\sqrt{\frac{4\pi G}{3V_{3}}}(a_{\sf f} a_{\sf i})^{\frac{p}{2}}\int_{-\infty}^{\infty} \frac{\rd E}{2\pi \hbar}\e^{\frac{i}{\hbar} E \mathscr{T}} \int_{0}^{s_{\sf f}} \rd s  \int \mathscr{D}a \exp \Bigg[\frac{i}{\hbar} \int_{0}^{s_{\sf f}} \rd s \Bigg( -\frac{3V_{3}}{8\pi G} \dot{a}^2 \nonumber\\
    & - a^{2p} \widetilde{\Delta V}_{Q}^{\sf PF} - E a^{-3\omega+1} \Bigg) \Bigg],
\end{align}
where the path integral measure is defined as
\begin{align}
    \mathscr{D}a \equiv \sqrt{\frac{(\frac{3V_{3}}{4\pi G})}{2\pi \hbar \Delta s^{(M)}}}  \prod_{j=1}^{M-1} \int_{0}^{\infty} \sqrt{\frac{(\frac{3V_{3}}{4\pi G})}{2\pi \hbar \Delta s^{(j)}}}  \rd a^{(j)},
\end{align}
where the non-uniform time interval is defined as $\Delta^{(j)} s = (a^{(j)})^{-p} (a^{(j-1)})^{-p} \epsilon $. However, in the following for formal manipulations, we shall not require the above discrete definition of the path integral measure. Notice that the above path integral problem, apart from the unusual negative sign for the kinetic term, is akin to that of a point particle moving in a potential given by
\begin{align}
    V = a^{2p} \widetilde{\Delta V}^{\sf PF}_{Q} + E a^{-3\omega+1} = - \frac{\hbar^2 \pi G}{6 V_{3}}  \frac{p(p+2)}{a^{2}} + E a^{-3\omega+1}.
\end{align}
Moreover, notice that $E$ can be interpreted as the energy of the system since we have the relation
\begin{align}
    \tilde{\mathscr{K}}(a_{\sf f},T_{\sf f} ; a_{\sf i},T_{\sf i}) = \int_{-\infty}^{\infty} \frac{\rd E}{2\pi\hbar} \e^{\frac{i}{\hbar}E\mathscr{T}} G_{E}(a_{\sf f};a_{\sf i}),
\end{align}
where the fixed energy propagator is defined as
\begin{align}\label{eq:propagator_intermediate}
    G_{E}(a_{\sf f};a_{\sf i}) = &  \Theta(\mathscr{T}) \sqrt{\frac{4\pi G}{3V_{3}}} (a_{\sf f}a_{\sf i})^{\frac{p}{2}} \int_{a_{\sf i}}^{a_{\sf f}} \mathscr{D}a \int_{0}^{\infty} \rd s_{\sf f} \exp \left\{\frac{i}{\hbar} \int_{0}^{s_{\sf f}} \rd s \Bigg( - \left(\frac{3V_{3}}{8 \pi G}\right)   \dot{a}^2 + \frac{\hbar^2 \pi G}{6 V_{3}}  \frac{p(p+2)}{a^{2}} - E a^{-3\omega+1} \Bigg)\right\}.
\end{align}
The above path integral is non-trivial because of the term involving arbitrary power of the scale factor, $-E a^{-3\omega+1}$ in the potential. We can, however, introduce a non-linear space-time transformation, $s\to\tau$ and $a(s)\to R(\tau)$, as in \cite{Steiner:1984uf,Grosche:1998yu} to change the problematic term. The specific transformation we need for this purpose is defined as follows
\begin{subequations}\label{eq:space-time_transformation}
\begin{align}\label{eq:coordinate_transform}
    a(s) = R(\tau)^{\frac{2}{3(1-\omega)}}, \qquad
    \rd s = \frac{4}{9(1-\omega)^2} a^{3\omega-1} \rd \tau = \frac{4}{9(1-\omega)^2} R(\tau)^{\frac{2(3\omega - 1)}{3(1-\omega)}} \rd \tau .
\end{align}    
\end{subequations}
With this transformation, the expression for the Feynman propagator changes to
\begin{align}
    \mathscr{K}(a_{\sf f},\mathscr{T};a_{\sf i},0) = &  \Theta(\mathscr{T})\frac{2}{3(1-\omega)} \sqrt{\frac{4\pi G}{3V_{3}}} (a_{\sf f}a_{\sf i})^{\frac{3\omega-1}{4}+\frac{p}{2}} \int_{R_{\sf i}}^{R_{\sf f}} \mathscr{D}R \int_{-\infty}^{\infty} \frac{\rd E}{2\pi\hbar} \e^{\frac{i}{\hbar}E\mathscr{T}} \int_{0}^{\infty} \rd s_{\sf f} \int_{0}^{\infty} \rd \tau_{\sf f}\int_{-\infty}^{\infty} \frac{\rd \lambda}{2\pi\hbar} \e^{\frac{i}{\hbar}\lambda s_{\sf f}}  \nonumber\\
    &  \exp \Bigg\{\frac{i}{\hbar} \int_{0}^{\tau_{\sf f}} \rd \tau \Bigg( - \left(\frac{3V_{3}}{8 \pi G}\right)   {R'}^2 + \frac{4}{9(1-\omega)^2}\left(\frac{\hbar^2 \pi G}{6 V_{3}}  \frac{p(p+2)}{R^{2}} - E - \lambda R^{\frac{2(3\omega-1)}{3(1-\omega)}} \right) + \frac{\hbar^2 L(L + 1)}{2 \left(\frac{3V_{3}}{4\pi G}\right) R^2} \Bigg)\Bigg\},
\end{align}
where we have defined $L = \frac{3\omega-1}{6(1-\omega)}$. The change in time variable required that we insert one more delta function enforcing the time interval constraint. Moreover, due to the specific coordinate transformations we have introduced in \ref{eq:coordinate_transform}, we have also included another $\mathcal{O}(\hbar^2)$ correction to the quantum potential. For the space and time coordinate transformation introduced of the above form, where $a=F(R)=R^{\frac{2}{3(1-\omega)}}$, the quantum correction to the potential is given by a Schwarzian derivative
%%%%%%%%%%%%%%%%%%%%%%%%%%%%%%%%%%%%%%%%%%%%%%%%%%%%%%%%%%%%%%%%%%%%%%%%%%%%%%%%%%%%%%%%%%%%%%%
\begin{align}
    \Delta V = \frac{\hbar^2}{8\left(-\frac{3V_{3}}{4\pi G}\right)} \left[ 3 \left(\frac{F''(R)}{F'(R)}\right)^2 - 2 \frac{F'''(R)}{F'(R)}\right] = - \frac{\hbar^2 L(L+1)}{2\left(\frac{3V_{3}}{4\pi G}\right)R^2}.
\end{align}
%%%%%%%%%%%%%%%%%%%%%%%%%%%%%%%%%%%%%%%%%%%%%%%%%%%%%%%%%%%%%%%%%%%%%%%%%%%%%%%%%%%%%%%%%%%%%%%%%%%%%%
The general proof of the above result can be found in \cite{Steiner:1984uf,Grosche:1998yu,Khandekar:1986ib} (notice that from the action in \ref{eq:propagator_intermediate}, we have identified the mass parameter $m$ with $-\frac{3V_{3}}{4\pi G}$). The inclusion of this $\mathcal{O}(\hbar^2)$ correction is essential in getting the correct propagator and the wave function from the path integral approach and cannot be ignored. Out of the above exercise, we have managed to absorb the $a^{-3\omega+1}$ factor as desired.

Now we consider the integration over $s_{\sf f}$, whose range is the half-line
\begin{align}
    \int_0^\infty \rd s_{\sf f} \, \e^{\frac{i}{\hbar}\lambda s_{\sf f}} = \pi \delta( \lambda/\hbar) + \mathcal{P}\left(\frac{i}{\lambda/\hbar}\right),
\end{align}
where $\mathcal{P}$ denotes the Cauchy principal value. Following this we can immediately perform the integration over $\lambda$, that is,
\begin{align}
    \int_{-\infty}^{\infty}\frac{\rd \lambda}{2\pi \hbar} \left( \pi \delta( \lambda/\hbar) + \mathcal{P}\left(\frac{i\hbar}{\lambda}\right)\right) f(\lambda/\hbar) = \frac{1}{2} f(0) + \frac{i}{2\pi} \mathcal{P}\left(\int_{-\infty}^{\infty} \frac{f(\lambda/\hbar)}{\lambda/\hbar} \frac{\rd \lambda}{\hbar}\right) = f(0),
\end{align}
where we have represented the integrand with $f(\lambda/\hbar)$ and noted the fact that the integrand exponentially decays in the lower-half complex plane. The result of these two consecutive integrations is to set $\lambda$ to zero, and thus we have
\begin{align}
    \mathscr{K}(a_{\sf f},\mathscr{T};a_{\sf i},0) = &  \Theta(\mathscr{T})\frac{2}{3(1-\omega)} \sqrt{\frac{4\pi G}{3V_{3}}} (a_{\sf f}a_{\sf i})^{\frac{3\omega-1}{4}+\frac{p}{2}} \int_{R_{\sf i}}^{R_{\sf f}} \mathscr{D}R \int_{-\infty}^{\infty} \frac{\rd E}{2\pi\hbar} \e^{\frac{i}{\hbar}E\mathscr{T}} \int_{0}^{\infty} \rd \tau_{\sf f} \exp \Bigg\{\frac{i}{\hbar} \int_{0}^{\tau_{\sf f}} \rd \tau \Bigg( \left(- \frac{3V_{3}}{8 \pi G}\right)   {R'}^2 \nonumber\\
    &  - \frac{4E}{9(1-\omega)^2} - \hbar^2\frac{\frac{(1+p)^2}{9(1-\omega)^2 }- \frac{1}{4}}{2\left(-\frac{3V_{3}}{4\pi G}\right) R^2} \Bigg)\Bigg\}.
\end{align}
The above radial path integral is already known in the literature and can be performed exactly (see, section 6.4 in \cite{Grosche:1998yu}, or \cite{peak1969summation,duru1985path,bhagwat1989new}) to yield
\begin{align}
    \mathscr{K}(a_{\sf f},a_{\sf i};\mathscr{T}) = & \Theta(\mathscr{T})\frac{2(a_{\sf f}a_{\sf i})^{\frac{3\omega-1}{4}+\frac{p}{2}}}{3(1-\omega)} \sqrt{\frac{4\pi G}{3V_{3}}}  \int_{-\infty}^{\infty} \frac{\rd E}{2\pi\hbar} \e^{\frac{i}{\hbar} E \mathscr{T}} \int_{0}^{\infty} \rd \tau_{\sf f} \, \e^{- \frac{i}{\hbar} \frac{4E}{9(1-\omega)^2}\tau_{\sf f}} i\frac{3 V_{3} \sqrt{R_{\sf f} R_{\sf i}}}{4\pi G\hbar \tau_{\sf f}} \exp\left[\frac{3 V_{3}}{8i\pi G \hbar \tau_{\sf f}} (R_{\sf f}^2 + R_{\sf i}^2 )\right] \nonumber\\
    & \times I_{\frac{|1+p|}{3(1-\omega)}} \left(i\frac{3 V_{3} R_{\sf f} R_{\sf i}}{4 \pi G \hbar \tau_{\sf f}}\right),
\end{align}
where we have implicitly assumed $\omega<1$ and $\omega\neq 1$. The integration over $E$ produces a Dirac delta function and subsequently the integration over $\tau_{\sf f}$ can be performed, in which case, the result is
\begin{align} \label{eq:propagator_final}
    \mathscr{K}(a_{\sf f},a_{\sf i};\mathscr{T}) & = i\Theta(\mathscr{T})\frac{2 (a_{\sf f}a_{\sf i})^{\frac{1+p}{2}}}{3 (1-\omega) \hbar\mathscr{T}} \sqrt{\frac{3 V_{3}}{4\pi G}} \exp\left[\frac{V_{3}}{6i\pi G \hbar (1-\omega)^2 \mathscr{T}} (a_{\sf f}^{3(1-\omega)} + a_{\sf i}^{3(1-\omega)} )\right] I_{\frac{|1+p|}{3(1-\omega)}} \left(i\frac{V_{3}(a_{\sf f}a_{\sf i})^{\frac{3(1-\omega)}{2}}}{3 \pi G \hbar (1-\omega)^2 \mathscr{T}}\right) \nonumber\\
    & =\frac{3}{2}(1-\omega) (a_{\sf f}a_{\sf i})^{\frac{1+p}{2}} \sqrt{\frac{4\pi G}{3V_{3}}} \int_{0}^{\infty} k \, \rd k \, J_{\frac{|1+p|}{3(1-\omega)}} \left(k a_{\sf f}^{\frac{3(1-\omega)}{2}}\right) J_{\frac{|1+p|}{3(1-\omega)}} \left(k a_{\sf i}^{\frac{3(1-\omega)}{2}}\right) \e^{\frac{i}{\hbar}\frac{3 \pi G \hbar^2 (1-\omega)^2 k^2}{2V_{3}}\mathscr{T}}, \quad (\mathscr{T}>0),
\end{align}
where the second equality follows from the integral identity \cite{NIST:DLMF}. We can read off the wave function, and the energy spectrum using the identity \ref{eq:spectral_decomposition}, from the second equality of \ref{eq:propagator_final}. It is evident that we have a continuous energy spectrum labeled by the wavenumber $k$ which, in turn, is related to the energy by the following definition
\begin{align}
    k & = \frac{\sqrt{2(-E_{\rm grav}) V_{3}}}{\hbar \sqrt{3 \pi G} (1-\omega) }.
\end{align}
The negative sign in front of $E_{\rm grav}$ requires elaboration. Recall that the Hamiltonian constraint enforces the vanishing of the total energy of the system, that is, $\mathcal{E}=E_{\rm grav}+E_{T}=0$. In our convention, we assume the energy of the clock fluid $E_{T}$ is given by $E>0$, and as a result, the compensating gravitational energy must be $E_{\rm grav}=-E$. Thus, the quantity $-E_{\rm grav}$ is actually positive.  The expression for the wave function read off from \ref{eq:propagator_final} is
\begin{align} \label{eq:wave_function}
    \psi_{E}(a) & = \frac{\left(2E\right)^{\frac{1}{4}}}{\sqrt{\hbar} } a^{\frac{1+p}{2}}  J_{\frac{|1+p|}{3(1-\omega)}} \left(\frac{2\sqrt{2 E \left(\frac{3V_{3}}{4\pi G}\right)}}{3\hbar (1-\omega) } a^{\frac{3(1-\omega)}{2}}\right).
\end{align}
The obtained wave function solves the following Wheeler-Dewitt equation
\begin{align}\label{WDW_equation_solvable}
    \left(\frac{1}{2} a^{-p} \frac{\rd}{\rd a} a^{-p} \frac{\rd}{\rd a} + \frac{3V_{3}E}{4\pi G \hbar^2} a^{-3\omega+1-2p}\right) \psi_{E}(a) = 0,
\end{align}
which indeed does correspond to the Laplace-Beltrami ordered Hamiltonian, that is,
\begin{align} \label{ew:WDW_equation}
    - \frac{\hbar^2}{2} \frac{1}{\sqrt{|\mathscr{G}|}}\partial_{a} \sqrt{|\mathscr{G}|} \mathscr{G}^{aa} \partial_{a} \Psi = i \hbar a^{-3\omega+1-2p} \frac{\partial\Psi}{\partial T},
\end{align}
where, $\Psi(a,T)=\psi_{E}(a)\varphi_{E}(T)$ and $(i\hbar\partial/\partial T)\varphi_{E} = E$. For $E>0$, there are two independent solution of the differential equation \ref{WDW_equation_solvable}, given by the Bessel $J_{\alpha}(x)$ and $Y_{\alpha}(x)$ functions. However, the functions $Y_{\alpha}(x)$ are not regular at the origin and thus can be ignored on physical grounds. For $E<0$, the argument of the Bessel function in \ref{eq:wave_function} is purely imaginary, and thus, for negative energies the wave function must be proportional to the modified Bessel function $I_{\alpha}(x)$.

It can be verified that with respect to the inner product \ref{eq:one_dimensional_inner_product}, the obtained wave functions \ref{eq:wave_function} labeled by $E$ are $\delta$-function orthogonal states, that is, we have
\begin{align}
\int_0^\infty \rd a\, \sqrt{\frac{3V_{3}}{4\pi G}} a^{-3\omega+1-p} \psi^*_{E}(a) \psi_{E'}(a) = \delta\left(\frac{2\sqrt{2 E \left(\frac{3V_{3}}{4\pi G}\right)}}{3\hbar (1-\omega) } - \frac{2\sqrt{2 E' \left(\frac{3V_{3}}{4\pi G}\right)}}{3\hbar (1-\omega) }\right).
\end{align}
Notice that the path integral computation has produced the stationary state with correct normalization.

Thus, we have arrived at the canonical structures introduced in Sec \ref{sec:CQG1}, starting from the path integral of the cosmological model. Here, it is apparent that the covariant nature under lapse rescaling of the wave function is broken for this 1-D minisuperspace. Therefore, the operator ordering ambiguity of canonical gravity \cite{Sahota:2022qbc,Sahota:2023kox} is a manifestation of the non-covariance of this system under lapse rescaling. 

For the specific case $p=-(3\omega-1)/2$, the Wheeler-DeWitt equation \ref{WDW_equation_solvable} has the usual Schr\"dinger form without any factor along with the time derivative term. In this case, the Wheeler-DeWitt equation reads
\begin{align}
    \left(\frac{1}{2} a^{(3\omega-1)/2} \frac{\rd}{\rd a} a^{(3\omega-1)/2} \frac{\rd}{\rd a} + \frac{3V_{3}E}{4\pi G \hbar^2}\right) \psi_{E}(a) = 0,
\end{align}
which is also the form of the operator ordering in the Wheeler-DeWitt equation suggested in \cite{PhysRevD.75.023516}. 

\subsection{Case II: Bianchi I spacetime with a free massless scalar field and a perfect fluid as clock}\label{sec:ESC2}

We have seen previously that the quantization of one-dimensional minisuperspace does not respect the conformal invariance present in the classical theory. Now we should look for a higher dimensional example of minisuperspace. The simplest extension from the above example would be to include a free massless scalar field, and the corresponding minisuperspace $(a,\phi)$ will have $D=2$. Even though this is a valid example, notice that in this case the conformal term $+\hbar^2 \frac{D-2}{8(D-1)}\mathcal{R}$ vanishes identically and no additional term is required for conformal invariance. For this reason, we shall give an example of another even higher dimensional minisuperspace so that the role and relevance of this potential-class term become clear. We shall, in particular, consider the Bianchi I spacetime, which describes a spatially flat, homogeneous, but anisotropic universe. Notice that the action in \ref{eq:mini_action_bianchi_rescaled}, can be obtained by starting with
\begin{align}\label{eq:intermediate_bianchi_action}
    S_{\sf Total} & = \int {\rm d}t \left[p_{a} \dot{a} + \boldsymbol{p_{\beta}} \cdot \dot{\boldsymbol{\beta}} + p_{T}\dot{T} + p_{\phi} \dot{\phi} - N \left(- \frac{2\pi G}{3V_{3}}  \frac{p_{a}^2}{a^{-3\omega+1}} + \frac{8\pi G}{3V_{3}} \frac{\boldsymbol{p_{\beta}}^2}{a^{-3\omega+3}} + \frac{p_{\phi}^2}{2V_{3}a^{-3\omega+3}} + p_{T} \right) \right],
\end{align}
and using lapse rescaling $\Omega^{-2} = a^{-3\omega-2p+1}$. Then from \ref{eq:propagator_conformal_transformation} we can write the Feynman propagator corresponding to the action \ref{eq:mini_action_bianchi_rescaled} as
\begin{align}\label{eq:propagator_bianchi}
    \tilde{\mathscr{K}}(\boldsymbol{q}_{\sf f},T_{\sf f} ; \boldsymbol{q}_{\sf i},T_{\sf i}) = & \Theta(\mathscr{T})\frac{32(G\pi)^{\frac{3}{2}}}{3\sqrt{3}V_{3}^{2}}(a_{\sf f} a_{\sf i})^{\frac{3\omega-2p-4}{2}} \int \mathscr{D}a \mathscr{D} p_{a} \int \mathscr{D}\boldsymbol{\beta} \mathscr{D} \boldsymbol{p_{\beta}} \int \mathscr{D}\phi \mathscr{D} p_{\phi} \exp \Bigg[\frac{i}{\hbar} \int_{0}^{\mathscr{T}} \rd t \Bigg( p_{a} \dot{a} + \boldsymbol{p_{\beta}} \cdot \boldsymbol{\dot{\beta}} + p_{\phi} \dot{\phi} \nonumber\\
    & + \frac{2\pi G}{3V_{3}}  \frac{p_{a}^2}{a^{-3\omega+1}} - \frac{8\pi G}{3V_{3}} \frac{\boldsymbol{p_{\beta}}^2}{a^{-3\omega+3}} - \frac{p_{\phi}^2}{2V_{3}a^{-3\omega+3}} - \frac{\hbar^2\pi G}{6V_{3}a^{-3\omega-3}} \Bigg) \Bigg],
\end{align}
where using the formula \ref{eq:quantum_potential} we have computed the total quantum potential due to the DeWitt metric corresponding to the action \ref{eq:intermediate_bianchi_action} in four-dimensional $(D=4)$ minisuperspace
\begin{align}
    \Delta V_{Q} = \Delta V^{\sf PF}_{Q} + \frac{\hbar^2}{12}\mathcal{R} = \frac{\hbar^2\pi G}{6V_{3}a^{-3\omega-3}} .
\end{align}
The extra term ``$+ \frac{\hbar^2}{12}\mathcal{R}$'' in the quantum potential is simply the additional potential-class correction term $+\hbar^2 \frac{D-2}{8(D-1)}\mathcal{R}$ evaluated in four dimensions. Moreover, the labels $\boldsymbol{q}_{\sf f/i}$ refers to the initial and final values $(a_{\sf f/i},\boldsymbol{\beta}_{\sf f/i},\phi_{\sf f/i})$. To compute the path integral \ref{eq:propagator_bianchi}, we again adopt the strategy of coordinate transformations such that it can be brought to a known solvable form.

First, we shall consider a time transformation which will decouple the path integrals over $(\boldsymbol{\beta},\boldsymbol{p_{\beta}})$ and $(\phi,p_{\phi})$ from that of over the scale factor. Specifically, we define a new time coordinate along with particular boundary conditions
\begin{subequations}
\begin{align}
    a^{3\omega-3} \rd t = \rd \xi,~~\text{or}~~ t(\xi)=\int_{0}^\xi \rd \xi' \, a(\xi')^{-3\omega+3},\\
    t(0) = 0,~ t(\xi_{\sf f}) = \mathscr{T},~~ a(t(0)) = a(0) = a_{\sf i}, ~~ a(t(\xi_{\sf f})) = a(\mathscr{T}) = a_{\sf f}.
\end{align}    
\end{subequations}
Like before, we achieve this constraint in the functional integral by means of introducing the Dirac delta and the expression for the propagator becomes
\begin{align}
    \tilde{\mathscr{K}}(\boldsymbol{q}_{\sf f},T_{\sf f} ; \boldsymbol{q}_{\sf i},T_{\sf i}) = & \Theta(\mathscr{T})\frac{32(G\pi)^{\frac{3}{2}}}{3\sqrt{3}V_{3}^{2}}(a_{\sf f} a_{\sf i})^{-p-\frac{1}{2}} \int \mathscr{D}a \mathscr{D} p_{a}   \int_{0}^{\infty} \rd \xi_{\sf f}  \int_{-\infty}^{\infty} \frac{\rd \mathcal{S}}{2\pi\hbar} \e^{\frac{i}{\hbar} \mathcal{S} \mathscr{T}}  \exp \left\{\frac{i}{\hbar} \int_{0}^{\xi_{\sf f}} \rd \xi \Bigg(p_{a} \dot{a} + \frac{2\pi G}{3V_{3}}   p_{a}^2 a^2 \right. \nonumber\\
    & \left.- a^{-3\omega+3} \mathcal{S} - \frac{\hbar^2 \pi G}{6 V_{3}} \Bigg)\right\} \times \int \mathscr{D}\boldsymbol{\beta} \mathscr{D} \boldsymbol{p_{\beta}} \exp \left\{\frac{i}{\hbar} \int_{0}^{\xi_{\sf f}} \rd \xi \Bigg( \boldsymbol{p_{\beta}} \cdot \dot{\boldsymbol{\beta}} - \frac{8\pi G}{3V_{3}}  \boldsymbol{p_{\beta}}^2\Bigg)\right\} \nonumber\\
    & \times \int \mathscr{D}\phi \mathscr{D} p_{\phi} \exp \left\{\frac{i}{\hbar} \int_{0}^{\xi_{\sf f}} \rd \xi \Bigg(p_{\phi} \dot{\phi} - \frac{p_{\phi}^2}{2V_{3}}\Bigg)\right\}.
\end{align}
Here, the overhead dot now means derivative with respect to the new time coordinate $\xi$. Because of this time transformation, the path integrals over the scalar and anisotropy degrees of freedom have become separable from that of the scale factor, as expected. Now, it is convenient to calculate the propagator such that the initial momenta for the anisotropic and scalar degrees of freedom are fixed (Neumann condition) instead of their coordinate values (Dirichlet condition). The propagator with Neumann and Dirichlet initial conditions are related to each other by means of Fourier transformations
\begin{align}
    \tilde{\mathscr{K}}(a_{\sf f}, \phi_{\sf f}, \boldsymbol{\beta}_{\sf f},\mathscr{T};a_{\sf i}, \mathsf{p}_{\phi}, \boldsymbol{\Pi},0) = \int  \frac{\rd\phi_{\sf i}}{\sqrt{2\pi\hbar}} \, \e^{-\frac{i}{\hbar} \phi_{\sf i}\mathsf{p}_{\phi}} \int \frac{\rd \boldsymbol{\beta}_{\sf i}}{2\pi\hbar} \, \e^{-\frac{i}{\hbar}\boldsymbol{\beta}_{\sf i}\cdot\boldsymbol{\Pi}} \tilde{\mathscr{K}}(a_{\sf f}, \phi_{\sf f}, \boldsymbol{\beta}_{\sf f},\mathscr{T};a_{\sf i}, \phi_{\sf i}, \boldsymbol{\beta}_{\sf i},0), 
\end{align}
where $\mathsf{p}_{\phi}$ and $\boldsymbol{\Pi}$ are the initial momenta corresponding to the scalar and anisotropic degrees of freedom respectively. The path integral over anisotropies can be performed in a straightforward manner (as these are Gaussian integrals) and the result is
\begin{align}
    \int \frac{\rd\boldsymbol{\beta}_{\sf i}}{2\pi\hbar} \, \e^{-\frac{i}{\hbar}\boldsymbol{\beta}_{\sf i}\cdot\boldsymbol{\Pi}}\int_{\boldsymbol{\beta}(0) = \boldsymbol{\beta}_{\sf i}}^{\boldsymbol{\beta}(\xi_{\sf f})=\boldsymbol{\beta}_{\sf f}} \mathscr{D} \boldsymbol{\beta} \mathscr{D} \boldsymbol{p_{\beta}} \exp\left\{\frac{i}{\hbar}\int_{0}^{\xi_{\sf f}} {\rm d}\xi \left[\boldsymbol{p_{\beta}} \cdot \boldsymbol{\dot{\beta}} - \left(\frac{8\pi G}{3V_{3}}\right) \boldsymbol{p_{\beta}}^2 \right]\right\} = \frac{1}{2\pi\hbar} \e^{-\frac{i}{\hbar}\frac{8\pi G}{3V_{3}}\boldsymbol{\Pi}^2\xi_{\sf f}-\frac{i}{\hbar}\boldsymbol{\Pi}\cdot\boldsymbol{\beta}_{\sf f}}.
\end{align}
We can also perform the path integral over $(\phi,p_{\phi})$ similarly. Then the expression for the propagator reads
\begin{align}
    \tilde{\mathscr{K}}(\boldsymbol{q}_{\sf f},T_{\sf f} ; \boldsymbol{q}_{\sf i},T_{\sf i}) = & \Theta(\mathscr{T})\frac{32(G\pi)^{\frac{3}{2}}}{3\sqrt{3}V_{3}^{2}(2\pi\hbar)^{\frac{3}{2}}} \e^{-\frac{i}{\hbar}\left(\boldsymbol{\Pi}\cdot\boldsymbol{\beta}_{\sf f} + \mathsf{p}_{\phi} \phi_{\sf f} \right)} (a_{\sf f} a_{\sf i})^{-p-\frac{1}{2}} \int \mathscr{D}a \mathscr{D} p_{a}   \int_{0}^{\infty} \rd \xi_{\sf f}  \int_{-\infty}^{\infty} \frac{\rd \mathcal{S}}{2\pi\hbar} \e^{\frac{i}{\hbar} \mathcal{S} \mathscr{T}}  \exp \left\{\frac{i}{\hbar} \int_{0}^{\xi_{\sf f}} \rd \xi \Bigg(p_{a} \dot{a} \right. \nonumber\\
    & \left. + \frac{2\pi G}{3V_{3}}   p_{a}^2 a^2 - a^{-3\omega+3} \mathcal{S} - \frac{\hbar^2 \pi G}{6 V_{3}} - \frac{8\pi G \boldsymbol{\Pi}^2}{3V_{3}} - \frac{\mathsf{p}_{\phi}^2}{2V_{3}} \Bigg)\right\}.
\end{align}
We have to now perform another time transformation so that the factor $a^{2}$ in the kinetic term can be removed. After the time transformation, $\rd s = a^{2} \rd \xi$, along with appropriate boundary conditions, we get
\begin{align}
    \tilde{\mathscr{K}}(\boldsymbol{q}_{\sf f},T_{\sf f} ; \boldsymbol{q}_{\sf i},T_{\sf i}) = & \Theta(\mathscr{T})\frac{32(G\pi)^{\frac{3}{2}}}{3\sqrt{3}V_{3}^{2}(2\pi\hbar)^{\frac{3}{2}}}\e^{-\frac{i}{\hbar}\left(\boldsymbol{\Pi}\cdot\boldsymbol{\beta}_{\sf f} + \mathsf{p}_{\phi} \phi_{\sf f} \right)} (a_{\sf f} a_{\sf i})^{-p-\frac{3}{2}} \int \mathscr{D}a \mathscr{D} p_{a}   \int_{0}^{\infty} \rd s_{\sf f}  \int_{-\infty}^{\infty} \frac{\rd E}{2\pi\hbar} \e^{\frac{i}{\hbar} E \mathscr{T}}  \exp \left\{\frac{i}{\hbar} \int_{0}^{s_{\sf f}} \rd s \Bigg(p_{a} \dot{a}  \right. \nonumber\\
    & \left. + \frac{2\pi G}{3V_{3}}   p_{a}^2- E a^{-3\omega+1} - \frac{\hbar^2 \pi G}{6 V_{3}a^2} - \frac{8\pi G \boldsymbol{\Pi}^2}{3V_{3}a^2} - \frac{\mathsf{p}_{\phi}^2}{2V_{3}a^2} \Bigg)\right\},
\end{align}
where, again, the overhead dot now means derivative with respect to $s$. After performing the momentum integrals $p_{a}$, we recognize that the form of the remaining path integral to be done has a similar form compared to \ref{eq:inverse_square_path_integral}, which we have already evaluated. Thus we do not repeat the steps here. We can follow the steps similar to \ref{eq:space-time_transformation} to \ref{eq:propagator_final} and obtain
\begin{align}
    \tilde{\mathscr{K}}(a_{\sf f}, \phi_{\sf f}, \boldsymbol{\beta}_{\sf f},\mathscr{T};a_{\sf i}, \mathsf{p}_{\phi}, \boldsymbol{\Pi},0) = & i\Theta(\mathscr{T})\frac{16 \sqrt{G\pi} (a_{\sf f}a_{\sf i})^{-(1+p)}}{3\sqrt{3} V_{3} (1-\omega) (2\pi\hbar)^{\frac{3}{2}} \hbar\mathscr{T}} \e^{-\frac{i}{\hbar}\left(\boldsymbol{\Pi}\cdot\boldsymbol{\beta}_{\sf f} + \mathsf{p}_{\phi} \phi_{\sf f} \right)} \exp\left[\frac{V_{3}}{6i\pi G \hbar (1-\omega)^2 \mathscr{T}} (a_{\sf f}^{3(1-\omega)} + a_{\sf i}^{3(1-\omega)} )\right] \nonumber\\
    & \times I_{\nu} \left(i\frac{V_{3}(a_{\sf f}a_{\sf i})^{\frac{3(1-\omega)}{2}}}{3 \pi G \hbar (1-\omega)^2 \mathscr{T}}\right)
\end{align}
where we have defined 
\begin{align}
    \nu = \pm i\frac{2\sqrt{2}}{3(1-\omega)}\sqrt{ 3\left(\frac{\mathsf{p}_{\phi}}{\sqrt{8\pi G}\hbar}\right)^2 + 2 \left(\frac{\boldsymbol{\Pi}}{\hbar}\right)^2}.
\end{align}
Therefore, the Feynman propagator with Dirichlet Boundary condition reads
\begin{align}
    \mathscr{K}(a_{\sf f}, \phi_{\sf f}, \boldsymbol{\beta}_{\sf f},\mathscr{T};a_{\sf i}, \phi_{\sf i}, \boldsymbol{\beta}_{\sf i},0) = & \int\frac{\rd \boldsymbol{\Pi}\rd \mathsf{p}_{\phi}}{(2\pi\hbar)^{\frac{3}{2}}}i\Theta(\mathscr{T})\frac{16 \sqrt{G\pi} (a_{\sf f}a_{\sf i})^{-(1+p)}}{3\sqrt{3} V_{3} (1-\omega) (2\pi\hbar)^{\frac{3}{2}} \hbar\mathscr{T}} \e^{-\frac{i}{\hbar}\left(\boldsymbol{\Pi}\cdot(\boldsymbol{\beta}_{\sf f}-\boldsymbol{\beta}_{\sf i}) + \mathsf{p}_{\phi} (\phi_{\sf f}-\phi_{\sf i}) \right)}  \nonumber\\
    & \times \exp\left[\frac{V_{3}}{6i\pi G \hbar (1-\omega)^2 \mathscr{T}} (a_{\sf f}^{3(1-\omega)} + a_{\sf i}^{3(1-\omega)} )\right]  I_{\nu} \left(i\frac{V_{3}(a_{\sf f}a_{\sf i})^{\frac{3(1-\omega)}{2}}}{3 \pi G \hbar (1-\omega)^2 \mathscr{T}}\right),
\end{align}
which can be re-written as
\begin{align} \label{eq:propagator_dirichlet}
    \mathscr{K}(a_{\sf f}, \phi_{\sf f}, \boldsymbol{\beta}_{\sf f},\mathscr{T};a_{\sf i}, \phi_{\sf i}, \boldsymbol{\beta}_{\sf i},0) = & \frac{16 (1-\omega)(\pi G)^{\frac{3}{2}} (a_{\sf f}a_{\sf i})^{-(1+p)}}{\sqrt{3} V^2_{3} (2\pi\hbar)^{\frac{3}{2}}} \int\frac{\rd \boldsymbol{\Pi}\rd \mathsf{p}_{\phi}}{(2\pi\hbar)^{\frac{3}{2}}} \int_{0}^{\infty} \rd k \, k\, \e^{-\frac{i}{\hbar}\left(\boldsymbol{\Pi}\cdot(\boldsymbol{\beta}_{\sf f}-\boldsymbol{\beta}_{\sf i}) + \mathsf{p}_{\phi} (\phi_{\sf f}-\phi_{\sf i}) \right)}    \nonumber\\
    & \times J_{\nu}  \left(k a_{\sf f}^{\frac{3(1-\omega)}{2}}\right) J_{\nu}  \left(k a_{\sf f}^{\frac{3(1-\omega)}{2}}\right) \e^{-\frac{i}{\hbar}\frac{-3\hbar^2\pi G (1-\omega)^2 k^2}{2V_{3}}\mathscr{T}},
\end{align}
where, again we have the relation $-E = \hbar^2\frac{3\pi G(1-\omega)^2}{2V_{3}}k^2$, with $E$ being the energy of the stationary state.
Then the corresponding wave function (for the initial states of the kind in \ref{eq:wave_function_bianchi}, i.e., states with fixed momentum for anistoropy and scalar sector) is
\begin{align}\label{eq:wave_function_bianchi1}
    \Psi_{E}(a,\phi,\boldsymbol{\beta}) & = \frac{4\left(2E\right)^{\frac{1}{4}}\sqrt{G\pi}}{\sqrt{3\hbar} V_{3}^{\frac{3}{4}} (2\pi\hbar)^{\frac{3}{2}}} \e^{\pm\frac{i}{\hbar}\left(\boldsymbol{\Pi}\cdot\boldsymbol{\beta} + \mathsf{p}_{\phi} \phi \right)}  a^{-(1+p)}  J_{\nu} \left(\frac{2\sqrt{2 E \left(\frac{3V_{3}}{4\pi G}\right)}}{3\hbar (1-\omega) } a^{\frac{3(1-\omega)}{2}}\right),
\end{align}
where we have introduced $\pm$ sign in the fixed momentum state for anisotropy and scalar field reflecting the fact that integral over $\boldsymbol{\Pi}$ and $\mathsf{p}_{\phi}$ in \ref{eq:propagator_dirichlet} is invariant under the transformations $\boldsymbol{\Pi} \to - \boldsymbol{\Pi}$ and/or $\mathsf{p}_{\phi}\to - \mathsf{p}_{\phi}$. We also note that the above wave function has the correct normalization with respect to the inner product \ref{BianchiInnerProduct}. Furthermore, this wave function, obtained from the path integral, satisfies the Wheeler-DeWitt equation \ref{eq:bianchi_wheeler_dewitt}. Therefore, the path integral and canonical approach produce consistent results. As expected, the wave function follows the scaling property under lapse rescaling \ref{eq:wave_function_conformal_transformation}, and the system is invariant under such rescalings as discussed for the canonical case in Sec. \ref{sec:CQG2}. Notice the fact that the quantum correction to the potential ($+\frac{D-2}{8(D-1)}\mathcal{R}$) that appeared in the Wheeler-DeWitt equation is crucial to this ambiguity-free description. It can be checked that without this additional correction, the index of the Bessel function depends on the ambiguity parameter $p$, making the predictions of the theory essentially dependent on the operator ordering through the parameter $p$.

Thus, the invariance of the systems with minisuperspace of dimension $D>2$ under lapse rescaling implies the operator ordering ambiguities (conformal and factor classes) leave no physical imprint for these systems. Notice that for this conclusion to hold the potential-class ambiguity term has to be fixed and is not arbitrary. Therefore, one may conclude that building quantum theories which preserve the symmetries of the classical theory, under point canonical transformation and lapse rescalings, has resulted into the ambiguities being immaterial for the physical predictions.
%%%%%%%%%%%%%%%%%%%%%%%%%%%%%%%%%%%%%%%%%%%%%%%%%%%%%%%%%%%%%%%%%%%%%%%%%%%%%%%%%%%%%%%%%%%%%%%%%%%%%%%%%%%%%%%%%%%%%%%%%%%%%
%let us compare the wave function \ref{eq:wave_function_bianchi} for $p=1/2$ and any other $p$ value (recall the choice $p=1/2$ corresponds to quantizing with the metric \ref{eq:metric_bianchi} without any lapse rescaling of the form $N\to N a^{-2p+1}$)
% \begin{align}
%     \frac{\Psi_{E,p}}{\Psi_{E,p=1/2}} = a^{\frac{1-2p}{2}}.
% \end{align}
% Identifying $\Omega^{-2} = a^{-2p+1}$ we can see that the above relation is consistent with the scaling property of the wave function in \ref{eq:wave_function_conformal_transformation}.

\section{Discussion and Conclusion}\label{sec:DAC}

In this work, we have dealt with the aspects of the problem of time and operator ordering ambiguity in quantum cosmology. The first issue stems from the fact that general relativity is a covariant theory (and the corresponding minisuperspace has time reparametrization invariance), while the second issue is caused by the fact that there can be, in principle, infinitely many quantum Hamiltonians that correspond to the same classical theory and we have no observational guidance to pick the correct Hamiltonian. To tackle the first problem, we have adopted the particular suggestion of adding a reference fluid to the system, which acts as the internal clock. As for the second problem, we remark that the operator ordering ambiguity is fundamental and cannot be resolved without additional assumptions regarding the system. For this purpose, following some prior works, we suggest here that the additional ingredient that may resolve the operator ordering ambiguity should be symmetry. The classical minisuperspace has two symmetries, namely the symmetry under arbitrary redefinitions of minisuperspace coordinates and the lapse function. We find that retaining these symmetries in the quantum theory as well, results in a potential resolution of the operator ordering ambiguities.

We see that the ambiguities in the Wheeler-DeWitt equation can be characterized by three distinct functions, which we have called conformal-class, factor-class, and potential-class ambiguities. Among these three classes, the factor-class only results in an overall factor in the wave function and gets canceled in the inner products, leaving no imprint in the physical predictions. Whereas the other two ambiguities require careful treatment. We further find that if we demand symmetry under point canonical transformations or the coordinate transformations in the minisuperspace, then the Laplacian in the Wheeler-DeWitt equation is defined with the Laplace-Beltrami operator in curved spaces, depending only on the metric of the minisuperspace. However, this does not fix any ambiguity in the Hamiltonian, as we still have the lapse rescaling freedom in the classical theory. We have shown that two equivalent classical theories, which differ by lapse rescaling, lead to two distinct and inequivalent quantum theories which differ by operator ordering ambiguity of the conformal-class. Therefore, resolving conformal-class ambiguity is equivalent to demanding conformal invariance in the quantum theory. This is achieved by choosing a particular form of the potential-class ambiguity term proportional to the Ricci scalar of the minisuperspace, more specifically $\displaystyle +\frac{D-2}{8(D-1)}\mathcal{R}$. Once this choice has been made, surprisingly, all the ambiguities of the conformal-class become irrelevant as these no longer leave any imprint on the inner products between quantum states. However, this symmetry principle does not work for $D=1,2$ case and only works for the cases $D>2$. Only a specific \textit{choice} for the potential-class term results in a conformally invariant theory in which conformal-class ambiguities become irrelevant. However, even a small perturbation of the potential-class term destroys the conformal invariance of the quantum theory and therefore, the ambiguities remain at large. Therefore, even though it is reasonable to preserve the classical symmetries in the quantum theory, and thereby arrive at a unique prescription for writing the quantum Hamiltonian, the operator ordering ambiguity cannot be resolved fully without empirical input.

We have derived the above results in both the canonical and path integral approaches. Noticing that operator ordering ambiguity and the problem of time in the path integral approach are rarely discussed in the literature, in this work, we have computed \textit{exact} path integrals for a flat and homogeneous quantum universe, both having isotropic FLRW metric and anisotropic Bianchi I metric and containing a single fluid with an arbitrary equation of state parameter $\omega$. The fluid acts as a reference clock and enables the unitary evolution of the quantum state of the universe. In the case of anisotropic cosmology, we have shown that a massless scalar field can easily be incorporated as well, and still, an exact path integral can be performed. As the path integrals are not of the Gaussian or canonical form, we had to introduce coordinate transformations to bring them into forms, for which the evaluation of the path integral is known. All the coordinate transformations required a careful treatment of the Jacobian factors in the integral and quantum corrections to the potential. We have further verified that the wave functions obtained from the path integral formalism, indeed, satisfy the corresponding Wheeler-DeWitt equations or the quantum Hamiltonian constraint. Moreover, the path integral computations produce wave functions with correct normalization. These concrete examples support our general derivations of the results stated above. Even though Halliwell noticed that the potential-class term could be wisely chosen to resolve the conformal-class ambiguities \cite{Halliwell:1988wc}, the author did not discuss the inner product between the quantum states, which we have done here. This question can be addressed in the timeless formalism of quantum cosmology considered in that analysis \cite{Halliwell:1988wc} by introducing the notion of Hilbert space and inner product by group averaging method \cite{Giulini_1999}. Moreover, our work is different in at least another key aspect: we have extended the proof of the lapse choice invariance to the case of the path integral formalism, which is absent in \cite{Halliwell:1988wc}.

The ordering ambiguity still remains in the case of $D=1$ and $D=2$ minisuperspace, and it cannot be removed, at least with the techniques we discussed in this article. More specifically, we have shown that the wave functions derived in both the isotropic and anisotropic cases have a Besselian form. In the case of the higher dimensional minisuperspace (in our example $D=4$), the index of the Bessel function is independent of the ambiguity parameter $p$, which only appears in the power of the scale factor multiplied with the Bessel function and does not appear in the inner products as this factor is canceled by appropriate integration measure. Therefore, the ambiguity is removed from the physical predictions. On the other hand, in the case of $D=1$ minisuperspace, the index of the Bessel function depends on the ambiguity parameter and cannot be removed by the integration measure. As a result, the physical predictions of the theory will always depend on the ordering choice of the Hamiltonian, as depicted in \cite{Sahota:2022qbc,Sahota:2023kox}.

On a broader note in the context of quantum cosmology, the operational approach to the observational aspects of quantum gravity is to quantize the background geometry with a reference clock that leads to a quantum-corrected spacetime \cite{Ashtekar:2009mb,Peter:2008qz}. The perturbations of the background observables are then quantized on this quantum geometry and their correlations give us the power spectrum that is central to the observational cosmology \cite{Agullo_2012,Peter:2005hm,Peter:2006hx}. However, these scenarios consider a $1+1$ dimensional minisuperspace, a flat FLRW universe with perfect fluid in \cite{Peter:2008qz} where the caveat discussed will have major repercussions and with a scalar field in \cite{Ashtekar:2009mb} a setup ideal for the further investigation of this question. 

In this light, the BKL conjecture \cite{Belinsky:1970ew,Belinsky:1981vdw,Belinsky:1982pk} presents an interesting outlook. It states that the dynamics of the spacetime near a generic spacelike singularity becomes local, oscillatory, and dominated by pure gravity with the spacetime following Mixmaster dynamics \cite{Misner_Mixmaster} with the geometry of Bianchi type IX. In other words, a minisuperspace of dimension $3$ is dynamically preferred, and the assumption of isotropy is not valid for spacetimes near the singularity. If one adopts this viewpoint and quantizes the Bianchi IX universe with a reference fluid, our analysis predicts an ambiguity-free quantum universe if the symmetries of the classical minisuperspace are preserved in the quantum theory. 

\section*{acknowledgements}

Research of KL is partially supported by SERB, Govt. of India through a MATRICS research grant no.
MTR/2022/000900. The research of VM is funded by the INSPIRE fellowship from the DST, Government of India (Reg. No. DST/INSPIRE/03/2019/001887). The research of HSS is supported by the Core Research Grant CRG/2021/003053 from the Science and Engineering Research Board, India. We are thankful to Sumanta Chakraborty for insightful discussions on the problem. VM acknowledges Kaustav Das for helpful discussions during the preparation of this manuscript. HSS and VM are grateful to the organizers of the workshop IAGRG School on Gravitation and Cosmology (code: ICTS/IAGRG2023/10) and the International Centre for Theoretical Sciences (ICTS) for hospitality.

\appendix

\section{Schutz fluid as clock}\label{sec:schutz_fluid}

Schutz introduced a relativistic treatment of the perfect fluid in terms of velocity potentials \cite{PhysRevD.2.2762,PhysRevD.4.3559}, where the fluid four-velocity is given in terms of non-canonical fields called velocity potentials
    \begin{align}
        U_\nu=\mu^{-1}\left(\phi_{,\nu}+\alpha\beta_{,\nu}+\theta S_{,\nu}\right),
    \end{align}
where $S$ is the specific entropy, $\alpha$ and $\theta$ are related to the fluid circulation or vorticity, and $\mu$ is termed the specific inertial mass, i.e., enthalpy defined via the zero-point energy density $\rho_0$, and the specific internal energy $\Pi$ as
    \begin{align}
        \mu=1+\Pi+\frac{p}{\rho_0}\label{pres}
    \end{align}
    with $p$ being the pressure. The total mass energy of the system is $\rho=\rho_{0}(1+\Pi)$. Then all thermodynamic quantities are expressed in terms of entropy and specific enthalpy. Normalization of four-velocity relates the enthalpy to the other velocity potentials through the equation
    \begin{align}
        \mu^2=-g^{\nu\sigma}\left(\phi_{,\nu}+\alpha\beta_{,\nu}+\theta S_{,\nu}\right)\left(\phi_{,\sigma}+\alpha\beta_{,\sigma}+\theta S_{,\sigma}\right).
    \end{align}
    The action for this fluid can be introduced using pressure $p$ as
    \begin{align}
        S_{\sf fluid}=\int \rd^4x\sqrt{-g} \, p.
    \end{align}
    Extremizing this action, we get the evolution equations for the velocity potentials
    \begin{subequations}
        \begin{align}
        U^\nu\phi_{,\nu}&=-\mu, \\
        U^\nu\alpha_{,\nu}&=0, \\
        U^\nu\beta_{,\nu}&=0, \\
        U^\nu\theta_{,\nu}&=\mathcal{T}, \\
        U^\nu S_{,\nu}&=0,\\
        (\rho_0 U^\nu)_{;\nu}&=0.
        \end{align}
    \end{subequations}
    
    Here $\mathcal{T}$ is the temperature associated with the fluid. The ADM decomposition of this system is required to arrive at the Hamiltonian formulation for this system. To find the relation between pressure and velocity potentials, we make use of the thermodynamic identity
    \begin{align}
        \mathcal{T}\rd S=\rd\Pi+p\rd(\rho_0^{-1}),
    \end{align}
    and Eq. \ref{pres}. Eliminating $\Pi$ and $\rho_0$ and using $\rho=\omega p$, we arrive at the expression
    \begin{align}
        p=\frac{\omega\mu^{1+1/\omega}e^{-S/\omega}}{(1+\omega)^{1+1/\omega}}.\label{Pressure}
    \end{align}
    For a homogeneous cosmological model, we assume the metric to be of the following form
    \begin{align}
        \rd s^2 = - N^2(t) \rd t^2 + h_{ij}(t) \rd x^i \rd x^j.
    \end{align}
    In such a spacetime, using the expression for the enthalpy
    \begin{align}
        \mu=\frac{1}{N}\left(\dot{\phi}+\alpha\dot{\beta}+\theta \dot{S}\right),
    \end{align}
    in Eq. \ref{Pressure} and dropping the potentials that correspond to the vorticity (irrotational fluid), the action of the fluid takes the form,
    \begin{align}
        S_{\sf fluid}=\int \rd t\left[V_{3} N^{-1/\omega}\sqrt{h}\frac{\omega\left(\dot{\phi}+\theta \dot{S}\right)^{1+1/\omega}e^{-S/\omega}}{(1+\omega)^{1+1/\omega}}\right],
    \end{align}
    where $V_{3}$ is the volume of the comoving hypersurface. The momentum conjugate to the velocity potentials takes the form
    \begin{align}
        p_S=\theta p_\phi=V_{3} \theta\sqrt{h} N^{-1/\omega}\frac{\left(\dot{\phi}+\theta \dot{S}\right)^{1/\omega}e^{-S/\omega}}{(1+\omega)^{1/\omega}}, \quad p_{\theta} = 0
    \end{align}
    leading to the canonical Hamiltonian of the form
    \begin{align}
        H_{\sf fluid}=V_3^{-\omega} N p_\phi^{\omega+1}h^{-\frac{\omega}{2}}e^{S}
    \end{align}
    After performing the canonical transformations (see for example \cite{Lapchinsky:1977vb,Alvarenga:2001nm}),
    \begin{align}
        T=-V_{3}^{\omega}p_Se^{-S}p_\phi^{-1-\omega},\quad P_T=V_{3}^{-\omega}p_\phi^{\omega+1}e^S,\\
        \phi'=\phi+(1+\omega)\frac{p_S}{p_\phi},\quad p_{\phi'}=p_\phi,
    \end{align}
    the Hamiltonian of the fluid takes the form 
    \begin{align}
        H_{\sf fluid}=\frac{N p_T}{\sqrt{h^\omega}}.
    \end{align}
%%%%%%%%%%%%%%%%%%%%%%%%%%%%%%%%%%%%%%%%%%%%%%%%%%%%%%%%%%%%%%%%%%%%%%%%%%%%%%%%%%%%%%%%%%%%%%%%%%%%%%%%%%%%%%%%%%%%%%%%%%%
Therefore, we find that the Hamiltonian of a perfect fluid in a homogeneous universe can be expressed in terms of canonical variables $(T,p_{T})$ such that the Hamiltonian depends linearly on its momentum. Notice that the above Hamiltonian differs from our discussion in the previous section by a factor of $h^{-\omega/2}$. This, however, does not alter any of the generic conclusions in the previous section and the factor can be absorbed with a lapse rescaling $N\to \tilde{N} h^{-\omega/2}$ or an equivalent time coordinate transformation as we shall see below.

\section{Conformal transformation of the quantum potential}\label{app:conformal_quantum_potential}

Note that the quantum correction to the potential in the lapse rescaled theory reads as 
\begin{align}\label{eq:quantum_correction_rescaled}
    \widetilde{\Delta V}_{Q}^{\sf PF} = \frac{\hbar^2}{8} \eta^{ab} \left[4 \tilde{e}^{A}_{a} (\partial_{A} \partial_{B} \tilde{e}^{B}_{b}) + 2 \tilde{e}^{A}_{a} \tilde{e}^{B}_{b} \frac{(\partial_{A} \partial_{B} \tilde{e})}{\tilde{e}} + 2 \tilde{e}^{A}_{a} \left((\partial_{A}\tilde{e}^{B}_{b})\frac{(\partial_{B} \tilde{e})}{\tilde{e}} + (\partial_{B}\tilde{e}^{B}_{b}) \frac{(\partial_{A} \tilde{e})}{\tilde{e}}\right) - \tilde{e}^{A}_{a}\tilde{e}^{B}_{b}\frac{(\partial_{A}\tilde{e})( \partial_{B}\tilde{e})}{\tilde{e}^2} \right],
\end{align}
where, the tetrads have undergone a scaling of the form
\begin{align}
    \tilde{e}^A_{a} = \Omega^{-1} e^{A}_{a}.
\end{align}
Now, we can try to express the new quantum correction \ref{eq:quantum_correction_rescaled} in terms of the old tetrads $e^a_{A}$. In the following, we write how each term in the square brackets can be re-expressed.

The first term re-expressed has the following form
\begin{align}
    4 \frac{\hbar^2}{8} \eta^{ab} \tilde{e}^{A}_{a} (\partial_{A} \partial_{B} \tilde{e}^{B}_{b}) & = \frac{\hbar^2}{2} \eta^{ab} \Omega^{-1} e^{A}_{a} (\partial_{A} \partial_{B} (\Omega^{-1} e^{B}_{b})) \nonumber\\
    & = \frac{\hbar^2}{2} \eta^{ab} \frac{e^{A}_{a}}{\Omega}   \left(\frac{\partial_{A} \partial_{B} e^{B}_{b}}{\Omega} -\frac{(\partial_{B} e^{B}_{b})(\partial_{A}\Omega)}{\Omega^2}  - \frac{(\partial_{A} e^{B}_{b})(\partial_{B}\Omega)}{\Omega^2} - e^{B}_{b} \frac{\partial_{A}\partial_{B}\Omega}{\Omega^2} + 2 e^{B}_{b} \frac{(\partial_{A} \Omega)(\partial_{B}\Omega)}{\Omega^3} \right).
\end{align}
The second term re-expressed:
\begin{align}
    2 \tilde{e}^{A}_{a} \tilde{e}^{B}_{b} \frac{(\partial_{A} \partial_{B} \tilde{e})}{\tilde{e}} & = \frac{2 e^A_a e^B_b}{\Omega^{2+D} e} \left(\partial_A\left(\Omega^D\partial_B e + D \Omega^{D-1} (\partial_{B}\Omega) e\right)\right) \nonumber\\
    & = \frac{2 e^A_a e^B_b}{\Omega^{2} e} \left(D \frac{(\partial_A \Omega) (\partial_{B} e)}{\Omega}  + (\partial_A \partial_B e) + D(D-1) \frac{(\partial_A \Omega) (\partial_{B}\Omega) e}{\Omega^2}  + D \frac{(\partial_A \partial_B \Omega) e}{\Omega}  + D \frac{(\partial_{B} \Omega) (\partial_{A} e)}{\Omega} \right).
\end{align}
The third term re-expressed:
\begin{align}
(\partial_{A}\tilde{e}^{B}_{b})\frac{(\partial_{B} \tilde{e})}{\tilde{e}} & = \left(\Omega^{-1}\partial_{A} e^{B}_{b} - e^B_b \frac{(\partial_{A} \Omega)}{\Omega^2}\right) \frac{(\partial_{B} e + D \Omega^{-1} e \partial_B \Omega)}{e} \nonumber\\
& = \frac{(\partial_A e^B_b)(\partial_{B} e)}{\Omega e} + D \frac{(\partial_A e^B_b)(\partial_{B}\Omega)}{\Omega^2} - e^B_b\frac{(\partial_A \Omega)(\partial_{B} e)}{\Omega^2 e} - D e^B_b \frac{(\partial_A \Omega)(\partial_B \Omega)}{\Omega^3}.
\end{align}
The fourth term re-expressed:
\begin{align}
(\partial_{B}\tilde{e}^{B}_{b})\frac{(\partial_{A} \tilde{e})}{\tilde{e}} & = \left(\Omega^{-1}\partial_{B} e^{B}_{b} - e^B_b \frac{(\partial_{B} \Omega)}{\Omega^2}\right) \frac{(\partial_{A} e + D \Omega^{-1} e \partial_A \Omega)}{e} \nonumber\\
& = \frac{(\partial_B e^B_b)(\partial_{A} e)}{\Omega e} + D \frac{(\partial_B e^B_b)(\partial_{A}\Omega)}{\Omega^2} - e^B_b\frac{(\partial_B \Omega)(\partial_{A} e)}{\Omega^2 e} - D e^B_b \frac{(\partial_B \Omega)(\partial_A \Omega)}{\Omega^3}.
\end{align}
And, finally, the fifth term re-expressed:
\begin{align}
     - \tilde{e}^{A}_{a}\tilde{e}^{B}_{b}\frac{(\partial_{A}\tilde{e})( \partial_{B}\tilde{e})}{\tilde{e}^2} & = - e^{A}_{a} e^{B}_{b}\frac{((\partial_{A} e) + D \Omega^{-1} (\partial_A \Omega) e)((\partial_{B} e) + D \Omega^{-1} (\partial_B \Omega) e)}{\Omega^{2} e^2} \nonumber\\
     & = - \frac{e^A_a e^B_b}{\Omega^2 e^2} \left((\partial_{A} e)(\partial_{B} e) + D \frac{(\partial_{A} e) (\partial_{B} \Omega) e}{\Omega} + D \frac{(\partial_{B} e) (\partial_{A} \Omega) e}{\Omega} + D^2 \frac{(\partial_{A} \Omega) (\partial_{B} \Omega) e^2}{\Omega^2} \right)
\end{align}
On the other hand the Ricci scalar transforms as
\begin{align}
    \xi \tilde{\mathcal{R}} & = \xi \mathcal{R} \Omega^{-2} - \xi (D-1)(D-4) \Omega^{-4} \eta^{ab} e^A_a e^B_b (\partial_A \Omega) (\partial_B \Omega) - 2 \xi (D-1) \Omega^{-3} \left(\frac{\eta^{ab}}{e}\partial_A (e e^A_a e^B_b \partial_B \Omega)\right) \nonumber \\
    & = \xi \mathcal{R} \Omega^{-2} - \xi (D-1)(D-4) \Omega^{-4} \eta^{ab} e^A_a e^B_b (\partial_A \Omega) (\partial_B \Omega) - 2 \xi (D-1) \Omega^{-3} \frac{\eta^{ab}}{e}\left( e^A_a e^B_b (\partial_A e) (\partial_B \Omega) \right. \nonumber\\
    & \left.+ e e^A_a (\partial_A  e^B_b) (\partial_B \Omega) + e  e^B_b (\partial_A e^A_a) (\partial_B \Omega) + e e^A_a e^B_b (\partial_A \partial_B \Omega)\right).
\end{align}
Collecting the terms coming from the quantum correction and the Ricci scalar together, we get
\begin{align}
     & \widetilde{\Delta V}_{Q}^{\sf PF} + \hbar^2 \xi \tilde{\mathcal{R}}= \Omega^{-2} \left(\Delta V_{Q}^{\sf PF} + \hbar^2 \xi \mathcal{R} \right) + \frac{\hbar^2}{2 \Omega^3} \left(\frac{D}{2} - 1 - 4 \xi (D-1) \right)  \eta^{ab} \Big( e^A_a (\partial_{B} e^{B}_{b})(\partial_{A}\Omega) + e^A_a (\partial_{A} e^{B}_{b})(\partial_{B}\Omega)  \nonumber \\
    & + e^A_a e^{B}_{b} (\partial_{A}\partial_{B}\Omega) + \frac{e^A_a e^B_b}{e}(\partial_A e)(\partial_B \Omega) \Big) + \frac{\hbar^2}{\Omega^4} \left(\frac{(D-4)(D-2)}{8} - \xi (D-1)(D-4) \right) \eta^{ab} e^A_a e^B_b (\partial_{A} \Omega)(\partial_{B}\Omega).
\end{align}
This is the desired expression in \ref{eq:quantum_correction_transformed}.

\section{Imprints of ambiguity parameter on the observables of the Bianchi I universe}\label{app:Bianchi}
In this section, we check whether the expectation values of the observables depend on the ambiguity parameter in this quantum model. We will work with a general wave packet constructed from the positive energy stationary states in \ref{eq:wave_function_bianchi}
\begin{align}
    \psi(a,\beta_1,\beta_2,\phi)=&\int_0^\infty dE\int_{-\infty}^{\infty}d\Pi_1\int_{-\infty}^{\infty}d\Pi_2\int_{-\infty}^{\infty}dp_\phi A(E,\Pi_1,\Pi_2,p_\phi)\Psi_E(a,\phi,\beta_1,\beta_2).
\end{align}
Since the ambiguity parameter appears only in the exponent of the scale factor in the stationary state, any wave packet will depend on the ambiguity parameter in the following form
\begin{align}
    \psi_p(a,\beta_1,\beta_2,\phi)=a^{-(1+p)}\psi(a,\beta_1,\beta_2,\phi).
\end{align}
With the inner product defined in \ref{BianchiInnerProduct}, the Hermitian representation of the momentum operator conjugate to scale factor is
\begin{align}
    \hat{p}_a=-i a^{-\frac{4+2p-3\omega}{2}}\frac{\partial}{\partial a}a^{\frac{4+2p-3\omega}{2}},
\end{align}
where we have ignored the factors of $\hbar$, $V_3$ and $G$ in this discussion. Since ambiguity dependence only comes through the isotropized scale factor, the expectation value of observables corresponding to the anisotropy and scalar field are independent of this ambiguity, as is seen from the expectation value of the operator corresponding to an arbitrary observable $O(\phi,\beta_1,\beta_2,p_\phi,p_{\beta_1},p_{\beta_2})$.
\begin{align}
    \braket{\hat{O}}&=\int_{-\infty}^\infty d\phi\int d^2\beta\int_{0}^\infty da\;a^{4+2p-3\omega}\psi_p^*(a,\phi,\beta_1,\beta_2,T)\hat{O}\psi_p(a,\phi,\beta_1,\beta_2,T)\\
    &=\int_{-\infty}^\infty d\phi\int d^2\beta\int_{0}^\infty da\;a^{2-3\omega}\hat{O}|\psi(a,\phi,\beta_1,\beta_2,T)|^2.
\end{align}
In the case of a general observable constructed from the isotropized scale factor, of the form $O=a^np_a^m$, we look at the expectation value of the operators symmetrized as $\hat{O}_1=\hat{a}^{n/2}\hat{p}_a^m\hat{a}^{n/2}$ and $\hat{O}_2=1/2\left(\hat{p}_a^{m}\hat{a}^n+\hat{a}^n\hat{p}_a^{m}\right)$.
\begin{align}
    \braket{O}_1=&(-i)^m\int_{-\infty}^\infty d\phi\int d^2\beta\int_{0}^\infty da\;a^{4+2p-3\omega}\psi_p^*(a,\phi,\beta_1,\beta_2,T)a^{\frac{n}{2}-\frac{4+2p-3\omega}{2}}\frac{\partial^{m}}{\partial a^m}\left(a^{\frac{n}{2}+\frac{4+2p-3\omega}{2}}\psi_p(a,\phi,\beta_1,\beta_2,T)\right)\\
    =&(-i)^m\int_{-\infty}^\infty d\phi\int d^2\beta\int_{0}^\infty da\;a^{\frac{n+4+2p-3\omega}{2}}a^{-(1+p)}\psi^*(a,\phi,\beta_1,\beta_2,T)\frac{\partial^{m}}{\partial a^m}\left(a^{\frac{n+4+2p-3\omega}{2}-1-p}\psi(a,\phi,\beta_1,\beta_2,T)\right)\\
    =&(-i)^m\int_{-\infty}^\infty d\phi\int d^2\beta\int_{0}^\infty da\;a^{\frac{n+2-3\omega}{2}}\psi^*(a,\phi,\beta_1,\beta_2,T)\frac{\partial^{m}}{\partial a^m}\left(a^{\frac{n+2-3\omega}{2}}\psi(a,\phi,\beta_1,\beta_2,T)\right),
\end{align}
which is independent of the ambiguity parameter. Similarly, the expectation value of $\hat{O}_2$ yields
\begin{align}
    \braket{O}_2=&\frac{1}{2}\int_{-\infty}^\infty d\phi\int d^2\beta\int_{0}^\infty da\;a^{4+2p-3\omega}\psi_p^*(a,\phi,\beta_1,\beta_2,T)\bigg[a^{n-\frac{4+2p-3\omega}{2}}\frac{\partial^{m}}{\partial a^m}\left(a^{\frac{4+2p-3\omega}{2}}\psi_p(a,\phi,\beta_1,\beta_2,T)\right)\nonumber\\
    &\qquad\qquad\qquad\qquad+a^{-\frac{4+2p-3\omega}{2}}\frac{\partial^{m}}{\partial a^m}\left(a^{n+\frac{4+2p-3\omega}{2}}\psi_p(a,\phi,\beta_1,\beta_2,T)\right)\bigg]\\
    =&\frac{1}{2}\int_{-\infty}^\infty d\phi\int d^2\beta\int_{0}^\infty da\;a^{\frac{4+2p-3\omega}{2}-1-p}\psi^*(a,\phi,\beta_1,\beta_2,T)\bigg[a^{n}\frac{\partial^{m}}{\partial a^m}\left(a^{\frac{4+2p-3\omega}{2}-1-p}\psi(a,\phi,\beta_1,\beta_2,T)\right)\nonumber\\
    &\qquad\qquad\qquad\qquad+\frac{\partial^{m}}{\partial a^m}\left(a^{n+\frac{4+2p-3\omega}{2}-1-p}\psi(a,\phi,\beta_1,\beta_2,T)\right)\bigg]\\
    =&\frac{1}{2}\int_{-\infty}^\infty d\phi\int d^2\beta\int_{0}^\infty da\;a^{\frac{2-3\omega}{2}}\psi^*(a,\phi,\beta_1,\beta_2,T)\bigg[a^{n}\frac{\partial^{m}}{\partial a^m}\left(a^{\frac{2-3\omega}{2}}\psi(a,\phi,\beta_1,\beta_2,T)\right)\nonumber\\
    &\qquad\qquad\qquad\qquad+\frac{\partial^{m}}{\partial a^m}\left(a^{n+\frac{2-3\omega}{2}}\psi(a,\phi,\beta_1,\beta_2,T)\right)\bigg],
\end{align}
which again is independent of the ambiguity parameter. Therefore, not only does the quantum cosmology model with minisuperspace of dimension $D=4$ is covariant with respect to the lapse rescaling, the observables of the quantum model are invariant under such rescalings.
%%%%%%%%%%%%%%%%%%%%%%%%%%%%%%%%%%%%%%%%%%%%%%%%%%%%%%%%%%%%%%%%%%%%%%%%%%%%%%
%\setlength{\bibsep}{0.0pt}
\bibliographystyle{utphys1}
\bibliography{reference}
\end{document}